\documentstyle[12pt,graphicx]{herm_mod}  
\begin{document}
\newfont{\ackn}{cmcsc10 scaled \magstep1}
\newfont{\aut}{cmcsc10 scaled \magstep1}
\def\ueberschrift{}
\def\autoren{}
\setcounter{chapter}{1}
\setcounter{page}{21}
\newcommand{\gam}{\mathop{\Gamma}}
\newcommand{\GB}{\raise 1pt\hbox{$\displaystyle\gam^{\lower
1pt\hbox{$\scriptstyle\leftarrow$}}$}}
\newcommand{\GF}{\raise 1pt\hbox{$\displaystyle\gam^{\lower
1pt\hbox{$\scriptstyle\rightarrow$}}$}}
\noindent
\chapter[Charge Tunneling Rates in Ultrasmall Junctions\hfill\break {\rm by} 
{\ackn G.-L.~Ingold} {\rm and} {\ackn Yu.~V.~Nazarov}]{Charge Tunneling Rates 
in Ultrasmall Junctions}
{Charge Tunneling Rates in Ultrasmall Junctions}{}
\def\autoren{G.-L.~Ingold and Yu.~V.~Nazarov}
\def\ueberschrift{Charge Tunneling Rates in Ultrasmall Junctions}
\noindent
{\halfrm GERT-LUDWIG INGOLD}

\vspace{5truept}
\noindent
{\sl Fachbereich Physik, Universit{\"a}t-GH Essen, 4300 Essen, Germany}

\vspace{12truept}
\noindent
{\halfrm and}

\vspace{12truept}
\noindent
{\halfrm YU.\ V. NAZAROV}

\vspace{5truept}
\noindent
{\sl Nuclear Physics Institute, Moscow State University, Moscow 119899 GSP,
USSR}

\vspace{0.075\textheight}
%
%
\arraycolsep 1.4pt
\section{Introduction}
%
%
\subsection{Ultrasmall tunnel junctions}
With the advances of microfabrication techniques in recent years it has become
possible to fabricate tunnel junctions of increasingly smaller dimensions and
thereby decreasing capacitance $C$. Nowadays one can study tunnel
junctions in a regime where the charging energy $E_c=e^2/2C$ is larger than the
thermal energy $k_BT$. Then charging effects play an important role and this
has been the subject of a by now large body of both theoretical and
experimental work.

To study ultrasmall tunnel junctions, that is tunnel junctions with
capacitances of $10^{-15}\,$F or less, one may either use metal-insulator-metal
tunnel junctions or constrictions in a two-dimensional electron gas formed by a
semiconductor
heterostructure. Due to the very different density of charge carriers in metals
and semiconductors the physics of metallic tunnel junctions and two-dimensional
electron gases with constrictions differ. In this chapter we restrict ourselves
to metallic tunnel junctions while Chap.~5 is devoted to single charging
effects in semiconductor nanostructures.

Metal-insulator-metal tunnel junctions produced by nanolithography are widely
studied today. Other systems where metallic tunnel junctions are present
include granular films, small metal particles embedded in oxide layers, crossed
wires, and the scanning tunneling microscope. The general setup consists
of two pieces of metal separated by a\vadjust{\break} thin insulating barrier as
shown in Fig.~1.
\begin{figure}
\begin{center}
\includegraphics{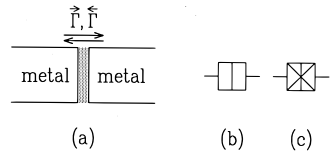}
\end{center}
\caption{
(a) Schematic drawing of a metal tunnel junction.
The arrows indicate forward and backward tunneling through the
barrier. (b) Symbol for an ultrasmall tunnel junction. The
capacitor-like shape emphasizes the role of the charging energy. (c)
Symbol for an ultrasmall superconducting tunnel junction.}
\end{figure}
The metal may either be normal or superconducting at low
temperatures. In the latter case, one may study Josephson 
junctions with
ultrasmall capacitances as well as normal tunnel junctions if a sufficiently
high magnetic field is applied.

Classically, there is no electrical transport through
the barrier and the junction will act like a capacitor of capacitance $C$. By
connecting a single junction to an external circuit it 
may be charged with a
charge $Q=CV$ where $V$ is the voltage applied to the junction. The charge $Q$
is an influence charge created by shifting the electrons in the two metal
electrodes with respect to the positive background charge. A very small shift
of the electrons will lead to a small change in $Q$. Therefore, the charge $Q$
is continuous even on the scale of the elementary charge. One finds that for
such small charges the interaction between the two charge distributions on
either side of the barrier may still be described by a charging energy
$Q^2/2C$.

Taking into account quantum effects there is a possibility of charge transport
through the barrier by tunneling of electrons indicated by the arrows in
Fig.~1. In contrast to the charge motion
in the electrodes this transport process involves discrete charges since only
electrons as an entity may tunnel. The typical change in energy for such a
process is therefore the charging energy $E_c = e^2/2C$.

This energy scale has to be compared with the other energy scale present in the
system, namely $k_BT$. Since our interest is in the discussion of charging
effects we require $e^2/2C \gg k_BT$. Otherwise thermal fluctuations will
mask these effects. As an example let us note that a junction with an
area of about $0.1\!\times\!0.1 \hbox{$\mu$m}^2$ and a typical oxide layer
thickness of 10\AA\ has a capacitance of about $10^{-15}\,{\rm F}$ corresponding
to a temperature of about $1{\rm K}$. For
decreasing capacitance which requires decreasing dimensions of the junction
this restriction for temperature becomes more relaxed.
%
%
\subsection{Voltage-biased tunnel junction}
\index{voltage bias}\index{junction!single|(}
In the following two sections we present two different pictures for the
behavior of tunnel junctions. We will not give very detailed derivations at
this point since similar calculations will be presented in subsequent sections.

Let us first consider a tunnel junction coupled to an ideal voltage source $V$.
In order to determine the current-voltage characteristic one needs to calculate
electron tunneling rates in both directions through the junction taking into
account the external bias. The tunneling process is described by a
tunneling Hamiltonian \cite{Tinkham}
\begin{equation}
H_T = \sum_{kq\sigma} T_{kq} c_{q\sigma}^{\dagger} 
c_{k\sigma}^{\phantom{\dagger}} + \mbox{H.c.}
\label{eq:I1}
\end{equation}
where the term written out explicitly describes the annihilation of an electron
with wave vector $k$ and spin $\sigma$
on the left electrode and the creation
of an electron with the same spin but wave vector $q$ on the right electrode
thereby transferring an electron from the left to the right side. The
corresponding matrix element is given by $T_{kq}$. The
Hermitian conjugate part describes the reverse process. We note that what we
call electrons here for simplicity, are really quasiparticles of the
many-electron system in the electrodes.

As a result of a golden
rule calculation treating $H_T$ as a perturbation and assuming an elastic
tunneling process one finds after some calculation for the average current
\begin{equation}
I(V) = {1\over e R_T}\int {\rm d}E \Big\{f(E)\left[1-f(E+eV)\right] -
\left[1-f(E)\right]f(E+eV)\Big\}.
\label{eq:I2}
\end{equation}
Here, $f(E) = 1/[1+\exp(\beta E)]$ is the Fermi function at inverse
temperature $\beta = 1/k_BT$. The integrand of (\ref{eq:I2}) can easily be
understood.
The first term gives the probability to find an electron with energy $E$ on the
left side and a corresponding empty state on the right side. The
difference in the Fermi energies on both sides of the barrier is
accounted for in the argument of the second Fermi function. This difference
is assumed to be fixed to $eV$ by the ideal voltage source. An analogous
interpretation may be given for the second term in the integrand which
describes the reverse process. The tunneling matrix element
and the densities of state were absorbed in $R_T$ which is proportional to
$1/\vert T_{kq}\vert^2$. The integral in (\ref{eq:I2}) will be calculated
explicitly in Sec.~3.2.\ where we will find independent of temperature
\begin{equation}
I(V) = V/R_T.
\label{eq:I3}
\end{equation}
Since this gives a current proportional to the applied voltage, the
current-voltage char\-ac\-ter\-is\-tic is of the same form as for an Ohmic
resistor. It is therefore suggestive to call $R_T$ a tunneling 
resistance.\index{tunnel resistance} We
stress, however, that the tunneling resistance should not be confused with an
Ohmic resistance because of the quite different nature of charge transport
through a tunnel junction and an Ohmic resistor. This becomes
apparent for example in the different noise spectra.\cite{LIBM88}
%
%
\subsection{Charging energy considerations}
In the previous section we have discussed a tunnel junction coupled to an ideal
voltage source. As a consequence, the charge on the junction capacitor is kept
fixed at all times. Now, we consider a different case where an ideal external
current $I$
controls the charge $Q$ on the junction. At zero temperature a tunneling
process leading from $Q$ to $Q-e$ is only possible
if the difference of charging energies before and after the
tunneling process is positive
\begin{equation}
\Delta E = {Q^2\over 2C} - {(Q-e)^2\over 2C} > 0.
\label{eq:I4}
\end{equation}
This condition is satisfied if $Q>e/2$ or the voltage across the junction $U >
U_c = e/2C$. Note, that we distinguish between the voltage $U$ across the
junction and an externally applied voltage $V$.

Assuming an ideal current-biased junction\index{current bias}, 
the following picture results.
Starting at a charge $\vert Q\vert < e/2$, the junction is charged by the
external current. At a charge $Q>e/2$ an electron may tunnel thereby decreasing
the charge on the junction below the threshold $e/2$. Then the cycle starts
again. This process which occurs with a frequency $f=I/e$ only
determined by the external bias current $I$ is called SET
oscillation.\cite{ALLOW86}\nocite{ALREV91}--\cite{SZPREP90}\index{SET
oscillation} One can show that 
the average voltage across the junction is proportional to $I^{1/2}$.

We may also use an ideal voltage source and feed a current to the junction
through a large resistor. Its resistance is assumed to be smaller than the
tunneling resistance of the junction but large enough to inhibit a fast
recharging of the capacitor after a tunneling event. According to the argument
given above
there will be no current if the external voltage is smaller than $e/2C$. Beyond
this voltage one finds at zero temperature that the average current is
determined by an
Ohmic current-voltage characteristic with resistance $R_T$ shifted in voltage
by $e/2C$. This
shift in the current-voltage characteristic is called the Coulomb 
gap\index{Coulomb gap} and the
phenomenon of suppression of the current below $U_c$ is referred to as
Coulomb
blockade. For the energy
consideration presented in this section it was important that the charge on the
capacitor is well defined and continuous even on the scale of an elementary
charge. Only
a junction charge less than $e/2$ together with the fact that tunneling always
changes this charge by $e$ gave rise to the possibility of a Coulomb gap.
%
%
\subsection{Local and global view of a single tunnel 
junction}\index{junction!single}
Comparing the discussions in the two previous sections we find that there are
different energy differences associated with the tunneling process, namely
$eV$ and $E_c$. As we will
argue now these two cases can be viewed as a local and a global description
\cite{ALREV91}--\cite{GSEPL89} of
a single tunnel junction\index{junction!single} coupled to an external 
circuit, at least at zero
temperature. In Sec.~1.3.\ we used the energy difference (\ref{eq:I4}) which
gives the
difference in charging energy of the junction before and immediately after the
tunneling process. This is called the local view since it only considers the
junction through which the electron is tunneling and ignores its interaction
with the rest of the world.

In contrast, in Sec.~1.2.\ the energy changes in the circuit were viewed
globally. After the tunneling process a nonequilibrium situation occurs since
the charge $Q-e$ on the junction and the charge $Q = CV$ imposed by the voltage
source are different. To reestablish equilibrium the voltage source transfers
an electron and recharges the junction capacitor to the charge $Q$. In the end
there is no change in charging energy. However, the work done by the voltage
source which amounts to $eV$ has to be taken into account. This is indeed the
case in (\ref{eq:I2}) where $eV$ appears as the difference between the Fermi
energies of the two electrodes.

Now, the question arises which one of these two descriptions, if any, is
correct. This problem cannot be solved by treating the single 
junction\index{junction!single} as
decoupled from the rest of the world or by replacing its surroundings by ideal
current or voltage sources. The discussion of the local and global view rather
suggests
that one has to consider the junction embedded in the electrical 
circuit.\cite{YuliT4}--\cite{Girvin} A
junction coupled to an ideal voltage source, for example, will always behave
according to the global rule since after the tunneling process charge
equilibrium is reestablished immediately. This is not the case if the voltage
source is attached to the junction with a large resistor in series. In the
sequel we will discuss the influence of the electrodynamic environment on
electron tunneling rates in ultrasmall tunnel junctions. We will learn
that the local and global rules are just limiting cases of the so-called
orthodox theory\index{orthodox theory} and find out to what
kind of circuit we have to couple the junction in order to observe
Coulomb blockade phenomena.
%
%
\section{Description of the environment}\index{environment!electromagnetic|(}
%
%
\subsection{Classical charge relaxation}
In this section we discuss the coupling of a tunnel junction to the external
circuit classically. That means that we forget about tunneling for the moment
and consider the junction just as a capacitor of capacitance $C$ carrying the
charge $Q = CU$ where $U$ is the voltage across the junction. The junction is
attached to the external circuit which we describe by its impedance
\begin{equation}
Z(\omega) = {V(\omega)\over I(\omega)}.
\label{eq:I5}
\end{equation}
The impedance gives the ratio between an alternating voltage of
frequency $\omega$ applied to the circuit and the
current which then is flowing through it if the junction capacitor is replaced
by a short. The external circuit shall contain an ideal dc voltage source $V$
in series with the impedance as
shown in Fig.~2.
\begin{figure}
\begin{center}
\includegraphics{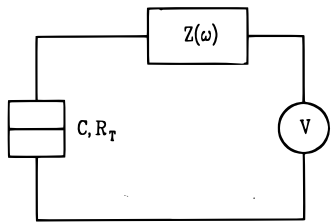}
\end{center}
\caption{
An ultrasmall tunnel junction with capacitance $C$
and tunneling resistance $R_T$ coupled to a voltage source $V$ via the
external impedance $Z(\omega)$.}
\end{figure}
As discussed in Chap.~1 and \cite{DEGPRL90,multizphys},
the assumption of a voltage source is reasonable even if in a
real experiment a current source is used. Generally, the leads attached to the
junction generate a capacitance which is so large compared to the junction
capacitance that a current source will charge this large capacitor which
then acts as an effective voltage source.

In equilibrium the average charge on the junction $Q_e=CV$ is determined by the
external voltage source. Let us assume now that at some initial time the
equilibrium is disturbed and the charge on the junction is $Q_0$. In the
following we will derive the relaxation from this initial condition back to
equilibrium. The information on the charge relaxation which depends only on the
external impedance and the junction capacitance will later be needed
to describe the influence of the environment.

To solve this initial
value problem it is convenient to work in the Laplace space. Then the Laplace
transform of the voltage across the impedance follows from (\ref{eq:I5}) as
\begin{equation}
\hat V(p) = \hat Z(p) \hat I(p)
\label{eq:I6}
\end{equation}
where the hat denotes the Laplace transform and
\begin{equation}
\hat Z(p) = Z(-ip).
\label{eq:I7}
\end{equation}
Applying the rule for the Laplace
transformation of a derivative, one finds for the current in terms of the charge
\begin{equation}
\hat I(p) = p\hat Q(p) - Q_0.
\label{eq:I8}
\end{equation}
Since the Laplace transform of the constant external voltage $V$ is $V/p$, we
get for the voltage balance in the circuit
\begin{equation}
{Q_e\over pC} = {\hat Q(p)\over C} + \hat Z(p)\left(p\hat Q(p) - Q_0\right).
\label{eq:I9}
\end{equation}
Solving this equation for $\hat Q(p)$, doing the inverse Laplace
transformation,
and rewriting the final result in terms of the original impedance $Z(\omega)$
we get for the relaxation of the charge
\begin{equation}
Q(t) = Q_e + (Q_0 - Q_e) R(t).
\label{eq:I10}
\end{equation}
Here, the Fourier transform of the charge relaxation function
\begin{equation}
\int_0^{\infty}{\rm d}t e^{-i\omega t}R(t) = CZ_t(\omega)
\label{eq:I11}
\end{equation}
is related to the total impedance
\begin{equation}
Z_t(\omega) = {1\over i\omega C + Z^{-1}(\omega)}
\label{eq:I12}
\end{equation}
of the circuit consisting of the capacitance $C$ in parallel with the external
impedance $Z(\omega)$. It will become clear in Sec.~6.2.\ that $Z_t(\omega)$ is
the effective impedance of the circuit as seen from the tunnel junction.
%
%
\subsection[Quantum mechanics of an $LC$-circuit]
{Quantum mechanics of an {\protect\boldmath $LC$}-circuit}
In the previous section we have found that the classical relaxation of charge
in a circuit can be described in terms of its impedance. We will now make a
first step towards the quantum mechanical treatment of a tunnel junction
coupled to an external circuit by discussing the most simplest case where the
environmental impedance of the circuit shown in Fig.~2 is given by an
inductance $Z(\omega)=i\omega L$. In the next section this will turn out to be
the fundamental building block for a general description of the environment.

For the following it is convenient to introduce the phase \cite{SPRB85}
\begin{equation}
\varphi(t) = {e\over \hbar}\int_{-\infty}^t {\rm d}t^{\prime} U(t^{\prime})
\label{eq:phase}
\end{equation}
where $U=Q/C$ is the voltage across the junction. The definition
(\ref{eq:phase}) becomes the Josephson relation for a superconducting tunnel
junction if we replace the electron charge $e$ by 
the charge of Cooper pairs
$2e$. In the superconducting case this phase is of course of great importance
as the phase of the order parameter.

To derive the Hamiltonian of a voltage-biased\index{voltage bias} $LC$-circuit 
let us first write
down the Lagrangian
\begin{equation}
{\cal L} = \frac{C}{2}\left(\frac{\hbar}{e}\dot\varphi\right)^2 -
\frac{1}{2L}\left(\frac{\hbar}{e}\right)^2\left(\varphi-\frac{e}{\hbar}
Vt\right)^2.
\label{eq:Lagr}
\end{equation}
The first term represents the charging energy of the capacitor which can easily
be verified by means of the definition of the phase (\ref{eq:phase}). The
magnetic
field energy of the inductor is given by the second term since up to a factor
$\hbar/e$ the flux through the inductor is given by the phase difference across
the inductor. The latter relation is obtained from the requirement that the
phase
differences at the capacitor and inductor should add up to the phase difference
$(e/\hbar)Vt$ produced by the voltage source according to (\ref{eq:phase}).

Switching to the Hamilton formalism we find that the charge $Q$ on the junction
is the conjugate variable to $(\hbar/e)\varphi$. In a quantum
mechanical description this results in the commutation relation
\begin{equation}
[\varphi,Q]=ie.
\label{eq:comm}
\end{equation}
Now, the phase $\varphi$, the voltage $U$ across the junction, and the charge
$Q$ are operators. Note that there is no problem with phase periodicity when
constructing the phase operator $\varphi$. Due to the continuous charge $Q$ the
spectrum of the phase operator is continuous on the interval from $-\infty$ to
$+\infty$.

From the Lagrangian (\ref{eq:Lagr}) we immediately get the Hamiltonian
\begin{equation}
H= \frac{Q^2}{2C} + \frac{1}{2L} \left(\frac{\hbar}{e}\right)^2 \left(\varphi -
\frac{e}{\hbar}Vt\right)^2.
\label{eq:HLC1}
\end{equation}
According to the equations of motion derived either from (\ref{eq:Lagr}) or
(\ref{eq:HLC1}) one finds that the average phase evolves in time like
$(e/\hbar)Vt$. The average charge on the capacitor is given by $CV$.
It is therefore convenient
to introduce the variables
\begin{equation}
\tilde\varphi(t) = \varphi(t) - \frac{e}{\hbar}Vt
\label{eq:phasfluk}
\end{equation}
and
\begin{equation}
\tilde Q = Q -CV
\label{eq:ladfluk}
\end{equation}
describing the fluctuations around the mean value determined by the external
voltage.\vadjust{\break} The commutator between the new variables is again
\begin{equation}
[\tilde\varphi,\tilde Q] = ie.
\label{eq:commtilde}
\end{equation}
Substituting $\varphi-(e/\hbar)Vt$ by $\tilde\varphi$ in the Hamiltonian
(\ref{eq:HLC1}) amounts to going into a ``rotating reference frame''. This
transformation results in an extra contribution $-(i/\hbar)QV$ to the time
derivative in the time-dependent Schr\"odinger equation. We thus obtain, up to
a term depending only on the external voltage,
\begin{equation}
H = \frac{\tilde Q^2}{2C} + \frac{1}{2L} \left( \frac{\hbar}{e}\tilde\varphi
\right)^2
\label{eq:HLC2}
\end{equation}
which demonstrates the equivalence between an $LC$-circuit and a harmonic
oscillator. Note that the influence of an external voltage is entirely
accounted for by the definitions (\ref{eq:phasfluk}) and (\ref{eq:ladfluk}).
%
%
\subsection{Hamiltonian of the environment}
The special environment of the previous section did not give rise to
dissipation so that there were no problems in writing down a Hamiltonian. On
the other hand, in general, an impedance $Z(\omega)$ will introduce
dissipation. At first sight, this is in contradiction to a Hamiltonian
description of the environment. However, after realizing that dissipation
arises from the coupling of the
degrees of freedom $Q$ and $\varphi$, in which we are interested, to other
degrees of freedom our task is not as hopeless anymore. We will introduce a
Hamiltonian for the system coupled to the environment which, after
elimination of the environmental degrees of freedom, describes a dissipative
system. One approach would be to start from a microscopic model. This will be
discussed in the appendix. Here, we represent the environment by a
set of harmonic oscillators which are bilinearly coupled to $\varphi$ and which
may be viewed as $LC$-circuits. These harmonic oscillators may in some cases be
justified microscopically. In most cases, however, this representation of the
environment is
introduced phenomenologically. It then has to fulfill the requirement that in
the classical limit the reduced dynamics is described correctly. We now can
write down the Hamiltonian for the environmental coupling
\begin{equation}
H_{\rm env} = {\tilde Q^2\over 2C} + \sum_{n=1}^N \left[{q_n^2\over 2C_n} +
\left({\hbar\over e}\right)^2 {1\over 2L_n}(\tilde\varphi-\varphi_n)^2\right]
\label{eq:I16}
\end{equation}
which is expressed in terms of the variables $\tilde\varphi$ and $\tilde Q$
defined in (\ref{eq:phasfluk}) and (\ref{eq:ladfluk}) thereby accounting for
an external voltage source.
The first term describes the charging energy of the junction
capacitor. In the second term we sum over the environmental degrees of freedom
represented by harmonic oscillators of frequency
$\omega_n=1/\sqrt{L_nC_n}$ which are bilinearly coupled
to the phase of the tunnel junction. In order to describe an effectively
dissipative environment the number $N$ of environmental degrees of freedom has
to be rather large. Usually, in practice the limit $N\to\infty$ has to be
performed. The model Hamiltonian (\ref{eq:I16}) is not new. Hamiltonians of
this form have been used in quantum optics for several decades.\cite{Haken} 
More recently,
Caldeira and Leggett \cite{CLANN83} introduced this description of the
environment in the context of macroscopic quantum tunneling.

To derive the reduced dynamics of the system described by (\ref{eq:I16}) we
write down the Heisenberg equations
of motion for the operators $\tilde Q, \tilde\varphi, q_n,$ and $\varphi_n$. It
is easy to solve
for $q_n$ and $\varphi_n$ by considering $\tilde\varphi$ as a given function of
time. After substituting the result into the equations of motion for $\tilde Q$
and $\tilde\varphi$ and solving these we obtain after a partial integration
\begin{equation}
\dot{\tilde Q}(t) + {1\over C}\int_0^t {\rm d}s\, Y(t-s) \tilde Q(s) = I_N(t).
\label{eq:I17}
\end{equation}
Here,
\begin{equation}
Y(t) = \sum_{n=1}^N {1\over L_n} \cos (\omega_n t).
\label{eq:I18}
\end{equation}
Note that an arbitrary function $Y(t)$ can be described in this way by an
adequate choice of the model
parameters $L_n$ and $C_n$. In general, for a given $Y(t)$ the sum has to be
replaced by an integral over a continuous distribution of harmonic
oscillators. The Fourier transform of $Y(t)$ is the admittance
$Y(\omega)=1/Z(\omega)$. The inhomogeneity $I_N(t)$ in (\ref{eq:I17})
is the quantum mechanical noise current and depends on the initial
conditions at $t=0$.
By Laplace transforming
the left-hand side of (\ref{eq:I17}) we recover (\ref{eq:I9}) which describes the
classical relaxation of the junction charge according to
the total impedance $Z_t(\omega)$ introduced in (\ref{eq:I12}). Therefore, the
Hamiltonian (\ref{eq:I16}) gives us an equivalent description of the environment
which
enables us to treat a tunnel junction coupled to
the external circuit quantum mechanically.

Sometimes it is useful to use a mechanical analogue of the model presented
above. The correspondence between the electrical and mechanical
quantities is given in table I.
\begin{table}[bt]
\caption{Correspondence between electrical and mechanical quantities}
\begin{center}
\begin{tabular}{ll}
\hline
\vrule width 0pt height 12pt depth 6pt
Electrical quantity              & Mechanical quantity  \\ \hline
\vrule width 0pt height 12pt depth 6pt
charge $Q$                       & momentum $p$         \\
\vrule width 0pt height 12pt depth 6pt
voltage $U=Q/C$                  & velocity $v=p/M$     \\
\vrule width 0pt height 12pt depth 6pt
capacitance $C$                  & mass $M$             \\
\vrule width 0pt height 12pt depth 6pt
phase $\varphi$                  & coordinate $x$       \\
\vrule width 0pt height 12pt depth 6pt
$[\varphi,Q] = ie$               & $[x,p] = i\hbar$     \\
\vrule width 0pt height 12pt depth 6pt
inductance $L$                   & spring constant $k$  \\
\vrule width 0pt height 12pt depth 6pt
$LC$-circuit                     & harmonic oscillator  \\ \hline
\end{tabular}
\end{center}
\end{table}
At zero bias the Hamiltonian (\ref{eq:I16}) may then be interpreted as
describing a free particle coupled to
$N$ harmonic oscillators forming the heat bath and (\ref{eq:I17}) is indeed the
equation of motion describing such a system.
%
%
\section{Electron tunneling rates for single tunnel 
junctions}\index{junction!single}
%
%
\subsection{Tunneling Hamiltonian}
In the previous section we have treated the tunnel junction as a capacitor
thereby neglecting the fact that electrons may tunnel through the junction. We
will now include tunneling to allow for a current through the junction. The
quasiparticles in the two metal electrodes are described by the Hamiltonian
\begin{equation}
H_{\rm qp} = \sum_{k\sigma} \epsilon_k c_{k\sigma}^{\dagger}
c_{k\sigma}^{\phantom{\dagger}}
+\sum_{q\sigma} \epsilon_q c_{q\sigma}^{\dagger} 
c_{q\sigma}^{\phantom{\dagger}}
\label{eq:Hqp1}
\end{equation}
where the first and second sum correspond to the left and right electrode,
respectively. $\epsilon_k$ and $\epsilon_q$ are the energies of quasiparticles
with wave vector $k$ and $q$ while $\sigma$ denotes their spin.

Tunneling is introduced by the Hamiltonian \cite{DEGPRL90,Odintsov,AveOdin}
\begin{equation}
H_T =\sum_{kq\sigma} T_{kq} c_{q\sigma}^{\dagger} 
c_{k\sigma}^{\phantom{\dagger}} e^{-i\varphi} + \mbox{H.c.}
\label{eq:HT1}
\end{equation}
This is the tunneling Hamiltonian (\ref{eq:I1}) presented in Sec.~1.2. apart
from the operator $\exp(-i\varphi)$. Using the mechanical analogue of Sec.~2.3.
the latter operator would correspond to a momentum shift operator.
Indeed according to
\begin{equation}
e^{i\varphi}Qe^{-i\varphi} = Q-e
\label{eq:I20}
\end{equation}
which follows from the commutator (\ref{eq:comm})
this new operator acts as a ``translation'' operator changing the charge
on the junction by an elementary charge $e$. In the Hamiltonian (\ref{eq:HT1})
we use
operators $c^{\dagger}$ and $c$ representing quasiparticles and in addition the
phase $\varphi$ which is conjugate to the charge $Q$. These operators may be
expressed in terms of true electron creation and annihilation operators.
In the following, we will assume that the quasiparticle operators commute with
the charge and
phase operators since a large number of quasiparticle states contribute to
these operators and the contribution of a single state is negligible. This is
closely related to the assumption of linearity of the electrodynamics
in the system under consideration. Although
quasiparticle and charge operators commute, the tunneling Hamiltonian
(\ref{eq:HT1}) now
establishes a coupling between the tunneling electron and the environment which
``sees'' the junction charge. It
will be shown below that this coupling makes the current-voltage 
characteristic of the junction nonlinear.

As in Sec.~2.2. we may introduce the phase $\tilde\varphi$ defined in
(\ref{eq:phasfluk}) into $H_T$. This will help to clarify the relation between
the two tunneling Hamiltonians (\ref{eq:I1}) and (\ref{eq:HT1}). Exploiting the
relation
\begin{equation}
e^{-i\alpha c^{\dagger}c} c e^{i\alpha c^{\dagger}c} = c e^{i\alpha}
\label{eq:Fermi1}
\end{equation}
we perform a time-dependent unitary transformation with

\break
\begin{equation}
U = \prod_{k\sigma}\exp\left[i\frac{e}{\hbar}Vtc_{k\sigma}^{\dagger}
c_{k\sigma}^{\phantom{\dagger}}\right].
\label{eq:unit1}
\end{equation}
The new tunneling Hamiltonian then reads
\begin{equation}
\tilde H_T = U^{\dagger}H_TU = \sum_{kq\sigma}T_{kq}c_{q\sigma}^{\dagger} 
c_{k\sigma}^{\phantom{\dagger}} e^{-i\tilde\varphi} + \mbox{H.c.}
\label{eq:HT2}
\end{equation}
Since the transformation (\ref{eq:unit1}) is time-dependent is also shifts
quasiparticle energies on the left electrode and we obtain from (\ref{eq:Hqp1})
\begin{eqnarray}
\tilde H_{\rm qp} &=& U^{\dagger} H_{\rm qp} U - i\hbar U^{\dagger}
\frac{\partial}{\partial t} U\nonumber\\
&=& \sum_{k\sigma}(\epsilon_k+eV) c_{k\sigma}^{\dagger} 
c_{k\sigma}^{\phantom{\dagger}} +
\sum_{q\sigma}\epsilon_q c_{q\sigma}^{\dagger} c_{q\sigma}^{\phantom{\dagger}}.
\label{eq:Hqp2}
\end{eqnarray}

In the absence of an environment, the operator $\exp(-i\tilde\varphi)$ in
(\ref{eq:HT2}) has no effect on the tunneling process. The Hamiltonian
(\ref{eq:HT2}) then becomes identical to the Hamiltonian (\ref{eq:I1}). The
phase factor $\exp[-(i/\hbar)eVt]$ which was present in (\ref{eq:HT1}) has
vanished. Instead, the energy levels on the left and right electrodes are
shifted by $eV$ relative to each other. This shift was taken into account in
the result (\ref{eq:I2}).

In the following we will use the tunneling Hamiltonian
in the form (\ref{eq:HT2}). Before starting with the calculation let us collect
the Hamiltonians describing the whole system. The total Hamiltonian
\begin{equation}
H = \tilde H_{\rm qp} + H_{\rm env} + \tilde H_T
\label{eq:Htot}
\end{equation}
contains the contributions of the quasiparticle Hamiltonian (\ref{eq:Hqp2}) for
the two electrodes, the Hamiltonian (\ref{eq:I16}) describing the environment
including the charge degree of freedom, and finally the tunneling Hamiltonian
(\ref{eq:HT2}) which couples the first two parts.
%
%
\subsection{Calculation of tunneling rates}
\subsubsection{Perturbation theory}
Starting from the Hamiltonian (\ref{eq:Htot}) we now calculate rates for
tunneling through
the junction. First we make two important assumptions. The tunneling
resistance $R_T$ shall be large compared to the resistance quantum $R_K=h/e^2$
which is a natural resistance scale (for Josephson 
junctions one often uses
$R_Q=h/4e^2$ to account for the charge $2e$ of Cooper pairs). Since $R_T$ is
inversely proportional to
the square of the tunneling matrix element, this implies that the states on the
two electrodes only mix very weakly so that the Hamiltonian (\ref{eq:Hqp2}) is
a good description of the quasiparticles in the electrodes. We then may
consider the tunneling Hamiltonian $\tilde H_T$ as a perturbation.
Here, we will restrict ourselves to the
leading order, i.e.\ we calculate the tunneling rate within the golden rule
approximation. We further assume that charge equilibrium is established before
a tunneling event occurs. This defines the states to be used in the
perturbation theoretical calculation as equilibrium states. On the other hand,
it means that the time
between two tunneling processes should be larger than the charge relaxation
time.

As mentioned above we will calculate the tunneling rates by means of the golden
rule
\begin{equation}
\Gamma_{i\to f} = {2\pi\over\hbar}\vert \langle f\vert \tilde H_T\vert
i\rangle\vert^2 \delta(E_i-E_f)
\label{eq:I22}
\end{equation}
which gives the rate for transitions between the initial state $\vert i\rangle$
and the final state $\vert f\rangle$. In the absence of the tunneling
Hamiltonian we may write the total state as a product of a quasiparticle state
and a charge state which in the following we call reservoir state because it is
connected with the coupling to the environment. Specifically, we set $\vert
i\rangle = \vert E\rangle \vert R\rangle$ and $\vert f\rangle = \vert E'\rangle
\vert R'\rangle$ where $\vert E\rangle$ and $\vert E'\rangle$ are quasiparticle
states of respective energy and $\vert R\rangle$ and $\vert R'\rangle$ are
reservoir states with energies $E_R$ and $E_R^{\prime}$. The matrix element in
(\ref{eq:I22}) then becomes
\begin{equation}
\langle f\vert \tilde H_T\vert i\rangle = \langle E'\vert H_T^e\vert E\rangle
\langle R'\vert e^{-i\tilde\varphi}\vert R\rangle + \langle E'\vert
{H_T^e}^{\dagger}\vert E\rangle \langle R'\vert e^{i\tilde\varphi}\vert
R\rangle
\label{eq:matfac}
\end{equation}
with
\begin{equation}
H_T^e = \sum_{kq\sigma} T_{kq} c_{q\sigma}^{\dagger} 
c_{k\sigma}^{\phantom{\dagger}}
\label{eq:I24}
\end{equation}
being the part of the tunneling Hamiltonian acting in the quasiparticle space.

To calculate the total rate for electron tunneling from left to right we have
to sum over all initial states weighted with the probability to find these
states and over all final states. We thus have to evaluate
\begin{eqnarray}
\GF(V) &=& \frac{2\pi}{\hbar}\int_{-\infty}^{+\infty} {\rm d}E{\rm d}E'
\sum_{R,R'}
\vert\langle E'\vert H_T^e\vert E\rangle\vert^2 \,\vert\langle R'\vert
e^{-i\tilde\varphi}\vert R\rangle\vert^2\nonumber\\
&&\hspace{5cm}\times P_{\beta}(E)P_{\beta}(R)\delta(E+eV+E_R-E'-E_R^{\prime}).
\label{eq:rate1}
\end{eqnarray}

Let us consider one term $T_{kq}c_{q\sigma}^{\dagger}
c_{k\sigma}^{\phantom{\dagger}}$ contained in
$H_T^e$. The only possible states with nonvanishing matrix element $\langle
E'\vert c_{q\sigma}^{\dagger}
c_{k\sigma}^{\phantom{\dagger}}\vert E\rangle$ are $\vert E\rangle =
\vert\ldots,1_{k\sigma},\ldots,0_{q\sigma},\ldots\rangle$ and $\vert E'\rangle
= \vert\ldots,0_{k\sigma},\ldots,1_{q\sigma},\ldots\rangle$ where this notation
means that
in $\vert E\rangle$ a quasiparticle with wave vector $k$ and spin $\sigma$ on
the left side of the barrier is
present while the state with wave vector $q$ and spin $\sigma$ on the right
side is unoccupied.
The occupation of states with other quantum numbers is arbitrary. Since
$P_{\beta}(E)$ factorizes we then obtain
\begin{eqnarray}
\GF(V) &=&\frac{2\pi}{\hbar}\int_{-\infty}^{+\infty} {\rm d}\epsilon_k {\rm
d}\epsilon_q \sum_{kq\sigma} \vert T_{kq}\vert^2 f(\epsilon_k)
[1-f(\epsilon_q)]\nonumber\\
&&\hspace{2.7cm}\times\sum_{R,R'}\vert\langle R'\vert e^{-i\tilde\varphi}\vert
R\rangle\vert^2
P_{\beta}(R)\, \delta(\epsilon_k+eV+E_R-\epsilon_q-E_R^{\prime}).
\label{eq:rate2}
\end{eqnarray}
Here, the probability to find the initial state is given by the product of
Fermi functions and $\epsilon_k+eV-\epsilon_q$ is the difference of
quasiparticle energies associated with the tunneling process since the
occupation of the other states remains unchanged. Note that $eV$ does not
appear in $f(\epsilon_k)$ since the Fermi level on the left side is shifted by
this amount. If the applied voltage is such that $eV$ is much smaller
than the Fermi energy we may assume\vadjust{\break} that all quasiparticle states involved have
energies close to the Fermi energy. Taking the tunneling matrix element to be
approximately independent of $\epsilon_k$ and $\epsilon_q$ we may replace
$\sum_{kq\sigma}\vert T_{kq}\vert^2$ by an averaged matrix element $\vert
T\vert^2$ which also accounts for the density of states at the Fermi energy.
We collect all constant terms in the tunneling resistance $R_T$.\index{tunnel
resistance} The rate
expression
(\ref{eq:rate2}) then becomes
\begin{eqnarray}
\GF(V) &=&
{1\over e^2 R_T} \int_{-\infty}^{+\infty}{\rm d}E{\rm d}E^{\prime}
f(E) [1-f(E^{\prime})]\nonumber\\
&&\hspace{2.7cm}\times\sum_{R,R^{\prime}}\vert\langle R\vert
e^{-i\tilde\varphi}\vert
R^{\prime}\rangle\vert ^2 P_{\beta}(R)\,
\delta(E+eV+E_R-E^{\prime}-E_R^{\prime})
\label{eq:I26}
\end{eqnarray}
where we have renamed the energies $\epsilon_k$ and $\epsilon_q$ into $E$ and
$E'$. The justification for calling $R_T$ a tunneling resistance will be given
below when we calculate current-voltage characteristics.
\subsubsection{Tracing out environmental states}
We now have to do the sum over $R$ and $R'$ in (\ref{eq:I26}).
The probability of finding the initial reservoir state $\vert R\rangle$ is
given by the corresponding matrix element
\begin{equation}
P_{\beta}(R) = \langle R\vert \rho_{\beta}\vert R\rangle
\label{eq:I28}
\end{equation}
of the equilibrium density matrix
\begin{equation}
\rho_{\beta} = Z_{\beta}^{-1} \exp(-\beta H_{\rm env})
\label{eq:I29}
\end{equation}
of the reservoir at inverse
temperature $\beta$. Here,
\begin{equation}
Z_{\beta} = \hbox{Tr}\Big\{\exp(-\beta H_{\rm env})\Big\}
\label{eq:I30}
\end{equation}
is the partition function of the environment. To proceed, it is useful to
rewrite the delta function in (\ref{eq:I26}) in terms of its Fourier transform
\begin{eqnarray}
\delta(E+eV+E_R-E^{\prime}-E_R^{\prime})&& \nonumber\\ 
&&\hspace{-2cm}={1\over 2\pi\hbar}
\int_{-\infty}^{+\infty} {\rm d}t \exp\left({i\over\hbar}
(E+eV+E_R-E^{\prime}-E_R^{\prime})t\right)
\label{eq:I31}
\end{eqnarray}
and to use the part containing the reservoir energies to introduce the time
dependent phase operator in the Heisenberg picture. We thus obtain
\begin{eqnarray}
\GF(V) &=&
{1\over e^2 R_T} \int_{-\infty}^{+\infty}{\rm d}E{\rm d}E^{\prime}\,
\int_{-\infty}^{+\infty}{{\rm d}t\over 2\pi\hbar}
\exp\left({i\over\hbar}(E-E^{\prime}+eV)t\right)f(E)
[1-f(E^{\prime})]\nonumber\\
&&\hspace{4.4cm}\times\sum_{R,R^{\prime}}P_{\beta}(R)\langle R\vert
e^{i\tilde\varphi(t)}\vert
R^{\prime}\rangle\langle R^{\prime}\vert e^{-i\tilde\varphi(0)}\vert R\rangle.
\label{eq:I32}
\end{eqnarray}
Since the reservoir states form a complete set we can do the sum over
$R^{\prime}$. Together with the definition (\ref{eq:I28}) of $P_{\beta}(R)$ we
find that
the reservoir part in the rate formula is given by the equilibrium correlation
function \begin{eqnarray}
\langle e^{i\tilde\varphi(t)} e^{-i\tilde\varphi(0)}\rangle &=&
\sum_R \langle R\vert e^{i\tilde\varphi(t)} e^{-i\tilde\varphi(0)}\vert R\rangle
P_{\beta}(R)\nonumber\\
&=& {1\over Z_{\beta}} \sum_R \langle R\vert e^{i\tilde\varphi(t)}
e^{-i\tilde\varphi(0)}
e^{-\beta H_{\rm env}}\vert R\rangle
\label{eq:I33}
\end{eqnarray}
so that we get from (\ref{eq:I32})
\begin{eqnarray}
\GF(V) &=&
{1\over e^2 R_T} \int_{-\infty}^{+\infty}{\rm d}E{\rm d}E^{\prime}
f(E)[1-f(E^{\prime})]\nonumber\\
&&\hspace{3.4cm}\times\int_{-\infty}^{+\infty}{{\rm d}t\over 2\pi\hbar}
\exp\left({i\over\hbar}(E-E^{\prime}+eV)t\right)
\langle e^{i\tilde\varphi(t)} e^{-i\tilde\varphi(0)}\rangle.
\label{eq:rate3}
\end{eqnarray}
\subsubsection{Phase-phase correlation function}
The correlation function defined in (\ref{eq:I33}) may be simplified.
According to (\ref{eq:I28}) the
probability $P_{\beta}(R)$ of the unperturbed reservoir
is given by the equilibrium density matrix of the environmental Hamiltonian
(\ref{eq:I16}). Since this Hamiltonian is harmonic, the equilibrium density
matrix in the $\tilde\varphi$-representation is a Gaussian
and therefore determined only by its first and second moments. Hence, it should
be possible to express the correlation function (\ref{eq:I33}) in terms of phase
correlation
functions of at most second order. This goal may be achieved by exploiting the
generalized Wick theorem for equilibrium correlation functions \cite{Louisell}
\begin{eqnarray}
\langle \psi_1 \psi_2 \ldots \psi_n\rangle &=& \langle \psi_1
\psi_2\rangle
\langle \psi_3 \psi_4 \ldots \psi_n\rangle + \langle \psi_1 \psi_3\rangle
\langle \psi_2 \psi_4 \ldots \psi_n\rangle + \ldots \nonumber\\
&&\hspace{5cm}+ \langle \psi_1 \psi_n
\rangle \langle \psi_2 \psi_3 \ldots \psi_{n-1}\rangle.
\label{eq:I34}
\end{eqnarray}
This theorem applies if the Hamiltonian of the system for which the thermal
average is performed may be represented in terms of independent harmonic
oscillators and if the operators $\psi_i$ are linear combinations of creation
and annihilation operators. The first condition is fulfilled since the
Hamiltonian (\ref{eq:I16}) may in principle be diagonalized. Due to the
linearity of the
equation of motion, $\tilde\varphi(t)$ is a linear combination of creation and
annihilation operators and thus also the second condition holds. After
expanding the exponentials on the right hand side of
\begin{equation}
{{\rm d}\over {\rm d}\alpha} \langle e^{i\alpha\tilde\varphi(t)}
e^{-i\alpha\tilde\varphi(0)}\rangle
= i\left[\langle \tilde\varphi(t) e^{i\alpha\tilde\varphi(t)}
e^{-i\alpha\tilde\varphi(0)}\rangle - \langle e^{i\alpha\tilde\varphi(t)}
\tilde\varphi(0) e^{-i\alpha\tilde\varphi(0)}\rangle\right]
\label{eq:I35}
\end{equation}
we may apply the generalized Wick theorem. The resulting sums may again be
expressed in terms of exponentials and we find
\begin{equation}
{{\rm d}\over {\rm d}\alpha} \langle e^{i\alpha\tilde\varphi(t)}
e^{-i\alpha\tilde\varphi(0)}\rangle = 2\alpha \langle
[\tilde\varphi(t)-\tilde\varphi(0)]\tilde\varphi(0)\rangle \langle
e^{i\alpha\tilde\varphi(t)}
e^{-i\alpha\tilde\varphi(0)}\rangle
\label{eq:I36}
\end{equation}
where we made use of $\langle\tilde\varphi(t)^2\rangle = \langle
\tilde\varphi(0)^2\rangle$
which is a consequence of the stationarity of\vadjust{\break} 
equilibrium correlation
functions. The differential equation (\ref{eq:I36}) may easily be
solved and we obtain with the correct initial condition at $\alpha = 0$ the
result for $\alpha = 1$
\begin{equation}
\langle e^{i\tilde\varphi(t)} e^{-i\tilde\varphi(0)}\rangle = e^{\langle
[\tilde\varphi(t)-\tilde\varphi(0)]\tilde\varphi(0)\rangle}.
\label{eq:I37}
\end{equation}
For later convenience we introduce the abbreviation
\begin{equation}
J(t) = \langle [\tilde\varphi(t)-\tilde\varphi(0)]\tilde\varphi(0)\rangle
\label{eq:Jt}
\end{equation}
for the
phase-phase correlation function.
In view of (\ref{eq:rate3}) it is useful to introduce the Fourier transform
of the correlation function (\ref{eq:I37})
\begin{equation}
P(E) = {1\over 2\pi\hbar}\int_{-\infty}^{+\infty}{\rm d}t \exp\left[J(t)
+ {i\over\hbar}E t\right].
\label{eq:I38}
\end{equation}
\subsubsection{Tunneling rate formula}
Using the definition of $P(E)$ we may
now rewrite the expression (\ref{eq:rate3}) for the forward tunneling rate
in the form
\begin{equation}
\GF(V) = {1\over e^2R_T} \int_{-\infty}^{+\infty}{\rm d}E{\rm d}E^{\prime}\,
f(E)[1-f(E^{\prime}+eV)] P(E-E^{\prime})
\label{eq:I39}
\end{equation}
which allows for a simple physical interpretation. As already pointed
out in Sec.~1.2.\ the Fermi functions describe the probability of finding an
occupied state on one side and an empty state on the other side of the barrier.
The difference in the Fermi energies due to the applied voltage is taken into
account in the argument of the second Fermi function. In the discussion of
Sec.~1.2.\ we had assumed an ideal voltage bias and no environmental modes
were present. Therefore, the energy conservation condition in the golden rule
(\ref{eq:I22})
applied directly to the tunneling electron. The expression (\ref{eq:I39}) is
more general
and takes into account the possibility of energy exchange between the tunneling
electron and the environment. We may interpret $P(E)$ as the probability to
emit the energy $E$ to the external circuit. Correspondingly, $P(E)$
for
negative energies describes the absorption of energy by the tunneling electron.

To further simplify (\ref{eq:I39}) we first calculate the integral over Fermi
functions
\begin{equation}
g(x) = \int_{-\infty}^{+\infty}{\rm d}E\, [f(E)-f(E+x)]
\label{eq:I40}
\end{equation}
which will also be of use later on.
The derivative of $g$ with respect to $x$ can easily be evaluated
yielding ${\rm d}g(x)/{\rm d}x = f(-\infty)-f(\infty)=1$.
Integration with the initial condition $g(0)=0$ then gives the formula
\begin{equation}
\int_{-\infty}^{+\infty}{\rm d}E\, [f(E)-f(E+x)] = x.
\label{eq:I41}
\end{equation}
By means of the relation
\begin{equation}
f(E)[1-f(E+x)] = \frac{f(E)-f(E+x)}{1-e^{-\beta x}}
\label{eq:ffrel}
\end{equation}
we find for the integral which we need in order to simplify (\ref{eq:I39})
\begin{equation}
\int_{-\infty}^{+\infty}{\rm d}E\,f(E)[1-f(E+x)] =
\frac{x}{1-e^{-\beta x}}.
\label{eq:I42}
\end{equation}
This together with (\ref{eq:I39}) finally gives for the forward tunneling rate
through a single junction\index{junction!single}
\begin{equation}
\GF(V) = {1\over e^2R_T}\int_{-\infty}^{+\infty}{\rm d}E\, {E\over 1-\exp(-\beta
E)} P(eV-E).
\label{eq:I43}
\end{equation}
A corresponding calculation can be done for the backward tunneling rate.
However, it is rather obvious from the symmetry of a voltage biased single
junction\index{junction!single} that
\begin{equation}
\GB(V) = \GF(-V)
\label{eq:I44}
\end{equation}
which is indeed the result one obtains from redoing the calculation.

For a further discussion of the tunneling rates and the current-voltage
characteristic of a single tunnel junction\index{junction!single} it is 
useful to know more about the
function $P(E)$ and the correlation function $J(t)$ by which it is determined.
We will be able to derive some general properties from which we will deduce a
few facts about rates and current-voltage characteristics. Further, we will
discuss the
limits of very low and very high impedance. For a realistic environment one
usually has to evaluate $P(E)$ numerically. We will present several examples
for impedances from which we learn more about how $P(E)$ is related to
properties of the environmental circuit.
%
%
\subsection{Phase-phase correlation function and environmental impedance}
In Sec.~2.3.\ we presented the Hamiltonian (\ref{eq:I16}) to describe the
electrodynamic environment and derived the operator equation of motion
(\ref{eq:I17}) for
the junction charge $\tilde Q$. According to (\ref{eq:phase}) the phase is
proportional to the time derivative
of the charge so that we immediately get the equation of motion for the phase
\begin{equation}
C\ddot{\tilde\varphi} +  \int_0^t{\rm d}s\,Y(t-s)
\dot{\tilde\varphi}(s) = \frac{e}{\hbar}I_N(t)
\label{eq:I45}
\end{equation}
where again $I_N(t)$ is the quantum mechanical noise current.
In terms of our mechanical analogue introduced in Table I, (\ref{eq:I45}) can be
interpreted as the equation of motion of a free Brownian particle.

The effect
of the environmental degrees of freedom on the charge and phase degrees
of freedom is twofold. They produce a damping term which depends on the
admittance $Y(\omega)$ and is responsible for the relaxation of the charge into
equilibrium. The relaxation of the mean charge is described by a
dynamical susceptibility which for this linear system is the same in the
classical and
the quantum case due to the Ehrenfest theorem. From our results in
Sec.~2.1.\ we obtain for the dynamical susceptibility
describing the response of the phase to the conjugate force $(e/\hbar)I(t)$
\begin{equation}
\chi(\omega) = \chi'(\omega) - i\chi''(\omega) =
\left({e\over\hbar}\right)^2 {Z_t(\omega)\over i\omega}.
\label{eq:I46}
\end{equation}

The second effect of the environment manifests itself in the noise current
$I_N(t)$ and appears in correlation functions as the one introduced in
(\ref{eq:Jt}). Since damping and fluctuations have the same microscopic origin
they are not independent of each other and in fact the so-called
fluctuation-dissipation theorem \cite{Kubo}
\begin{equation}
\tilde C(\omega) = {2\hbar \over 1-e^{-\beta\hbar\omega}}\chi''(\omega)
\label{eq:I47}
\end{equation}
relates the absorptive part $\chi''(\omega)$ of the dynamical susceptibility
(\ref{eq:I46}) to the Fourier transform
\begin{equation}
\tilde C(\omega) = \int_{-\infty}^{+\infty}{\rm d}t\, e^{-i\omega t} \langle
\tilde\varphi(0) \tilde\varphi(t) \rangle
\label{eq:I48}
\end{equation}
of the equilibrium phase-phase correlation function.
The fluctuation-dissipation theorem may be proven
in the framework of linear response theory which becomes exact if a linear
system is treated as is the case here. From (\ref{eq:I47}) and (\ref{eq:I48})
together with the stationarity of equilibrium correlation functions we
then immediately get
\begin{equation}
\langle \tilde\varphi(t)\tilde\varphi(0)\rangle = 2\int_{-\infty}^{+\infty}
{{\rm d}\omega\over\omega} {\hbox{Re} Z_t(\omega)\over R_K}{e^{-i\omega t}\over
1-e^{-\beta\hbar\omega}}.
\label{eq:I49}
\end{equation}
Since in general the real part of the impedance $Z_t$ at $\omega = 0$ does not
vanish, the
correlation function (\ref{eq:I49}) does not exist due to an infrared
divergence. This
can easily be understood within our mechanical picture of a free Brownian
particle.\cite{report}
In the absence of a confining potential the variance of the position of the
particle in equilibrium should diverge. There are, however, no problems with
the correlation function $J(t)$ in which according to its definition
(\ref{eq:Jt}) the diverging static
correlation $\langle \tilde\varphi^2\rangle$ is subtracted off. Since the
Fourier transform of the impedance has to be real, the real part $\hbox{Re}
Z_t(\omega)$ of the total impedance is even and together with the identity
\begin{equation}
{1\over 1-e^{-\beta\hbar\omega}} = {1\over 2} + {1\over 2}\coth\!\left({1\over
2}\beta\hbar\omega\right)
\label{eq:I50}
\end{equation}
we finally get for the correlation function appearing in the definition
(\ref{eq:I38}) of $P(E)$
\begin{equation}
J(t) = 2\int_0^{\infty} {{\rm d}\omega\over\omega} {\hbox{Re} Z_t(\omega)\over
R_K} \left\{\coth\!\left({1\over 2}\beta\hbar\omega\right)[\cos(\omega t)-1]
-i\sin(\omega t)\right\}.
\label{eq:I51}
\end{equation}
%
%
\subsection[General properties of $P(E)$]
{General properties of {\protect\boldmath $P$($E$)}}
With the expression (\ref{eq:I51}) for the correlation function $J(t)$ derived
in the
last section one may calculate the probability $P(E)$ for energy exchange
between the tunneling electron and the environment once the external impedance
is known. In general it is not possible to calculate $P(E)$ analytically for a
given impedance except for some special cases which we will discuss later. On
the other hand, there are general properties of $P(E)$ which are
independent of the actual impedance.

Recalling the definition (\ref{eq:I38}) of $P(E)$ we find a first sum rule
\begin{equation}
\int_{-\infty}^{+\infty}{\rm d}E\,P(E) = e^{J(0)} = 1
\label{eq:I52}
\end{equation}
since $J(0) = 0$ which follows directly from the definition (\ref{eq:Jt}).
Eq.~(\ref{eq:I52}) confirms our interpretation of $P(E)$ as a probability. A
second
sum rule is obtained by taking the time derivative of $\exp[J(t)]$ resulting in
\begin{equation}
\int_{-\infty}^{+\infty}{\rm d}E\,EP(E) = i\hbar J'(0) = E_c.
\label{eq:I53}
\end{equation}
To prove this relation we have to calculate
$J'(0)$ which can be done by a short time expansion of (\ref{eq:I51}) which
yields
\begin{equation}
J'(0) = -i\int_{-\infty}^{+\infty} {\rm d}\omega\, {Z_t(\omega)\over R_K}.
\label{eq:I54}
\end{equation}
Here, we made use of the fact that the imaginary part of the impedance is
antisymmetric in $\omega$. The integral in (\ref{eq:I54}) can be evaluated by
integrating
(\ref{eq:I11}) over $\omega$. Since the response function $R(t)$
jumps from zero to one at $t=0$ we get
\begin{equation}
\int_{-\infty}^{+\infty} {\rm d}\omega\, Z_t(\omega) = {\pi\over C}
\label{eq:I55}
\end{equation}
which together with (\ref{eq:I54}) proves the last equality in (\ref{eq:I53}).
It should
be noted that we
assumed that there is no renormalization of the tunnel capacitance by the
environment which means that the first derivative of the charge with
respect to time in the equation
of motion (\ref{eq:I17}) stems from the charging energy of the tunnel junction.
Otherwise,
one would have to replace $C$ by an effective tunnel capacitance defined by
(\ref{eq:I55}).

Another important property of $P(E)$ concerns a relation between the
probabilities to emit and to absorb the energy $E$. We make use of the two
identities
\begin{equation}
\langle e^{i\tilde\varphi(t)} e^{-i\tilde\varphi(0)}\rangle = \langle
e^{-i\tilde\varphi(t)} e^{i\tilde\varphi(0)}\rangle
\label{eq:I56}
\end{equation}
and
\begin{equation}
\langle e^{i\tilde\varphi(t)} e^{-i\tilde\varphi(0)}\rangle = \langle
e^{-i\tilde\varphi(0)} e^{i\tilde\varphi(t+i\hbar\beta)}\rangle.
\label{eq:I57}
\end{equation}
One may convince oneself that (\ref{eq:I56}) is correct by substituting
$\tilde\varphi$ by
$-\tilde\varphi$ in (\ref{eq:I37}). To prove the second identity one writes the
correlation function as a trace
\begin{equation}
\langle e^{i\tilde\varphi(t)} e^{-i\tilde\varphi(0)}\rangle =
\hbox{Tr}\left(e^{-\beta H}e^{{i\over\hbar}Ht}
e^{i\tilde\varphi}e^{-{i\over\hbar}
Ht} e^{-i\tilde\varphi}\right)/\hbox{Tr}\left(e^{-\beta H}\right)
\label{eq:I58}
\end{equation}
and exploits the invariance of the trace under cyclic permutations. With
(\ref{eq:I56}) and
(\ref{eq:I57}) one finds from the definition (\ref{eq:I38}) of $P(E)$ the
so-called detailed balance symmetry
\begin{equation}
P(-E) = e^{-\beta E}P(E)
\label{eq:I59}
\end{equation}
which means that the probability to excite environmental modes compared to the
probability to absorb energy from the environment is larger by a Boltzmann
factor. Another consequence is that at zero temperature no energy can be
absorbed from the environment. $P(E)$ then vanishes for negative
energies.

At zero temperature the asymptotic behavior of $P(E)$ for larges energies may
be obtained from an integral equation which is also useful for numerical
calculations. We will now derive the integral equation following an idea of
Minnhagen \cite{Minnhagen}. From (\ref{eq:I49}) we find for the phase-phase
correlation function at zero temperature
\begin{equation}
J(t) = 2\int_0^{\infty}{{\rm d}\omega\over\omega}{\hbox{Re} 
Z_t(\omega)\over R_K} (e^{-i\omega t}-1).
\label{eq:I60}
\end{equation}
Taking the derivative of $\exp[J(t)]$ with respect to time we get
\begin{equation}
{{\rm d}\over {\rm d}t}\exp[J(t)] = -2i\exp[J(t)]\int_0^{\infty}{\rm d}\omega
{\hbox{Re}Z_t(\omega)\over R_K} e^{-i\omega t}.
\label{eq:I61}
\end{equation}
Since we are interested in $P(E)$, we take the Fourier transform which on the
left hand side results in a term proportional to $EP(E)$ and a convolution
integral on the right hand side. Using the fact that $P(E)$ at zero
temperature vanishes for negative energies we finally get
\begin{equation}
EP(E) = 2\int_0^E{\rm d}E'\,{\hbox{Re}\left[
Z_t\left({\displaystyle{E-E'\over\hbar}}\right)\right]\over R_K} P(E').
\label{eq:I62}
\end{equation}
This enables us to calculate $P(E)$ numerically by
starting from an arbitrary $P(0)$ and subsequently normalizing the result.
For finite
temperatures energy can also be absorbed from the environment. Then
an inhomogeneous integral equation may be derived which is more
complicated.\cite{IGEPL91}

We now consider the integral equation (\ref{eq:I62}) for large energies so 
that the integral
on the right hand side covers most of the energies for which $P(E)$ gives a
contribution. For these large energies we may neglect $E'$ with respect to $E$
in the argument of the impedance and end up with the normalization integral for
$P(E)$. For large energies and zero temperature $P(E)$ therefore decays
according to \cite{FBSEPL91}
\begin{equation}
P(E) = {2\over E}{\hbox{Re} Z_t(E/\hbar)\over R_K}\ \ \ \ \ \hbox{for}\
E\to\infty.
\label{eq:I63}
\end{equation}
For the limits of low and high energies one may often approximate the external
impedance $Z(\omega)$ by a constant. In this case we can apply the results 
(\ref{eq:I99}) and (\ref{eq:I102}) which will be derived in Sec.~4.2.\ for an 
Ohmic environment.
%
%
\subsection{General properties of current-voltage characteristics}
The detailed balance relation (\ref{eq:I59}) is useful to derive a simple
formula for the
current-voltage characteristic of a single tunnel 
junction\index{junction!single}. The total current
through the junction is given by the transported charge $e$ times the
difference of the forward and backward tunneling rates
\begin{equation}
I(V) = e(\GF(V)-\GB(V)).
\label{eq:I65}
\end{equation}
The backward tunneling rate may be obtained from the forward tunneling rate
(\ref{eq:I43})\vadjust{\break} 
by means of the symmetry (\ref{eq:I44}). Together with the
detailed balance relation (\ref{eq:I59}) one obtains
\begin{equation}
I(V) = {1\over eR_T}(1-e^{-\beta eV})\int_{-\infty}^{+\infty} {\rm d}E\, {E\over
1-e^{-\beta E}}P(eV-E).
\label{eq:I66}
\end{equation}
This formula has the property $I(-V) = -I(V)$ as one would expect.

We now consider the limit of zero temperature and assume that $V>0$. Taking
into account that $P(E)$ then vanishes for negative energies, we obtain from
(\ref{eq:I66})
\begin{equation}
I(V) = {1\over eR_T} \int_0^{eV}{\rm d}E\,(eV-E)P(E).
\label{eq:I67}
\end{equation}
It is no surprise that, in contrast to the finite temperature case,
at zero temperature the current at a
voltage $V$ depends only on the probability to excite environmental modes
with a total energy less than $eV$ since this is the maximum energy at the
disposal of the
tunneling electron. According to (\ref{eq:I67}) the current at the
gap voltage $e/2C$ depends on $P(E)$ at all energies up to the charging energy
$E_c$. In view of the integral equation (\ref{eq:I62}) this means that the
environmental impedance up to the frequency $E_c/\hbar$ (which is of the order
of $20\,$GHz for $C=10^{-15}\,$F) is relevant. The general behavior of an
impedance up to high frequencies is discussed in Chap.~1.
Another consequence of the zero temperature result
(\ref{eq:I67}) is
that the probability $P(E)$ directly determines the second derivative of the
current-voltage characteristic of normal tunnel junctions
\begin{equation}
{{\rm d}^2I\over {\rm d}V^2} = {e\over R_T}P(eV).
\label{eq:I68}
\end{equation}

The sum rules derived in the last section can be used to
determine the current-voltage characteristic at very large voltages. We assume
that $eV$ is much larger than energies for which $P(E)$ gives a noticeable
contribution and that $eV\gg k_BT$. Then the expression (\ref{eq:I66})
becomes
\begin{equation}
I(V) = {1\over eR_T} \int_{-\infty}^{+\infty}{\rm d}E\,(eV-E)P(E)
\label{eq:I69}
\end{equation}
which together with the sum rules (\ref{eq:I52}) and (\ref{eq:I53}) yields
\begin{equation}
I(V) = {V-{e/2C}\over R_T}
\label{eq:I70}
\end{equation}
for very large positive voltages. The slope of $I(V)$ confirms the
interpretation of $R_T$ as a tunneling resistance. The shift in voltage by
$e/2C$ represents the Coulomb gap.\index{Coulomb gap} In Sec.~4.2.\ we will 
discuss in more detail
for the Ohmic model how the asymptotic current-voltage characteristic is
approached for large voltages.
%
%
\subsection{Low impedance environment}
A special case of an environment is when the impedance is so low that one may
effectively set $Z(\omega)=0$. This will be a good approximation if the
impedance is much less than the resistance quantum $R_K$. Since then the 
phase fluctuations described by
$J(t)$ vanish, we find $P(E) = \delta(E)$. This corresponds to the fact that in
the absence of environmental modes only elastic tunneling processes are
possible. From (\ref{eq:I43}) we immediately get for the forward tunneling rate
\begin{equation}
\GF(V) = {1\over e^2 R_T} {eV\over 1-\exp(-\beta eV)}.
\label{eq:I71}
\end{equation}
According to Sec.~1.4.\ this is the global rule result
which was already discussed in Sec.~1.2.\ where we introduced the
voltage-biased tunnel junction\index{voltage bias}. The appearance 
of the global rule in this limit is easy to
understand. The external voltage source keeps the voltage across the junction
fixed at any time. Therefore, after the tunneling process the electron has to
be transferred through the circuit immediately to restore the charge on the
junction capacitor. The work $eV$ done by the voltage source is thus the only
energy which can appear in the rate expressions.

We remark that the second sum rule (\ref{eq:I53}) is violated if the impedance
vanishes.
As a consequence, in the absence of an external impedance we do not find a
Coulomb gap even at highest voltages. On the other hand, for a small but finite
impedance the sum rule (\ref{eq:I53}) is valid although the
current-voltage characteristic will show a clear Coulomb gap only at
very large voltages (cf.\ Eq.~(\ref{eq:I103})).
%
%
\subsection{High impedance environment}
We now consider the limit of a very high impedance environment, i.e.\ the
impedance is much larger than $R_K$. Then the
tunneling electron may easily excite modes. This situation is described by a
spectral density of the environmental modes which is sharply peaked at $\omega
= 0$. To check this we consider the case of Ohmic damping, i.e. $Z(\omega) =
R$. Then the real part of the total impedance is given by $R/(1+(\omega
RC)^2)$. For very large resistance this becomes
$(\pi/C)\delta(\omega)$. The prefactor is consistent with our result
(\ref{eq:I55})
for the integral over the total impedance. For the correlation function $J(t)$
this concentration of environmental modes at low frequencies means that the
short time expansion
\begin{equation}
J(t) = -{\pi\over CR_K} \left( it + {1\over \hbar\beta}t^2\right)
\label{eq:I72}
\end{equation}
applies for all times. Inserting this result into the definition (\ref{eq:I38})
of $P(E)$ one gets a Gaussian integral which may easily be evaluated yielding
\begin{equation}
P(E) = {1\over \sqrt{4\pi E_c k_B T}}\exp\!\left[-\frac{(E-E_c)^2}{4E_ck_BT}
\right].
\label{eq:I73}
\end{equation}
This result obviously satisfies the sum rules (\ref{eq:I52}) and (\ref{eq:I53})
derived
earlier. For very low temperatures $k_BT\ll E_c$ the probability to excite
environmental modes reduces to
\begin{equation}
P(E) = \delta(E-E_c)
\label{eq:I74}
\end{equation}
so that each electron transfers to the environment an amount of energy 
corresponding to the charging energy $E_c$. The expression (\ref{eq:I73}) 
may be used to
calculate tunneling rates and current-voltage characteristics in the high
impedance limit. The broadening of the Gaussian distribution with respect to
the delta function (\ref{eq:I74}) describes the washout of the Coulomb 
blockade\index{Coulomb blockade} at finite
temperatures. For zero temperature the expression (\ref{eq:I67}) for the current
together with (\ref{eq:I74}) yields
\begin{equation}
I(V) = {eV-E_c\over eR_T}\Theta(eV-E_c)
\label{eq:I75}
\end{equation}
where $\Theta(E)$ is the unit step function. Since according to (\ref{eq:I74}) a
tunneling
electron always transfers the energy $E_c$ to the environment, tunneling
becomes
possible only if the energy $eV$ at disposal exceeds $E_c$. We
thus find the Coulomb gap as we did in Sec.~1.3.\ by considering only the
charging energy of the junction. To make this connection clearer we note 
that the energy difference (\ref{eq:I4}) of the local rule
appears in (\ref{eq:I75}) since
\begin{equation}
{Q^2\over 2C} - {(Q-e)^2\over 2C} = eV-E_c
\label{eq:I76}
\end{equation}
if $V$ is the voltage across the junction before the tunneling event.

We conclude the discussion of the last two sections by noting that the answer
to whether one should use the global or local rule to determine the behavior
of a tunnel junction is as follows. In general, neither rule is valid and
the rate depends on the external circuit to which the
junction is coupled. For impedances very low compared to the resistance quantum
we find that the global rule
leads to a correct description whereas for a high impedance environment and
very low temperatures the local rule is correct. In all other cases $P(E)$ has
to be calculated for the specific environment present in order to get the
correct current-voltage characteristic. In the following section we will
present various examples for external impedances and discuss how they affect
tunneling rates and current-voltage characteristics.
%
%
\section{Examples of electromagnetic environments}
%
%
\subsection{Coupling to a single mode}
As a first example let us study the coupling of a tunnel junction to one single
environmental mode which comes from a resonance in the lead impedance or might
be associated with a molecule in the barrier. This model is
so simple that analytical solutions are available for arbitrary 
temperatures.\cite{DEGPRL90,onemode}
In addition, the simplicity of the model will allow us to learn important facts
about how properties of the environment show up in the probability for energy
transfer between the tunneling electron and the external circuit.

The coupling of the tunnel junction to one environmental mode may be
accomplished by putting just one inductance $L$ into the external circuit. In
this special case our model for the environment introduced in Sec.~2.2.\ may
be taken rather literally. With the impedance $i\omega L$ of an inductor we
find for the total impedance
\begin{equation}
Z_t(\omega) = {1\over C} {i\omega \over \omega_s^2-(\omega-i\epsilon)^2}
\label{eq:I77}
\end{equation}
where we introduced the frequency
\begin{equation}
\omega_s = {1\over\sqrt{LC}}
\label{eq:I78}
\end{equation}
of the environmental mode. The small imaginary part $\epsilon$ is necessary to
obtain the correct result for the real part. In the limit $\epsilon\to 0$ we
obtain
\begin{equation}
\hbox{Re} Z_t(\omega) = {\pi\over 2C}[\delta(\omega-\omega_s) +
\delta(\omega+\omega_s)]
\label{eq:I79}
\end{equation}
which is what we expected since only the mode with frequency $\omega_s$ should
be present and the prefactor satisfies (\ref{eq:I55}).

Due to the delta functions in (\ref{eq:I79}) we get the correlation function
$J(t)$
simply by substituting $\omega$ by $\omega_s$ in (\ref{eq:I51}). Inserting the
result into (\ref{eq:I38}) we then find
\begin{equation}
P(E) = {1\over 2\pi\hbar} \int_{-\infty}^{+\infty} {\rm d}t\, \exp\!\left[\rho
\Big\{\!\coth({\beta\hbar\omega_s\over
2})(\cos(\omega_st)-1)-i\sin(\omega_st)\Big\}+{i\over\hbar}Et\right].
\label{eq:I80}
\end{equation}
Here, we have introduced the parameter
\begin{equation}
\rho = {\pi\over CR_K\omega_s} = {E_c\over \hbar\omega_s}
\label{eq:I81}
\end{equation}
which should be of relevance since it compares the single electron charging
energy with the mode excitation energy. This parameter determines the size of
charge fluctuations
\begin{equation}
\langle\tilde Q(t) \tilde Q(0)\rangle = -\left({\hbar C\over e}\right)^2
\ddot J(t).
\label{eq:I82}
\end{equation}
Using (\ref{eq:I82})
which is obtained from the relation (\ref{eq:phase}) between the phase and
the charge together with (\ref{eq:phasfluk}), (\ref{eq:ladfluk}), and the
stationarity of equilibrium correlation functions
\begin{equation}
\langle A(t) B(0)\rangle = \langle A(0)B(-t)\rangle,
\label{eq:I83}
\end{equation}
we find
\begin{equation}
\langle \tilde Q^2\rangle = {e^2\over 4\rho} \coth({\beta\hbar\omega_s\over
2})
\label{eq:I84}
\end{equation}
so that at zero temperature charge fluctuations will only be small compared to
the elementary charge if $\rho\gg 1$.

We now proceed with the calculation of the current-voltage characteristic.
Using the equality
\begin{equation}
\cos(\omega_s t) \coth\left({\beta\hbar\omega_s\over 2}\right) -i
\sin(\omega_s t) =
{\cosh\left({\displaystyle {\beta\hbar\omega_s\over 2}}-i\omega_s
t\right)\over\sinh\left({\displaystyle {\beta\hbar\omega_s\over 2}}\right)}
\label{eq:I85}
\end{equation}
we can take advantage of the generating function \cite{Abramowitz}
\begin{equation}
\exp[{y\over 2}(z + {1\over z})] = \sum_{k=-\infty}^{+\infty} z^k I_k(y)
\label{eq:I86}
\end{equation}
of the modified Bessel function $I_k$ for $z = \exp(x)$. Then the integral over
time in (\ref{eq:I80}) can easily be done leading to 
\begin{eqnarray}
P(E) &=& \exp\!\left(-\rho\coth({\beta\hbar\omega_s\over 2})\right)\nonumber\\
&&\hspace{3cm}\times\sum_{k=-\infty}^{+\infty}\!I_k\!\Bigg({\rho\over
\sinh({\displaystyle{\beta\hbar\omega_s\over
2}})}\Bigg)\exp\!\left(k{\beta\hbar\omega_s\over 2}\right)
\delta(E-k\hbar\omega_s).
\label{eq:I87}
\end{eqnarray}
Although this expression for $P(E)$ is rather complicated it has a simple
physical origin. This becomes particularly apparent at zero temperature 
where we find 
\begin{equation}
P(E) = e^{-\rho} \sum_{k=0}^{\infty}{\rho^k\over k!} \delta(E-k\hbar\omega_s)
= \sum_{k=0}^{\infty}p_k\delta(E-k\hbar\omega_s).
\label{eq:I88}
\end{equation}
Here $p_k$ is the probability to emit $k$ oscillator quanta. Comparing the
second and the third expression in (\ref{eq:I88}), one sees that $p_k$ obeys a
Poissonian
distribution. Therefore, the quanta are emitted independently. The way of
reasoning may now be reversed. Making the assumption of independent emission,
(\ref{eq:I88}) may of course immediately be obtained. But we also get the
expression (\ref{eq:I87}) for finite temperatures. Introducing the Bose
factor $N =1/
[\exp(\hbar\beta\omega_s)-1]$, the probability to emit a quantum is given by
$\rho_e = \rho(1+N)$ and the probability for absorption is $\rho_a = \rho N$.
The probability to absorb $m$ quanta and to emit $n$ quanta will then be
$\exp[-(\rho_a+\rho_e)] \rho_a^m \rho_e^n/(m!n!)$ so that
\begin{equation}
P(E) = \exp[-(\rho_a+\rho_e)] \sum_{m,n} {\rho_a^m\rho_e^n\over m!n!}
\delta(E-(n-m)\hbar\omega_s).
\label{eq:I89}
\end{equation}
Doing the sum over the variable $l=m+n$ and using the ascending series of
the modified Bessel function \cite{Abramowitz}
\begin{equation}
I_k(z) = \left({z\over 2}\right)^k \sum_{l=0}^{\infty} {(z^2/4)^l\over
l!(k+l)!}
\label{eq:I90}
\end{equation}
one is left with a sum over the difference $k=n-m$ which is our finite
temperature result (\ref{eq:I87}). We note that the argument given here can be
generalized to the case of two or three modes and finally to infinitely many
modes. The representation (\ref{eq:I89}) of $P(E)$ points clearly to its 
physical significance. It is apparent that $P(E)$ gives the quantity describing
the probability to exchange the energy $E$ with the environment.

With the form (\ref{eq:I87}) for $P(E)$ the convolution integral appearing in 
the expression
(\ref{eq:I66}) for the current-voltage characteristic can easily be evaluated
yielding \cite{DEGPRL90,onemode}

\break
\begin{eqnarray}
I(V) &=& {1\over eR_T}\sinh\!\left({\beta eV\over 2}\right)\exp\!\left(-\rho
\coth({\beta\hbar\omega_s\over 2})\right)\nonumber\\
&&\hspace{5cm}\times\sum_{k=-\infty}^{+\infty}\!
I_k\!\Bigg({\rho\over {\displaystyle \sinh({\beta\hbar\omega_s\over 2}})}\Bigg)
{\epsilon_k \over {\displaystyle \sinh({\beta\epsilon_k\over 2}})}.
\label{eq:I91}
\end{eqnarray}
Here, we introduced the energy
\begin{equation}
\epsilon_k = eV-k\hbar\omega_s
\label{eq:I92}
\end{equation}
left to the electron after having excited $k$ quanta $\hbar\omega_s$.
In the limit of zero temperature and for positive voltages (\ref{eq:I91})
becomes
\begin{equation}
I(V) = {1\over eR_T} e^{-\rho}
\sum_{k=0}^{n}
{\rho^k\over k!}(eV-k\hbar\omega_s)
\label{eq:I93}
\end{equation}
where $n$ is the largest integer smaller or equal to $eV/\hbar\omega_s$.
This result has a simple interpretation. The sum runs over all possible
numbers of excited quanta where the maximum number of modes which can be
excited is given by $n$. The factor $\exp(-\rho)$ determines the slope at zero
voltage since at very low voltages only the term with $k=0$ contributes to
the sum.
As we expected from our discussion of $\rho$, this quantity is important for
the occurrence of the Coulomb blockade. For small $\rho$
there is no Coulomb blockade and the conductance
at zero voltage is about $1/R_T$. Only for large enough $\rho$ a
Coulomb blockade becomes apparent in the small
factor $\exp(-\rho)$. According to the definition (\ref{eq:I81}), large $\rho$
means that
the mode energy $\hbar\omega_s$ is small compared to the charging energy $E_c$
which indicates a high impedance environment as discussed earlier. So again
this example shows that Coulomb blockade can only be found if the environmental
impedance is large enough.
Fig.~3
\begin{figure}
\begin{center}
\includegraphics{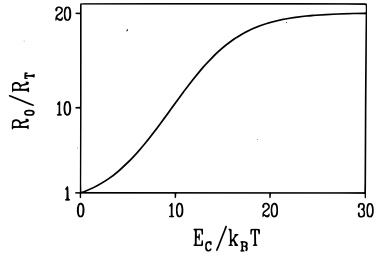}
\end{center}
\caption{
Zero-bias differential resistance as a function
of temperature for $\rho=3$.}
\end{figure}
presents the differential resistance $R_0 = {\rm d}V/{\rm d}I$ at
zero bias $V=0$ for $\rho=3$. For large temperatures the Coulomb blockade
is lifted by thermal fluctuations and $R_0$ is of the order of $R_T$. As
the temperature is decreased a Coulomb gap forms and $R_0/R_T$ approaches
$\exp(\rho)$ for vanishing temperature.

So far we have discussed the small voltage behavior. But it is also the
current-voltage characteristic at finite voltages which contains information
about the environment. Every time the voltage becomes an integer multiple of
the mode energy $\hbar\omega_s$ the slope of the current-voltage characteristic
changes. This becomes even more apparent in the differential current-voltage
characteristic where we find steps at voltages $k\hbar\omega_s/e$ when new
inelastic channels are opened.
This is in agreement with (\ref{eq:I68}) according to which the second
derivative at zero
temperature is $P(E)$ for which we know that it is a series of delta functions
at voltages $k\hbar\omega_s/e$. Thus derivatives of the current-voltage
characteristic contain information about the structure of the environment. As
an example we show in Fig.~4
\begin{figure}
\begin{center}
\includegraphics{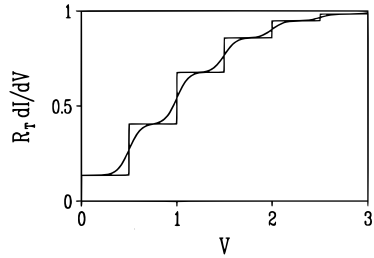}
\end{center}
\caption{
Differential current-voltage characteristic for $\rho=2$.
The voltage is given in units of $e/2C$. The step-like curve corresponds to
zero temperature while the smooth curve is for $k_BT = 0.04~E_c$.}
\end{figure}
the differential current-voltage characteristic
for $\rho=2$. At zero temperature one gets the steps as expected. At finite
temperature, however, the sharp resonances in $P(E)$ are washed out and
therefore the steps are smoothed.

To end this section let us apply the mechanical analogue of Table I to point
out the relation between the M\"o\ss{}bauer effect in solid state physics and
the environmental effects on single charge tunneling. For the M\"o\ss{}bauer
effect one considers a radioactive nucleus embedded in a crystal. When a
$\gamma$~quant is emitted there are two ways to satisfy momentum conservation.
The first possibility is to excite phonons in the crystal, i.e.\ momentum is
transferred to the emitting nucleus and the energy of the $\gamma$~quant is
reduced. In the second possiblity, the so-called M\"o\ss{}bauer transition, the
recoil momentum is transferred to the whole crystal. This will be more likely
if it is difficult to excite phonons. Due to the large mass of the crystal the
energy of the $\gamma$~quant and, more important for us, the momentum of the
nucleus then remain unchanged.

In ultrasmall tunnel junctions the emission of a $\gamma$~quant corresponds to
the tunneling of an electron. According to Table I the momentum of the nucleus
is related to the charge of the junction. The question is whether a tunneling
process changes the junction charge or not. If this charge is kept fixed we do
not find Coulomb blockade.\index{Coulomb blockade} This corresponds to the 
M\"o\ss{}bauer transition.
In both cases no environmental modes are excited. For the occurrence of Coulomb
blockade we need a change of the junction charge. This is analogous to a
non-M\"o\ss{}bauer transition and requires the excitation of environmental
modes. We conclude from this analogy that Coulomb blockade is only possible if
there are low frequency environmental modes which are coupled strongly to the
tunneling electron, i.e.\ a high impedance environment is needed. This is in
agreement with our previous findings. The analogy with the M\"o\ss{}bauer
effect allows us also to interpret the factor $\exp(-\rho)$ in (\ref{eq:I88})
as a Debye-Waller factor giving the possibility for electron tunneling without
the excitation of environmental modes.
%
%
\subsection{Ohmic impedance}
For the impedance caused by an external circuit a more realistic choice than
the single mode model
would be an ideal Ohmic resistor described by the frequency-independent
impedance $Z(\omega) =
R$. We introduce the dimensionless parameter
\begin{equation}
g={R_K\over R}
\label{eq:I94}
\end{equation}
which is proportional to the lead conductance.

The mechanical analogue of this problem is very well studied since this
special impedance results in a Fourier transform of the admittance which is
proportional to a delta function. Reinterpreting (\ref{eq:I45})
we then find the equation of motion for a free Brownian particle
which contains a damping term proportional to the velocity of the
particle. Without a confining potential the Brownian particle undergoes a
diffusive motion. From our knowledge of classical diffusion we conclude that
for long times the correlation function $J(t)$ should increase proportional to
time $t$. This classical result holds also for low temperatures. At zero
temperature, however, the environment cannot provide the diffusing particle
with energy and the correlation function will increase
somewhat slower, namely proportional
to $\ln(t)$.\cite{report} In general, it is not possible to obtain analytical 
results for the
Ohmic model. We therefore restrict ourselves to the case of zero temperature
and consider the limits of low and high energies in $P(E)$. This allows us
to find explicit expressions for the current-voltage characteristics at small
and
large voltages. Numerical calculations bridge the gap between the two limits.

We first discuss the low energy behavior of $P(E)$, which is determined by the
long time behavior of the correlation function $J(t)$. This case is of general
importance since according to (\ref{eq:I62}) at low voltages and very low
temperatures $P(E)$ is governed by the impedance at low frequencies. As long 
as the impedance at
zero frequency is nonvanishing the Ohmic model with $R=Z(0)$ will apply in
this regime.

In order to avoid lengthy calculations we determine the low energy behavior of
$P(E)$ from the integral equation (\ref{eq:I62}) which is valid at zero
temperature.
Since this integral equation is homogeneous it will allow us to determine
$P(E)$ only up to a multiplicative constant which depends on the behavior of
$P(E)$ at all energies.
To solve the integral equation we need the real
part of the total impedance
\begin{equation}
{\hbox{Re}Z_t(\omega)\over R_K} = 
\frac{1}{R_K}\hbox{Re}\left[\frac{1}{i\omega C + 1/R}
\right] = {1\over g} {1\over 1+(\omega/\omega_R) ^2}
\label{eq:I95}
\end{equation}
where the frequency
\begin{equation}
\omega_R = {1\over RC} = \frac{g}{\pi}\frac{E_c}{\hbar}
\label{eq:I96}
\end{equation}
describes an effective cutoff for the total impedance due to the junction
capacitance.
At energies small compared to $\hbar\omega_R$ we may approximate the real
part of the total impedance by a constant. Taking the derivative of
(\ref{eq:I62}) with respect to energy we get the differential equation

\break
\begin{equation}
{{\rm d}P(E)\over {\rm d}E} = \left({2\over g}-1\right){P(E)\over E}
\label{eq:I97}
\end{equation}
which may easily be solved yielding
\begin{equation}
P(E)\sim  E^{(2/g-1)}
\label{eq:I98}
\end{equation}
for small positive energies. For negative energies $P(E)$ vanishes since we
consider the case of zero temperature.
With a more complete analysis of $J(t)$ and $P(E)$ one may determine the
normalization constant. One finds \cite{DEGPRL90}
\begin{equation}
P(E) = {\exp(-2\gamma/g)\over \Gamma(2/g)}
{1\over E} \left[\frac{\pi}{g}\frac{E}{E_c}\right]^{2/g},
\label{eq:I99}
\end{equation}
where $\gamma = 0.577\ldots$ is the Euler constant. The factors appearing in
(\ref{eq:I99}) may be motivated by the behavior of the correlation function
$J(t)$ for large times \cite{report}
\begin{equation}
J(t) = -{2\over g}[\ln(\omega_R t) + i{\pi\over 2} +\gamma] \ \ \ \
\hbox{for\ } t\to\infty
\label{eq:I100}
\end{equation}
so that the offset of the logarithmic divergence appears in the result
(\ref{eq:I99}).
From (\ref{eq:I67}) it is straightforward to calculate the current-voltage
characteristic for small voltages
\begin{equation}
I(V) = {\exp(-2\gamma/g)\over \Gamma(2+2/g)}
{V\over R_T} \left[\frac{\pi}{g}\frac{e\vert V\vert}{E_c}\right]^{2/g}
\label{eq:I101}
\end{equation}
which leads to a zero-bias anomaly of the 
conductance\index{conductance!zero-bias anomaly} ${\rm d}I/{\rm d}V \sim
V^{2/g}$.\cite{YuliT4,DEGPRL90,Girvin,AveOdin,PanZaik} This result remains 
valid for a more general environment with a
finite zero-frequency impedance $Z(0)$. The power law exponent is then given by
$2/g=2Z(0)/R_K$ but the prefactor in (\ref{eq:I101}) depends on the
high-frequency behavior of the impedance.

Besides the behavior at low voltages it is also of interest how fast the
current-voltage characteristics for finite lead conductance $g$ approach the
high impedance asymptote (\ref{eq:I70}). To answer this we need to know the
high energy behavior of $P(E)$ which for an Ohmic impedance follows from
(\ref{eq:I63}) and (\ref{eq:I95}) as
\begin{equation}
P(E) = {2g\over \pi^2}{E_c^2\over E^3}\ \ \
\ \ \hbox{for}\ E\to\infty.
\label{eq:I102}
\end{equation}
Inserting this into the expression (\ref{eq:I67}) for the current at zero
temperature one finds
\begin{equation}
I(V) = {1\over R_T}\left[V - {e\over 2C} + {g\over \pi^2}{e^2\over
4C^2}{1\over V}\right]\ \ \ \ \ \hbox{for}\ V\to\infty.
\label{eq:I103}
\end{equation}
As expected the corrections to (\ref{eq:I70}) for finite lead conductance 
are positive and for a
given voltage they become smaller with decreasing conductance of the external
resistor. The voltage at which the corrections become negligible will increase
with $\sqrt{g}$ as the lead conductance is increased. In the limit
$g\to\infty$ no crossover will occur and the Ohmic current-voltage
characteristic (\ref{eq:I3}) will be correct for all voltages.

In Fig.~5
\begin{figure}
\begin{center}
\includegraphics{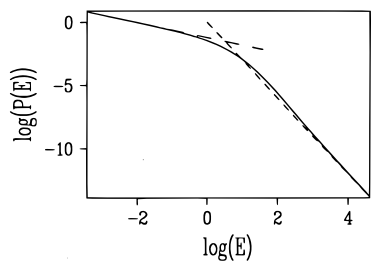}
\end{center}
\caption{
Log-log plot of $P(E)$ at zero temperature for the Ohmic model with $g=5$.
Also shown are the low energy asymptote according
to (\protect\ref{eq:I99}) (long-dashed) and the
high energy asymptote according to
(\protect\ref{eq:I102}) (short-dashed). Energy is taken in units of $E_c$.}
\end{figure}
we present a $P(E)$ for zero temperature and an Ohmic lead conductance $g=5$
together with its low and high energy asymptotes. The data were
obtained numerically by solving the integral equation (\ref{eq:I62}).
The dependence of $P(E)$ and of the corresponding current-voltage
characteristics on the lead conductance is shown in Fig.~6
\begin{figure}
\begin{center}
\includegraphics{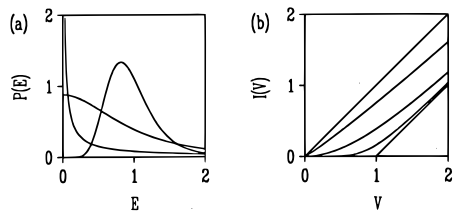}
\end{center}
\caption{
(a) $P(E)$ at zero temperature for the Ohmic model with
lead conductances $g=0.2$ (peaked around $E_c$), $g=2$ (zero slope at $E=0$),
and $g=20$ (diverging at $E=0$). Energy is taken in units of $E_c$. (b) Zero
temperature current-voltage characteristics for the Ohmic model with $g=
\infty, 20, 2, 0.2,$ and $0$ from top to bottom.}
\end{figure}
for three different values of $g$.
According to (\ref{eq:I98}) the $P(E)$ depicted in Fig.~6a has a
singularity for the large conductivity $g=20$, starts with zero slope for $g=2$
and is peaked around $E_c$ for the small conductivity $g=0.2$. The
current-voltage characteristics of Fig.~6b demonstrate that quantum
fluctuations destroy the Coulomb blockade.
Again, a clear Coulomb blockade is obtained for a high impedance environment.
As a criterion for the occurrence of a Coulomb blockade\index{Coulomb blockade} 
one may require that for
vanishing voltage the curvature of the current-voltage characteristic goes to
zero. Since the curvature is given by $P(E)$ we find that this criterion is
fulfilled if the dimensionless lead conductance is sufficiently small
($g<2$). This is related to the fact that at this lead conductance $P(E)$
switches
from a divergent behavior for $E\to 0$ to a regime where $P(E)$ vanishes in
this limit. This singular behavior of $P(E)$ for $g>2$ disappears for finite
temperatures but then thermal fluctuations also contribute to the destruction
of the Coulomb blockade.
%
%
\subsection{A mode with a finite quality factor}
We now combine the two models considered previously and discuss the case of
finite temperatures. As in the first model we start with a single mode. But now
we allow for a finite quality factor which means that the
resonance is broadened. Technically this is achieved by putting a resistor in
series with the inductor of the single mode model. We may keep the notation of
the previous sections where we introduced the mode frequency $\omega_s =
(LC)^{1/2}$, the inverse relaxation time $\omega_R = 1/RC$, and the
lead conductance $g = R_K/R$. However, it is useful to introduce the quality
factor
$Q = \omega_R/\omega_s$ which measures the broadening of the resonance or
equivalently how fast an oscillation decays with respect to the oscillation
period. The single mode case then corresponds to $Q = \infty$ while the
Ohmic case is approached for $Q\to 0$. By varying the quality factor, we are
able to change qualitative features of the environment. For an
environment with a resistor and an inductor in series we get for the total
impedance
\begin{equation}
{Z_t(\omega)\over R_K} = {1\over g} {1+iQ^2(\omega/\omega_R) \over 1+
i(\omega/\omega_R) -Q^2(\omega/\omega_R)^2}.
\label{eq:I104}
\end{equation}

For the calculation of $P(E)$ and of finite temperature current-voltage
characteristics one has to resort to numerical methods. The results presented
in Fig.~7
\begin{figure}
\begin{center}
\includegraphics{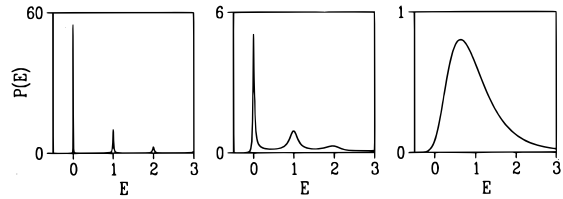}
\end{center}
\caption{
$P(E)$ at $k_BT = 0.05~E_c$ for the total impedance (\protect\ref{eq:I104}).
The quality factor decreases from left to right $Q=50, 5, 0.25$ and
$\hbar\omega_s=E_c$. Energy is taken in units of $E_c$.}
\end{figure}
were obtained by means of an inhomogeneous integral equation which
is a generalization of the integral equation (\ref{eq:I62}). An inhomogeneous
term, which allows for a simple recursive algorithm, was obtained by
splitting off the Ohmic long time behavior of the correlation function
$J(t)\sim t$ discussed in the last section.\cite{IGEPL91}

In Fig.~7 we have chosen the mode energy $\hbar\omega_s$ equal to the charging
energy $E_c$. The quality factors range from 50 which gives a very good
resonance over $Q=5$ showing a considerable broadening to the rather low value
of 0.25. The temperature $k_BT = 0.05~E_c$ is very low so that for
negative energies $P(E)$ is strongly suppressed as can be seen very clearly
from
the figure. The $P(E)$ for the high quality factor reflects the sharp 
resonance in
the environmental impedance and also describes the possibility of exciting
more than one quantum according to (\ref{eq:I87}). The broadening of the lines
which is
connected to additional bath modes is clearly seen for $Q$ = 5. For $Q$=0.25
one finds a broad distribution for $P(E)$ resembling the one found for
the pure
Ohmic model with a broad frequency range of environmental oscillators. It is
obvious from this discussion that $P(E)$ contains a lot of information about
the environment to which the junction is coupled. According to (\ref{eq:I68})
$P(E)$ for normal tunnel junctions is proportional to the second derivative of
the current-voltage characteristic and therefore rather difficult to measure.
However, we will show in Sec.~5 that the Josephson current in ultrasmall
Josephson junctions at $T=0$ is related directly to $P(E)$. For normal tunnel
junctions one may measure the first derivative of the current-voltage
characteristic.
For a single bath mode we had
already seen that differential current-voltage characteristics show more
details than the $I$-$V$ curve itself. As an example, Fig.~8
\begin{figure}
\begin{center}
\includegraphics{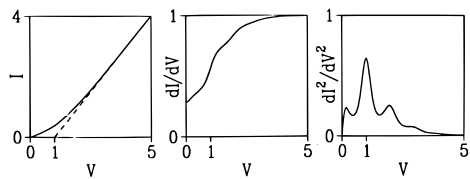}
\end{center}
\caption{
Current-voltage characteristic and its first and second
derivatives as calculated from $P(E)$ for $Q=5$ given in Fig.~7. The dashed
line in the $I$-$V$ characteristic indicates the ideal Coulomb blockade
characteristic. Currents are taken in units of $e/2CR_T$ and voltages in
units of $e/2C$.}
\end{figure}
presents results
for the case $Q = 5$. It is difficult to distinguish this current-voltage
characteristic from the characteristic of the Ohmic model. However,
in the first derivative with respect to voltage we find a step-like structure
which we know is due
to the resonance in the environment. It is smeared because of the finite
quality factor and thermal fluctuations. In the second derivative we almost
reproduce $P(E)$. According to (\ref{eq:I68}) this would be exact
at zero temperature.
For finite temperatures the second derivative is roughly given by the 
antisymmetric part of $P(E)$. The antisymmetric part ensures that the current 
vanishes at zero voltage. This leads to noticeable deviations from $P(E)$ at
low voltages as seen in Fig.~8.
%
%
\subsection{Description of transmission lines}
So far we have treated only impedances which can be described by at most two
lumped circuit elements like a resistor and an inductor. To model a real
experiment, however, this is often not sufficient. Thinking for example
of
wires attached to the junction, one has to model the environment by distributed
resistors, inductors, and capacitors characterized by the three parameters
$R_0$, $L_0$, and $C_0$ which are resistance, inductance, and capacitance per
unit length, respectively. Before discussing two special cases of such
transmission lines, let us first derive the impedance for a more general 
transmission
line. We describe two wires by segments containing a resistor and an inductor
in series with a capacitive coupling between the wires as shown in Fig.~3 of
Chap.~1. We neglect a
conductance between the wires which is sometimes also taken into account. The
voltage drop along the line is connected with the current flowing through the
wire and the impedance per unit length via the differential equation
\begin{equation}
{\partial V\over \partial x} = - I(x) (R_0 + i\omega L_0)
\label{eq:I105}
\end{equation}
where we assumed that the time dependence of the current and the voltage is
given
by $\exp(i\omega t)$. This equation is complemented by the continuity equation
\begin{equation}
i\omega q(x) + {\partial I\over\partial x} = 0
\label{eq:I106}
\end{equation}

\break\noindent
where $q(x) = C_0 V(x)$ is the charge sitting on the capacitor at position
$x$. This
charge can only change if current flowing through the wires charges the
capacitor. Equations (\ref{eq:I105}) and (\ref{eq:I106}) describe the
dynamics of
the transmission line. Eliminating the current we obtain
\begin{equation}
{\partial^2 V\over \partial x^2} = -k^2 V(x)
\label{eq:I107}
\end{equation}
where we introduced the wave number
\begin{equation}
k = \sqrt{\omega(-i R_0 C_0 + \omega L_0 C_0)}
\label{eq:I108}
\end{equation}
which indeed has the dimension of an inverse length since the parameters $R_0$,
$L_0$, and $C_0$ are taking per unit length. We note that in general $k$ is not
real so that only for an $LC$ transmission line ($R_0 = 0$) the propagation of
undamped waves becomes possible. It is straightforward to solve (\ref{eq:I107})
for the voltage yielding
\begin{equation}
V(x) = A e^{-ikx} + B e^{ikx}.
\label{eq:I109}
\end{equation}
We make use of (\ref{eq:I105}) to obtain the current
\begin{equation}
I(x) = {ik\over R_0 + i\omega L_0} (A e^{-ikx} - B e^{ikx}).
\label{eq:I110}
\end{equation}
If we attach a semi-infinite transmission line to the right of the point $x=0$
we only have waves traveling to the right, i.e.\ $B=0$. Then the impedance of
the transmission line at $x=0$ is
\begin{equation}
Z_{\infty}(\omega) = \sqrt{R_0 + i\omega L_0\over i\omega C_0}\ .
\label{eq:I111}
\end{equation}

In reality, a transmission line has a finite length $\ell$. Let us determine the
impedance $Z$ at $x=0$ for a transmission line terminated at $x=\ell$ by a load
impedance $Z_L$. This leads to the boundary condition $V(\ell)=Z_LI(\ell)$ at
the end of the line. From (\ref{eq:I109}) and (\ref{eq:I110}) we then get
\begin{equation}
V(x) = \frac{I(\ell)}{2} \left[(Z_{\infty}+Z_L) e^{-ik(x-\ell)} +
(Z_L-Z_{\infty}) e^{ik(x-\ell)}\right]
\label{eq:voltterm}
\end{equation}
and
\begin{equation}
I(x) = \frac{I(\ell)}{2Z_{\infty}} \left[(Z_{\infty}+Z_L) e^{-ik(x-\ell)} -
(Z_L-Z_{\infty}) e^{ik(x-\ell)}\right].
\label{eq:currterm}
\end{equation}
The impedance at $x=0$ is given by $Z=V(0)/I(0)$ for which we find
\begin{equation}
Z = Z_{\infty} {e^{2ik\ell} - \lambda \over e^{2ik\ell} + \lambda}\ .
\label{eq:I112}
\end{equation}
Here, we introduced the reflection coefficient

\break
\begin{equation}
\lambda = {Z_{\infty} - Z_L\over Z_{\infty} + Z_L}
\label{eq:I113}
\end{equation}
which is obtained from (\ref{eq:voltterm}) as the negative of the ratio between
the voltages at $x=\ell$ of the reflected and
incident waves. For $Z_{\infty}\gg Z_L$ we have a short at the end of the
line and the voltage vanishes there. In the opposite limit $Z_L\gg Z_{\infty}$
the line is open at its end and the voltage has a maximum.

According to the form of the impedance (\ref{eq:I111}) we may
distinguish between two cases of relevance as far as the effect on Coulomb
blockade phenomena is concerned. If the relevant frequencies
of order $E_c/\hbar$
are much larger than $R_0/L_0$ we may neglect the resistance in
(\ref{eq:I111}) and consider an $LC$ transmission line. If, on the other hand,
the relevant frequencies are much smaller than $R_0/L_0$ we may neglect the
inductance and end up with an $RC$ transmission line. Typical experimental
values for the capacitance and inductance per unit length
are of the order of $C_0\simeq 10^{-16} \hbox{F/$\mu$m}$ and
$L_0\simeq 10^{-13} \hbox{H/$\mu$m}$. Therefore, the adequate model depends to
a large extent on the specific resistance of the wire material. For a pure
metal like aluminium the wire resistance is typically of the order of
$R_0\simeq 10^{-3}
\hbox{$\Omega/\mu$m}$ and the crossover frequency is then
$R_0/L_0 \simeq 10^{10}
\hbox{Hz}$. For capacitances in the fF-range this frequency is below
$E_c/\hbar$ and the $LC$ transmission line model is applicable.
On the other hand, for wires
made of high resistive alloys, $R_0$ may be larger than $10\hbox{$\Omega/\mu$m}$
and the crossover frequency then exceeds $10^{14}\hbox{Hz}$. In this case
the $RC$ line will render a reasonable description. In the following two
sections we discuss the influence of these transmission lines on charging
effects more specifically.
%
%
\subsection[$LC$ transmission line]
{{\protect\boldmath $LC$} transmission line}
The limit of an $LC$ transmission line is obtained from the case considered in
the previous section by setting $R_0 = 0$.  The
wave number of the solutions (\ref{eq:I109}) and (\ref{eq:I110}) becomes
$k=\omega(L_0C_0)^{1/2}$
and thus describes waves propagating along the line with velocity $u =
1/(L_0C_0)^{1/2}$. From (\ref{eq:I111}) it follows that the
impedance of an infinite line is purely Ohmic, i.e.\ $Z_{\infty} =
(L_0/C_0)^{1/2}$.
The line impedance $Z_{\infty}$ varies with
geometry and typically ranges between 10 and a few 100 $\Omega$. Hence, it is
of the order of the
free space impedance $(\mu_0/\varepsilon_0)^{1/2} = 377\Omega$, that is much
smaller than the quantum resistance $R_K$. As discussed in the previous section
we may terminate the line at $x=\ell$ with a load resistor $Z_L$ and get
for the external impedance
\begin{equation}
Z(\omega) = Z_{\infty} {\exp(2i\omega\ell/u)-\lambda\over\exp(2i\omega\ell/u) +
\lambda}.
\label{eq:I114}
\end{equation}
This impedance exhibits resonances at $\omega_n = \pi n u/\ell$ for $Z_{\infty}
\ll Z_L$ and at $\omega_n = \pi(n-1/2)u/\ell$ for $Z_{\infty}\gg Z_L$. From
our experience with the single mode model we expect
these features to show up in the differential current-voltage characteristic as
steplike increases of the dynamic conductance ${\rm d}I/{\rm d}V$ at the voltages
$\hbar\omega_n / e$. Every step corresponds to a new inelastic channel which is
opened as the voltage increases.

To see this more explicitly we assume
$Z_{\infty},Z_L \ll R_K$ which is frequently the case. Under these conditions
we may expand $\exp[J(t)]$ in the definition
(\ref{eq:I38}) of $P(E)$.
Keeping the first two terms we get by virtue of (\ref{eq:I66}) for the
current-voltage characteristic \cite{Yuli16}
\begin{eqnarray}
I(V) &=& {1\over eR_T}\left[eV + \int_{-\infty}^{+\infty}{{\rm d}E\over E}
{1\over 1-e^{-\beta E}} {{\rm Re} Z_t(E/\hbar)\over R_K}\right.\nonumber\\
&&\hspace{5cm}\left.\times\left({(eV-E)(1-e^{-\beta eV})\over 
1-e^{-\beta(eV-E)}}-eV \right) \right].
\label{eq:I115}
\end{eqnarray}
As mentioned earlier, the environmental effect becomes more apparent in
derivatives of the current-voltage characteristic.
Fig.~9
\begin{figure}
\begin{center}
\includegraphics{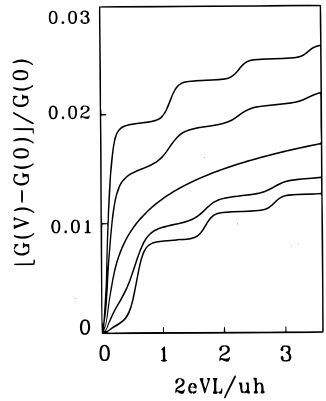}
\end{center}
\caption{
Results of numerical calculations of the differential
conductance of a tunnel junction attached to an $LC$-line of finite length
$L$. The Ohmic conductance $G(0)=V/R_T$ has been subtracted off. The line
impedance $Z_{\infty}=50\,\Omega$, and the ratio
$Z_L/Z_{\infty} = 10, 3, 1, 1/3, 1/10$ from the upper to the lower curve. The
temperature is 4.2 K and the voltage is taken in units of $uh/2eL=12\,$mV. In
this figure taken from Ref.\ \protect\cite{Yuli16} the length of the line 
is denoted by $L$ instead of $\ell$ used in the text.}
\end{figure}
shows numerical results for the differential conductance $G(V) = {\rm d}I/{\rm
d}V$ for various ratios $Z_L/Z_{\infty}$ and finite temperature $T=4.2\,$K.
For simplicity, the difference between $Z_t(\omega)$ and $Z(\omega)$ was
neglected which is appropriate if the junction capacitance is very small.
The expected
steps can be seen very clearly except for the case $Z_L = Z_{\infty}$ where the
terminating resistance matches the line impedance and thus no resonances are
present.
%
%
\subsection[$RC$ transmission line]
{{\protect\boldmath $RC$} transmission line}
We now consider the $RC$ transmission line which is obtained in the limit
$L_0=0$ from the more general model discussed above. For the impedance of an
infinite line we obtain from (\ref{eq:I111})
\begin{equation}
Z(\omega) = \sqrt{R_0\over i\omega C_0}
\label{eq:I116}
\end{equation}
so that the impedance increases with decreasing frequency. For a finite line
there will always be a cutoff and $Z(\omega)$ remains finite for $\omega\to 0$.
From (\ref{eq:I116}) the total impedance of an infinite line takes the form
\begin{equation}
Z_t(\omega) = {1\over i\omega C + \sqrt{i\omega C_0/R_0}}\ .
\label{eq:I117}
\end{equation}
Since the influence of the environment depends on the ratio between
$Z_t(\omega)$ and $R_K$, the relevant
dimensionless parameter is $\kappa = (R_0 C/C_0)/R_K $. This
gives the resistance of a piece of wire whose capacitance equals the
capacitance of the tunnel junction.

In the limit  $\kappa\rightarrow \infty$ we approach the high impedance limit
and a classical Coulomb blockade picture emerges. The
tunneling is completely suppressed for $V < e/2C$. At higher voltages we find
the shifted linear characteristic $I=V-e/2C$. If $\kappa$ is finite but large
the sharp curve is smoothed and an exponentially small tunneling current
appears for voltages below $e/2C$. This is in accordance with our earlier
findings.

It is surprising that there is also a substantial suppression of tunneling
in the opposite case $\kappa \ll 1$. In this limit we see from (\ref{eq:I117})
that at frequencies of order $E_c/\hbar$
the effective shunt resistance is much smaller than
$R_K$. Hence, there is no blockade in this region. However, at lower
frequencies the impedance increases and reaches $R_K$ for
frequencies of the order of $eV_c/\hbar$ where $V_c=2e R_0/(C_0R_K) = 2\kappa
e/C$. Provided the line is sufficiently long, so that $Z_t(\omega)$ does not
yet saturate at $\omega\simeq eV_c/\hbar$, tunneling is strongly suppressed
for voltages smaller than $V_c$.
It is worth noting that this phenomenon does not depend on the junction
capacitance.

Thus, for a long $RC$ line environment two types of Coulomb blockade
exist.\cite{YuliT4}\index{Coulomb blockade}
Let us consider in more detail the second type of blockade occurring in the
limit $\kappa\ll 1$ for voltages below $V_c$. For small frequencies the total 
impedance is approximately given by the line impedance (\ref{eq:I116}). At zero
temperature we then may calculate $P(E)$ for small energies and find
\begin{equation}
P(E) = \sqrt{\frac{eV_c}{4\pi E^3}}\exp(-eV_c/4E).
\label{eq:perc}
\end{equation}
According to (\ref{eq:I68}) the second derivative of the current-voltage
characteristic with respect to the voltage is proportional to $P(E)$ and
therefore its low voltage behavior is determined by (\ref{eq:perc}). As a
consequence the current is suppressed exponentially like $\exp(-V_c/4V)$ at
very small voltages.
The current-voltage characteristic of a junction coupled to an $RC$-line
together with its second derivative is presented in Fig.~10.
\begin{figure}
\begin{center}
\includegraphics{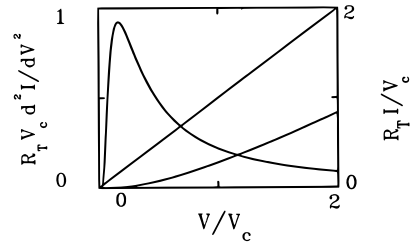}
\end{center}
\caption{
Coulomb blockade of the second type in a tunnel junction
attached to an $RC$-line. The current-voltage characteristic shows a significant
deviation from the straight line representing Ohm's law at voltages $V\simeq
V_c$. On the other hand, the current is suppressed exponentially only for voltages up to
about $0.05 V_c$ as can be seen from the second derivative of the $I$-$V$
characteristic which rises sharply above this voltage.}
\end{figure}
The current is noticeably
suppressed for voltages $\simeq 2 V_c$ whereas it becomes exponentially small
only for voltages below about 0.05$V_c$. This is quite unusual for a 
one-parameter behavior.
Temperatures of the order of $eV/k_B$ wash out the
suppression of tunneling at voltages less than $eV$ and Ohm's law is
restored for $eV\ll k_BT$. Finally, we note
that at large voltages $V\gg e/(\kappa C)$ the effect of the junction
capacitance  dominates resulting in an offset of the $I$-$V$ curve by $e/2C$.
\index{environment!electromagnetic|)}
 
%
%
\section{Tunneling rates in Josephson junctions}\index{junction!Josephson|(}
%
%
\subsection{Introduction}
So far we have studied the effect of an external circuit on
electron tunneling rates in normal tunnel junctions. It is also interesting to
consider Josephson junctions. In this case we have 
two kinds of charge
carriers, namely Cooper pairs and quasiparticles. While the concepts developed
for normal junctions are still valid, it will turn out that the
influence of the environment on Cooper pair tunneling is even simpler to
describe than its effect on electron tunneling in normal junctions considered
so far. 
Furthermore, the
supercurrent provides a more direct mean to measure the environmental
influence. The experimental relevance of superconducting 
junctions is also due
to the fact that most metals
used to fabricate tunnel junctions become superconducting at sufficiently low
temperatures. Often one even has to apply an external magnetic field to drive 
these junctions normal. 

As in the previous sections, we will concentrate on the environmental influence
on single charge tunneling. For other aspects of ultrasmall Josephson
junctions we refer the reader to Chap.~4 and to 
the review provided by
Ref.~\cite{SZPREP90}. In the previous sections, the concept of a phase proved 
to be very useful to
determine the current-voltage characteristics of normal tunnel junctions. It is
clear that the phase will be even more important in the superconducting
case where it has a non-vanishing expectation value due to the
long-range order in the superconducting leads.
In contrast to the phase which is usually introduced by means of the
Josephson relation we keep the phase as defined in (\ref{eq:phase}). For
quasiparticle tunneling this is the adequate choice. The factor of two
explicitly appearing in expressions for the supercurrent will always remind us
of the fact that Cooper pairs carry twice the electron charge. According
to the commutator (\ref{eq:comm}), the operator $\exp(-2i\varphi)$ leads to
a change of the junction charge by $2e$ associated with the tunneling of a
Cooper pair.

For normal tunnel junctions we have seen that the relevant energy scales were
the charging energy $E_c$ and the thermal energy $k_BT$. In Josephson 
junctions
an additional energy scale appears in form of the Josepson coupling energy
$E_J$. One may distinguish the regime $E_c\gg E_J$, which means that the
charge is well defined, from the regime $E_J\gg E_c$ where the phase 
fluctuates only little. We will calculate the tunneling rates for weak
Josephson coupling and then present a duality transformation which relates the
weak coupling regime to the strong coupling regime.
In the following we
treat the tunneling of Cooper pairs and of quasiparticles separately thereby
neglecting effects which couple the two processes. 
%
%
\subsection{Tunneling of Cooper pairs}
In this section we consider the tunneling of Cooper pairs in an ultrasmall
Josephson junction. We neglect quasiparticle 
excitations which is a good
approximation at temperatures very low compared to the critical temperature 
and voltages below the gap voltage $2\Delta/e$, where $2\Delta$ is the 
superconducting gap.
Before we start calculating the tunneling rates, we need to discuss the
main differences between Cooper pair tunneling and quasiparticle tunneling. In
Sec.~3.1.\ we had decomposed the total Hamiltonian (\ref{eq:Htot}) into
contributions of the quasiparticles and the environment, and both were coupled 
by the tunneling Hamiltonian. In contrast, for Cooper pair tunneling we only 
have a Hamiltonian acting in the Hilbert space of $Q$, $\varphi$ and the
environmental degrees of freedom. The Cooper pairs form a condensate and
therefore do not lead to additional dynamical degrees of freedom. The only consequence
of the coupling between the superconducting leads is the Josephson energy given
by the second term in the total Hamiltonian
\begin{equation}
H = H_{\rm env} + E_J\cos(2\varphi).
\label{eq:Htotsuper}
\end{equation}
The environmental Hamiltonian was defined in (\ref{eq:I16}) and remains
unchanged. Rewriting the Josephson term as
\begin{equation}
	E_J\cos(2\varphi) = \frac{E_J}{2}e^{-2i\varphi} + \hbox{H.c.}
\label{eq:Joseph}
\end{equation}
we see that it replaces the electron tunneling Hamiltonian $H_T$ defined in
(\ref{eq:HT1}). The operator $e^{-2i\varphi}$ changes the charge $Q$ on the
junction by $2e$. This process is connected with the tunneling of a Cooper
pair, although the Cooper pairs appear in the Hamiltonian only through the
phase difference between the condensate wave functions on both sides of the
barrier.  The Hamiltonian (\ref{eq:Htotsuper}) is similar to the
total Hamiltonian (\ref{eq:Htot}) for
electron tunneling except that there are no electronic degrees of freedom. This
allows us to calculate tunneling rates for Cooper pairs in the spirit of
Sec.~3.2. However, the steps performed in Sec.~3.2.1.\ are now obsolete and we
can start the calculation by tracing out the environment. 
Considering
forward tunneling, the expression analogous to (\ref{eq:I26}) reads
\begin{equation}
\GF(V) = \frac{\pi}{2\hbar}E_J^2\sum_{R,R'}\vert\langle R\vert e^{-2i\varphi}
\vert R'\rangle\vert^2P_{\beta}(R)\delta(E_R-E_R^{\prime}).
\label{eq:srate1}
\end{equation}
This is just the golden rule rate with (\ref{eq:Joseph}) as perturbation
averaged over an equilibrium distribution of initial states.
So far we have kept the dependence on the external voltage in the phase. The
trace over the environmental degrees of freedom is performed like in
Sec.~3.2.2. We then arrive at
\begin{equation}
\GF(V) = \frac{E_J^2}{\hbar^2}\int_{-\infty}^{+\infty}{\rm d}t\,\exp\!\left(
\frac{2i}{\hbar}eVt\right)\langle e^{2i\tilde\varphi(t)} e^{-2i\tilde\varphi(0)}
\rangle
\label{eq:srate2}
\end{equation}
where we introduced $\tilde\varphi(t)$ according to (\ref{eq:phasfluk}). We
may again exploit the generalized Wick theorem and express the correlation
function in (\ref{eq:srate2}) in terms of the phase-phase correlation function
$J(t)$ given by (\ref{eq:I51}). In analogy to (\ref{eq:I38}) we define
\begin{equation}
P'(E)=\frac{1}{2\pi\hbar}\int_{-\infty}^{\infty} {\rm d}t 
\exp\!\left[4J(t)+\frac{i}{\hbar}Et\right]
\label{eq:psvone}
\end{equation}
and get for the forward tunneling rate for Cooper pairs
\begin{equation}
\GF(V)=\frac{\pi}{2\hbar}E_J^2P'(2eV).
\label{eq:srate3}
\end{equation}
The probability $P'(E)$ differs from the probability $P(E)$ introduced for
normal junctions only by a factor 4 in front of the phase-phase correlation 
function $J(t)$ which arises from the fact 
that the charge of Cooper pairs is twice the electron charge. In view of the
relation (\ref{eq:I51}) for the correlation function $J(t)$ we may absorb this
factor into the ``superconducting resistance quantum'' $R_Q=h/4e^2=R_K/4$.
Since the total impedance must now be compared with $R_Q$
the influence of the environment on Cooper pair tunneling rates,
and thus on the supercurrent, is more
pronounced than for the current through a normal junction. 

Before calculating the supercurrent let us briefly discuss the range of
validity of the perturbative result (\ref{eq:srate3}). Since the Josephson
coupling was considered as a perturbation, the Josephson energy $E_J$ has to be
small. From an analysis of higher order terms, one finds that our lowest order
result is correct if $E_JP'(2eV)\ll 1$. Obviously, this condition depends on
the voltage and on the environmental impedance. To be more specific let us
choose an Ohmic environment and zero temperature. If the impedance $Z$ is of
the order of the resistance quantum $R_Q$, the 
probability $P'(2eV)$ will be a broad distribution with a maximum height of the
order of the inverse charging energy $E_c^{-1}$ (cf.\ Fig.~6a). Then our rate 
expression is 
correct if $E_J\ll E_c$. On the other hand, for a high impedance environment, 
$P'(2eV)$ is peaked around $E_c$. Now, $P'(E_c)$ is found to be of order
$(1/E_c)(Z/R_Q)^{1/2}$ for $Z\gg R_Q$ and the rate formula (\ref{eq:srate3})
holds provided the condition $E_J\ll E_c(R_Q/Z)^{1/2}$ is satisfied. This
latter condition is more restrictive.  In the 
opposite case of a low impedance environment $P'(2eV)$ is sharply peaked at 
$V=0$ and decreases with increasing voltage. The condition $E_J\ll 1/P'(2eV)$
is then always violated at sufficiently low voltages.

From the rate expression (\ref{eq:srate3}) together with the symmetry
$\GB(V)=\GF(-V)$ we immediately get for the supercurrent \cite{Ysuper}
\begin{equation}
I_S(V) = 2e\Big(\GF(V)-\GB(V)\Big) = \frac{\pi e E_J^2}{\hbar}\Big(P'(2eV) -
P'(-2eV)\Big)
\label{eq:is1}
\end{equation}
where we accounted for the charge $2e$ which each tunneling process transports. 
This result reflects the fact that a Cooper pair tunneling in
the direction of the applied voltage carries an energy $2eV$. This energy 
has to be transferred to the environment since the
Cooper pairs have no kinetic energy that could absorb a part of $2eV$. 
The probability for this transfer of energy is $P'(E)$.
Since the supercurrent depends directly on the probability $P'(E)$, it enables 
one to measure properties of the environment more directly. For normal 
junctions it was necessary to
measure the second derivative of the current-voltage characteristic which is
more complicated. On the other hand, this relation in principle provides
a possibility to check the consistency of the theory. Of course, one always has
to account for the relative factor of 4 in the definitions of $P(E)$ and
$P'(E)$. 

In Sec.~3.4.\ we had derived some general properties of the probability $P(E)$.
The sum rules discussed there now become sum rules for supercurrent-voltage
characteristics at zero temperature. Since at $T=0$ the probability $P'(E)$ 
vanishes for negative energies, we may directly employ (\ref{eq:I52}) and
(\ref{eq:I53}) yielding
\begin{equation}
\int_0^{\infty} {\rm d}V I_S(V) = \frac{\pi E_J^2}{2\hbar}
\label{eq:isum1}
\end{equation}
and
\begin{equation}
\int_0^{\infty} {\rm d}V\,VI_S(V) = \frac{\pi e E_J^2}{2\hbar C}.
\label{eq:isum2}
\end{equation}

For specific environments one may derive further properties of the
supercurrent-voltage characteristics in accordance with our discussion of
$P(E)$ for normal junctions. Here, we concentrate on an Ohmic impedance
$Z(\omega)=R=R_K/g$ and consider first the case of zero temperature. As for
normal junctions we find a zero-bias anomaly which is now given by $I_S\sim
V^{2/g-1}$. This behavior is shown in Fig.~11 
\begin{figure}
\begin{center}
\includegraphics{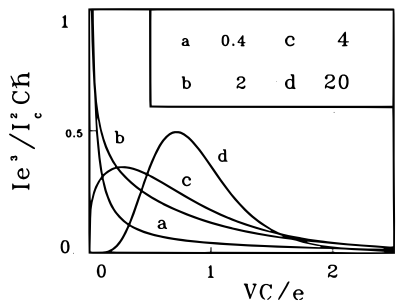}
\end{center}
\caption{Zero-temperature supercurrent-voltage characteristics for a Josephson
junction in the limit $E_J\ll E_c$. The junction is coupled to Ohmic resistors
with four different values given in k$\Omega$. $I_c=2eE_J/\hbar$ is the
critical current.}
\end{figure}
where supercurrent-voltage
characteristics are shown for various values of the dimensionless conductance
$g$. For $g>2$ the supercurrent is peaked at $V=0$ and decreases with
increasing voltage. On the other hand, for $g<2$ the supercurrent increases
with voltage for small $V$, thereby leading to a peak at finite
voltage. For rather small conductances a well marked gap is present at small
voltages.  Fig.~11 of course corresponds to Fig.~6a which shows $P(E)$ at zero
temperature for a normal junction coupled to an Ohmic environment.

Let us now have a closer look at the peak developing at $V=e/C$ for low
conductance $g$. Thermal and quantum fluctuations broaden this peak. Its
shape close to the maximum is given by the Gaussian
\begin{equation}
I_S(V) = I_{\rm max}\exp\left[-\frac{(V-e/c)^2}{W}\right]
\label{eq:ishi}
\end{equation}
with the width \cite{YuliS5}
\begin{equation}
W = \left\{
         \begin{array}{ll}
           {\displaystyle \frac{e^2}{2\pi^2C^2}g}
                          &\ \ \ \mbox{for $k_BT\ll{\displaystyle 
                                            \frac{e}{C}}g$}\\
                          \vrule height 2pt width 0pt&\\
           {\displaystyle \frac{2}{C}k_BT}&\ \ \ \mbox{for 
                             $k_BT\gg{\displaystyle \frac{e}{C}}g$}.\\
                         \end{array}\right.
\label{eq:width}
\end{equation}
In the first case, for very low temperatures, the peak is broadened by quantum
fluctuations which decrease as the conductance is decreased. The second case
describes the thermal broadening in analogy to the result (\ref{eq:I73})
derived for normal tunnel junctions in the high impedance limit where  
the conductance $g\ll k_BTC/e$. 
%
%
\subsection{Charge-phase duality and incoherent tunneling of the phase}
\index{duality!charge-phase}
In the previous subsection we have discussed the case of weak Josephson
coupling where the charge on the junction is well defined. For the following 
discussion it
is convenient to express the Hamiltonian (\ref{eq:Htotsuper}) in terms of
charge states $\vert N\rangle$, for which $Q=2eN$. As mentioned before, the
operator $e^{-2i\varphi}$ changes the charge $Q$ by $2e$. From
(\ref{eq:Joseph}) we then find the equivalence
\begin{equation}
E_J\cos(2\varphi) \leftrightarrow \frac{E_J}{2}\sum_N\left( \vert
N+1\rangle\langle N\vert + \vert N\rangle \langle N+1\vert\right).
\label{eq:qpequiv}
\end{equation}
In the charge representation the Hamiltonian may be written as 
\begin{equation}
H = \frac{E_J}{2}\sum_N \left(\vert N+1\rangle\langle N\vert + \vert N\rangle
\langle N+1\vert\right) + 2e(V+\tilde V)\sum_N N\vert N\rangle \langle N\vert
\label{eq:qHamil}
\end{equation}
where the environment couples to the charge via the external voltage $V$ and a
voltage operator $\tilde V$ describing the voltage fluctuations at the junction
induced by the environment. The Hamiltonian (\ref{eq:qHamil}) could
alternatively be used to derive the expressions for the tunneling rates.

In the limit of large Josephson coupling $E_J\gg E_c$ the phase is well defined
and localized in one of the wells of the Josephson potential. We introduce
phase states $\vert n\rangle$ where the phase is given by $\varphi=\pi n$.
Using these states we may write the Hamiltonian as
\begin{equation}
H = \sum_n \Delta_0 \left(\vert n+1\rangle \langle n\vert + \vert n\rangle
\langle n+1\vert\right) + \frac{\pi\hbar}{e}(I+\tilde I)\sum_n n\vert n\rangle 
\langle n\vert.
\label{eq:tight}
\end{equation}
The first term describes tunneling of the phase from one well to a neighboring
one and $\Delta_0$ is the tunnel matrix between adjacent ground states in the
wells. The
second term couples the phase to an external current $I$ and an operator 
$\tilde I$ describing a fluctuating current through the Josephson junction
caused by the environment. The tight-binding Hamiltonian
(\ref{eq:tight}) makes sense if only the ground states in the wells can be
occupied. The excitation energy is related to the oscillation frequency in the
well given by
$(2E_JE_c)^{1/2}/\hbar$, and we thus find that this approach is valid if
frequency, current, and temperature fulfill the requirements $\omega, I/e,
k_BT/\hbar \ll (E_JE_c)^{1/2}/\hbar$.

Since phase and charge have to be exchanged when going from (\ref{eq:qHamil})
to (\ref{eq:tight}), i.e.\ from the weak coupling
limit to the strong coupling limit, the influence of the environment is now
described by the charge-charge correlation function $\langle[\tilde Q(t)-
\tilde Q(0)]\tilde Q(0)\rangle$ replacing the phase-phase correlation function
$J(t)$. The charge $\tilde Q$ is related to the fluctuating current $\tilde I$
by
\begin{equation}
\tilde Q(t) = \int_{-\infty}^t {\rm d}t'\, \tilde I(t').
\label{eq:qstilde}
\end{equation}
The correlation function of the fluctuating current is determined by the
environmental admittance $Y(\omega)$ as
\begin{equation}
\langle \tilde I(t) \tilde I(0)\rangle = \int_0^{\infty} \frac{{\rm d}\omega}
{\pi}\, \hbar\omega {\rm Re}[Y(\omega)]
\left\{\coth\left(\frac{1}{2}\beta
\hbar\omega\right)\cos(\omega t)-i\sin(\omega t)\right\}.
\label{eq:icorr}
\end{equation}
From (\ref{eq:qstilde}) we then get for the charge-charge correlation
function 
\begin{eqnarray}
\langle[\tilde Q(t)-\tilde Q(0)]\tilde Q(0)\rangle &=& \frac{\hbar}{\pi}
\int_0^{\infty} \frac{{\rm d}\omega}{\omega}\,{\rm Re}[Y(\omega)]\nonumber\\
&&\hspace{2cm}\times\left\{\coth\left(\frac{1}{2}\beta
\hbar\omega\right)[\cos(\omega t)-1]-i\sin(\omega t)\right\}.
\label{eq:qcorr}
\end{eqnarray}
Now we are able to transform
results obtained for weak coupling into results for the strong
coupling case by means of simple replacements. Comparing (\ref{eq:qHamil}) 
and (\ref{eq:tight}) we see that we
have to replace the Josephson coupling by the tunnel splitting, the voltage by
the current, and the Cooper pair charge by the flux quantum. Furthermore,
according to (\ref{eq:phase}) and (\ref{eq:qstilde}) the charge replaces the
phase, and according to (\ref{eq:I51}) and (\ref{eq:qcorr}) we have to
substitute the environmental admittance for the total impedance. Thus, we 
arrive at the well-known phase-charge duality transformations \cite{SZPREP90} 
\begin{equation}
\frac{E_J}{2} \Leftrightarrow \Delta_0\ \ \ \ \ 
V \Leftrightarrow I\ \ \ \ \ 
2e \Leftrightarrow \frac{h}{2e}\ \ \ \ \ 
\varphi \Leftrightarrow \frac{\pi}{2e}Q\ \ \ \ \ 
Z_t(\omega) \Leftrightarrow Y(\omega).
\label{eq:subst}
\end{equation}
The process dual to the tunneling of Cooper pairs in the weak coupling limit is
incoherent tunneling of the phase in the strong coupling limit.

In Fig.~12
\begin{figure}
\begin{center}
\includegraphics{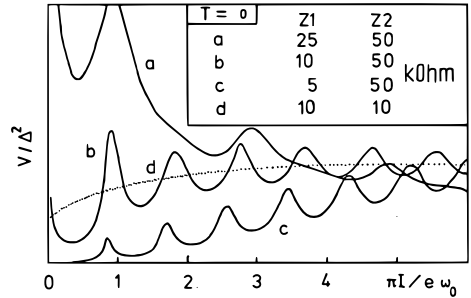}
\end{center}
\caption{Voltage across a strongly coupled Josephson 
junction at zero
temperature. The junction is connected to a finite $LC$ transmission line of
length $\pi u/\omega_0$ where $u$ is the wave propagation velocity. The line
impedance $Z_1$ and the load impedance $Z_2$ take different values given in the
insert. In this figure taken from Ref.~\protect\cite{Ysuper} $Z_1$ and $Z_2$
correspond to $Z_{\infty}$ and $Z_L$ as defined in Sec.~4.4., respectively.}
\end{figure}
we give an example of the voltage as a function of the bias current of a
Josephson junction in the strong coupling regime as calculated from the
equation dual to (\ref{eq:is1}). The environment is described by 
an $LC$ transmission line of length
$\ell$ as discussed in Sec.~4.5. For $Z_{\infty}\ll R_K \ll Z_L$ one 
finds peaks separated by current intervals $e\omega_0/\pi$ where 
$\omega_0=\pi u/\ell$ and $u$ is the wave propagation velocity. If one
averages over the oscillations of the characteristic one finds the
corresponding characteristic for an Ohmic resistor $Z_{\infty}$, which for 
curve b is curve d. Such an averaging occurs when the temperature becomes of the
order of $\hbar I/e$.  

%
%
\subsection{Tunneling of quasiparticles}\index{quasiparticle!tunneling}
In Josephson junctions apart from Cooper pairs 
also quasiparticles may tunnel. Basically, we have to treat 
quasiparticle tunneling in a superconducting junction like quasiparticle
tunneling in a
normal junction. Therefore, most of the calculations performed in Sec.~3.2.\
can be taken over to the superconducting case. There is however one important
difference. In Sec.~3.2.1.\ we had assumed that the density of states at the
Fermi surface is constant. In a superconductor the quasiparticle
density of states close to the gap
depends very strongly on energy. Within the BCS-theory one finds for the
reduced quasiparticle density of states \cite{Tinkham}
\begin{equation}
\frac{N_S(E)}{N(0)} = \left\{
                         \begin{array}{ll}
                          {\displaystyle \frac{\vert E\vert}
                              {(E^2-\Delta^2)^{1/2}}} &\ \ \ \mbox{for
                                                    $\vert E\vert>\Delta$}\\
                          \vrule height 2pt width 0pt\\
                          0 &\ \ \ \mbox{for $\vert E\vert<\Delta$.}
                         \end{array}\right.
\label{eq:superdos}
\end{equation}
The density of states is taken relative to the density of states in the 
normal metal at an energy in the middle of the gap. $2\Delta$ is again the 
size of the superconducting gap within which the quasiparticle density of 
states 
vanishes. For the forward tunneling rate we then have as an extension of
(\ref{eq:I39})
\begin{eqnarray}
\GF(V) &=& \frac{1}{e^2R_T}\int_{-\infty}^{+\infty}\!{\rm d}E{\rm d}E'\,
\frac{N_S(E)N_S(E'+eV)}{N(0)^2}\nonumber\\ 
&&\hspace{4cm}\times f(E)[1-f(E'+eV)]P(E-E').
\label{eq:qprate}
\end{eqnarray}
Here, the probability to exchange energy with the environment is given by
$P(E)$ since quasiparticles carry the charge $e$. 

As for the normal tunnel junction we use the symmetry $\GB(V)=\GF(-V)$ to
obtain from the rate expression (\ref{eq:qprate}) the quasiparticle current
\begin{eqnarray}
I_{\rm qp}(V) &=& \frac{1}{eR_T} \int_{-\infty}^{+\infty}{\rm d}E {\rm d}E'\,
\frac{N_S(E)N_S(E')}{N(0)^2} \Big[f(E)[1-f(E')]P(E-E'+eV) \nonumber\\
&&\hspace{5.7cm}-f(E')[1-f(E)]P(E'-E-eV)\Big].
\label{eq:qpcurr1}
\end{eqnarray}
Using the detailed
balance symmetry (\ref{eq:I59}) of $P(E)$ and the relation (\ref{eq:ffrel}) 
for Fermi functions, 
this equation may be rewritten in a more convenient way as

\break
\begin{eqnarray} 
I_{\rm qp} &=& \frac{1}{eR_T}\int_{-\infty}^{+\infty}{\rm d}E{\rm d}E'\,
\frac{N_S(E')N_S(E'+E)}{N(0)^2} \frac{1-e^{-\beta eV}}{1-e^{-\beta E}}
\nonumber\\ 
&&\hspace{6cm}\times P(eV-E) [f(E')-f(E'+E)].
\label{eq:qpcurr2}
\end{eqnarray}
In the absence of an external impedance we recover the familiar quasiparticle
current of a voltage biased Josephson 
junction
\begin{equation}
I_{\rm qp,0}(V) = \frac{1}{eR_T}\int_{-\infty}^{+\infty}{\rm d}E
\frac{N_S(E)N_S(E+eV)}{N(0)^2}[f(E)-f(E+eV)].
\label{eq:qpcurr0}
\end{equation}
For zero temperature the integral may be evaluated yielding \cite{Werthamer}
\begin{equation}
I_{\rm qp,0}(V) = \frac{\Delta}{eR_T}\left[2xE(m)-\frac{1}{x}K(m)\right]\ \ \ \
\mbox{for $x>1$}
\label{eq:giaever}
\end{equation}
where $m=1-1/x^2$ with $x=eV/2\Delta$. $K(m)$ and $E(m)$ are the complete
elliptic integrals of the first and second kind, respectively.\cite{Abramowitz}
For voltages below $2\Delta/e$ the quasiparticle
current vanishes as a consequence of the energy gap $2\Delta$. 
We may use (\ref{eq:qpcurr0}) to express the quasiparticle current in the 
presence of an environment as \cite{FBSEPL91}
\begin{equation}
I_{\rm qp}(V) = \int_{-\infty}^{+\infty}{\rm d}E \frac{1-e^{-\beta eV}}
{1-e^{-\beta E}} P(eV-E) I_{\rm qp,0}(\frac{E}{e}).
\label{eq:qpcurr}
\end{equation}
This expression is rather general. For example, if we insert for $I_{\rm
qp,0}(V)$ the Ohmic current-voltage characteristic of a normal tunnel junction
we directly get our earlier result (\ref{eq:I66}).

For an Ohmic environment and zero temperature we may calculate the 
current-voltage
characteristic for voltages slightly above $2\Delta/e$ by inserting the low
energy behavior (\ref{eq:I99}) of $P(E)$ into (\ref{eq:qpcurr}). Expanding
(\ref{eq:giaever}) we obtain to leading order 
\begin{equation}
I_{\rm qp}(V) = \frac{\pi g \Delta}{4eR_T} \frac{e^{-2\gamma/g}}{\Gamma(2/g)}
\left[\frac{\pi}{gE_c}(eV-2\Delta)\right]^{2/g}
\label{eq:gappot}
\end{equation}
where $g=R_K/R$ is the Ohmic lead conductance and $\gamma$ is again the Euler
constant. As for normal junctions and the
supercurrent in Josephson junctions we find 
an anomaly $I_{\rm qp} \sim
(eV-2\Delta)^{2/g}$ \cite{FBSEPL91}, which now is shifted in voltage 
by $2\Delta/e$. Fig.~13 shows
\begin{figure}
\begin{center}
\includegraphics{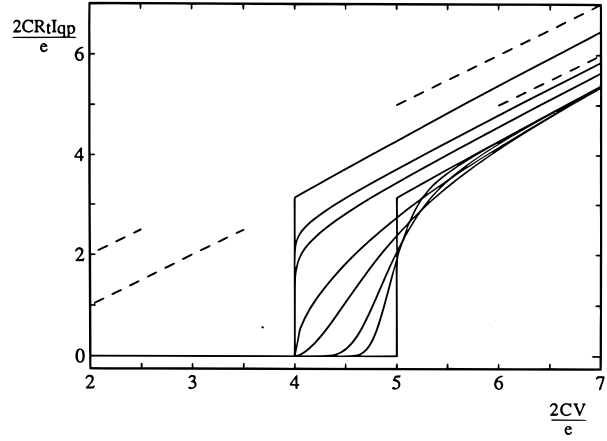}
\end{center}
\caption{The quasiparticle current-voltage characteristic at zero temperature
is shown for an Ohmic environment with $g=R_K/R=\infty, 40, 20, 4, 1.2, 0.2,
0.05,$ and 0 from left to right. The superconducting gap is choosen as $\Delta =
2E_c$. The two dashed lines represent the large voltage asymptotes for
$g=\infty$, i.e.\ for $I_{qp,0}$, and for the other values of $g$. Here, the
tunneling resistance is denoted by $R_t$. Taken from \protect\cite{FBSEPL91} 
with permission.}
\end{figure}
the formation of a Coulomb gap\index{Coulomb gap} with decreasing lead 
conductance $g$ in
accordance with the power law (\ref{eq:gappot}). For any nonvanishing
lead conductance the current-voltage characteristic for large voltages
approaches $I_{\rm qp,0}$ shifted in voltage by $e/2C$. As for normal junctions
the high voltage behavior exhibits a Coulomb gap for $g\ne 0$ even though the
gap might not be apparent at voltages close to $2\Delta/e$.

Finite temperature current-voltage characteristics for Ohmic environments with
different conductances are shown in Fig.~14.
\begin{figure}
\begin{center}
\includegraphics{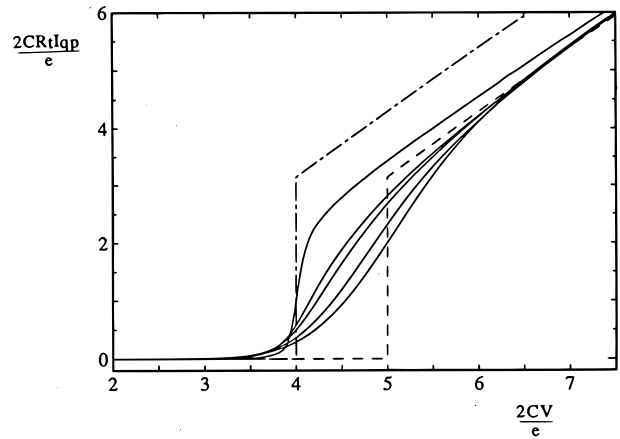}
\end{center}
\caption{The quasiparticle current-voltage characteristic for finite
temperature $k_BT/E_c=0.25$ is shown for an Ohmic environment with $g=R_K/R=
20, 4.8, 3.2, 1.2,$ and 0.2 from left to right. The superconducting gap is
choosen as $\Delta = 2E_c$. The dash-dotted line and the dashed line are
$I_{qp,0}$ and the same curve shifted by $E_c/e$, respectively, taken at the
same temperature. Here, the tunneling resistance is denoted by $R_t$. Taken
from \protect\cite{FBSEPL91} with permission.}
\end{figure}
Due to the finite temperature the gap is smeared. Interestingly, for voltages 
below
$2\Delta/e$ one finds an increase of the current due to the environmental
coupling. In contrast to the behavior at high voltages and the current in
normal tunnel junctions, the quasiparticle tunnel current increases with 
decreasing lead conductance for low voltages.

In the next section we consider multijunction systems and restrict the
discussion to normal junctions. For superconducting double junction systems the
combined tunneling of Cooper pairs and quasiparticles leads to new effects. For
details we refer the reader to the literature.\cite{Maassen}
\index{junction!Josephson|)}\index{junction!single|)}
 
%
%
\section{Double junction and single electron 
transistor}\index{junction!double|(}\index{SET transistor}
%
%
\subsection{Island charge}
In this section we discuss circuits of tunnel junctions. As we shall see, as
far as the calculation of tunneling rates is concerned, most of the new
features arising when several tunnel junctions are combined are already present
in a double junction setup. Hence, we shall mainly 
discuss two-junction systems
and briefly address more complicated circuits at the end of this section.
Systems containing two tunnel junctions in series as shown in Fig.~15
\begin{figure}
\begin{center}
\includegraphics{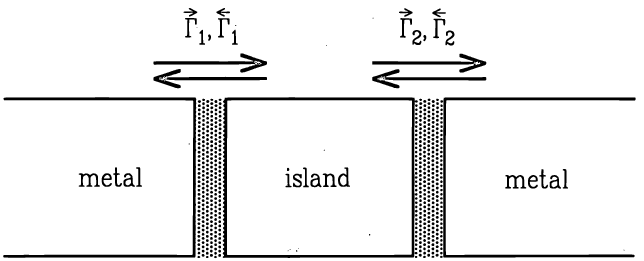}
\end{center}
\caption{
Schematic drawing of a metallic double junction system.
The arrows indicate forward and backward tunneling through the two insulating
barriers.}
\end{figure}
differ
significantly from a single junction\index{junction!single} because of the 
metallic island between the
two junctions.\cite{ALREV91,Kulik} While earlier work has entirely disregarded
the influence of the electromagnetic environment we shall take it into account
here following the line of reasoning in \cite{multizphys,triest}. The external 
circuit sees the two tunnel 
junctions with capacitance $C_1$ and $C_2$ as a capacitor of total capacitance
\begin{equation}
C = {C_1 C_2\over C_1 + C_2}.
\label{eq:D1}
\end{equation}
Since the voltage across the two junctions is $U = Q_1/C_1 + Q_2/C_2$ the
total charge seen from the outside is
\begin{equation}
Q = CU = {C_2Q_1 + C_1Q_2\over C_1 + C_2}.
\label{eq:D2}
\end{equation}
As for a single junction\index{junction!single} this charge is to be 
considered as a continuous
variable. The metallic island carries the charge
\begin{equation}
Q_1 - Q_2 = ne
\label{eq:D3}
\end{equation}
which may change only by tunneling of electrons to or from the island. This
leads to the quantization of the island charge which will turn out to be
very important for the behavior of double junction
systems. To describe the
charges on the capacitors one may either use $Q_1$ and $Q_2$ or $Q$ and $ne$.
The corresponding charging energy reads
\begin{equation}
{Q_1^2\over 2C_1} + {Q_2^2\over 2C_2} = {Q^2\over 2C} + {(ne)^2\over
2(C_1+C_2)}.
\label{eq:D4}
\end{equation}
Compared with the single junction the charging energy now contains a
contribution arising from the island charge.

In real double junction systems charged defects 
are frequently present in the
vicinity of the junction. They lead to an effective island charge where the
discrete set of island charges $ne$ is shifted by an offset 
charge.\index{offset charge}
To influence the effective island charge in a controlled way one frequently
uses the single electron transistor setup presented in Fig.~16
\begin{figure}
\begin{center}
\includegraphics{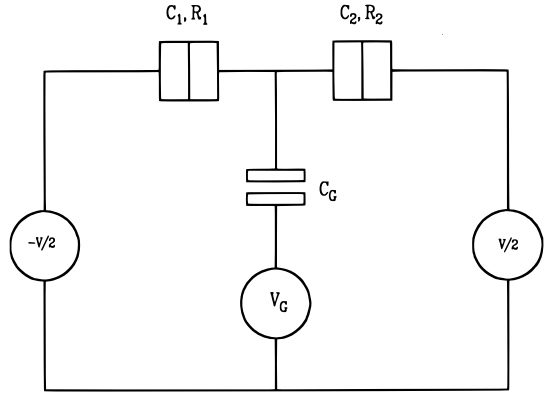}
\end{center}
\caption{The single electron transistor setup consisting of a double junction
with a control voltage source coupled capacitively to the island.}
\end{figure}
where a gate
voltage $V_G$ is coupled capacitively to the island. We will show now that
the voltage source $V_G$ together with the capacitance $C_G$ effectively leads
to a shift of the island charge by $Q_0 = C_G V_G$. To this end we first
determine the average charges on the capacitors in electrostatic equilibrium
for given applied voltages $V$ and $V_G$ and given island charge
\begin{equation}
ne = Q_1-Q_2-Q_G.
\label{eq:D5}
\end{equation}
Using Kirchhoff's law for two loops we find
\begin{eqnarray}
Q_1 &=& {C_1\over C_{\Sigma}}\left[(C_2+{C_G\over 2})V + C_GV_G + ne\right]
\label{eq:D6}\\
Q_2 &=& -{C_2\over C_{\Sigma}}\left[-(C_1+{C_G\over 2})V + C_GV_G + ne\right]
\label{eq:D7}\\
Q_G &=& -{C_G\over C_{\Sigma}}\left[{1\over 2}(C_2-C_1)V - (C_1+C_2)V_G + ne 
\right]
\label{eq:D8}
\end{eqnarray}
where we introduced the capacitance
\begin{equation}
C_{\Sigma} = C_1 + C_2 + C_G.
\label{eq:D9}
\end{equation}
We suppose now that an electron has tunneled through the left junction onto
the island thereby changing $Q_1$ into $Q_1-e$ and $ne$ into $(n-1)e$. The new
charges $Q_1-e$, $Q_2$, and $Q_G$ do no longer satisfy electrostatic
equilibrium
since the replacement of $n$ by $n-1$ in (\ref{eq:D6}-\ref{eq:D8}) does not
result in a change of $Q_1$ by $e$.
Equilibrium is reestablished by a transfer of charge through the
voltage sources leading to the following difference of charges before and after
the tunneling process
\begin{eqnarray}
\delta Q_1 &=& -(C_1/C_{\Sigma}) e = -e + \delta Q_2 + \delta Q_G
\label{eq:D10}\\
\delta Q_2 &=& (C_2/C_{\Sigma}) e
\label{eq:D11}\\
\delta Q_G &=& (C_G/C_{\Sigma}) e.
\label{eq:D12}
\end{eqnarray}
Apart from the energy transfer to or from the environmental modes the energy
determining the tunneling rates is the difference in electrostatic energy of
the entire circuit. In contrast to the case of a single junction, this energy 
difference now not
only consists of contributions from the work done by the various voltage
sources. It also has to account for the change in charging energy. The
total charging energy
may be decomposed into a contribution depending on the voltages $V$ and $V_G$
which does not change and a contribution $(ne)^2/2C_{\Sigma}$ depending on the
island charge. Thus the change 
in charging
energy is entirely due to the change of the island charge. We finally obtain
for the difference in electrostatic energy associated with the tunneling of an 
electron through the first junction onto the island
\begin{eqnarray}
&&{(ne)^2\over 2C_{\Sigma}}-{[(n-1)e]^2\over 2C_{\Sigma}} - {V\over 2} 
(\delta Q_1 + e) + {V\over 2}\delta Q_2 + V_G\delta Q_G\nonumber\\
&&\hspace{4.0truecm}= {e\over C_{\Sigma}}\left[(C_2 + {C_G\over 2})V + C_G V_G 
+ ne - {e\over 2}
\right].
\label{eq:D13}
\end{eqnarray}
The extra elementary charge added to $\delta Q_1$ is due to the fact that an
electron has tunneled through the first junction and therefore the charge
transferred by the voltage source is diminished by $-e$.
From the right hand side of (\ref{eq:D13}) it becomes clear now that the work
done by the
gate voltage source leads indeed to an effective island charge $q=ne+Q_0$ with
$Q_0=C_GV_G$.

If the gate capacitance $C_G$ is small compared to the junction capacitances
$C_1$ and $C_2$ and if no other impedance is present in the gate branch, the
only effect of $C_G$ is the shift of the effective island charge which we just 
discussed. To
retain this shift we may let the gate capacitance go to zero. However, we have
to keep the work done by the voltage source finite. In this limit the
charge on the gate capacitor (\ref{eq:D8}) is negligible as is the charge
(\ref{eq:D12}) transferred
after tunneling. On the other hand the gate voltage is assumed
to be sufficiently large to cause an offset charge $C_GV_G$.

In the following, we
will restrict ourselves to this limit where $C_G$ may be neglected. In the
literature the reader will find a more complete discussion of the effect of
gate and stray capacitances and of gate impedances.\cite{Wyrowski,OFSPRB91} 
In the limit we are
considering here, the single electron transistor is reduced to a double
junction with an offset of the island charge. We will 
therefore concentrate on
the double junction and discuss the effects arising due to the offset whenever
appropriate.
%
%
\subsection{Network analysis}
The system we are considering in the sequel is the double 
junction shown in
Fig.~17
\begin{figure}
\begin{center}
\includegraphics{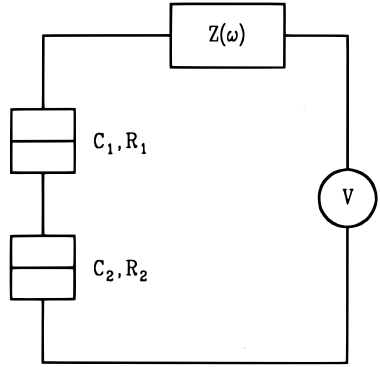}
\end{center}
\caption{
A double junction system with capacitances $C_1$, $C_2$
and tunneling resistances $R_1$, $R_2$ coupled to a voltage source $V$ via the
external impedance $Z(\omega)$.}
\end{figure}
which is
coupled to a voltage source via an external impedance $Z(\omega)$.
It would be straightforward to carry out a golden rule
calculation as we have done for the single junction\index{junction!single}. 
However, it turns out
that this is not necessary because some simple network
considerations yield the same result.\cite{multizphys} Furthermore, they will 
give us some
additional insight. Before starting we would like to mention an important
assumption underlying our approach. Second order perturbation theory or golden
rule is only sufficient if tunneling through both junctions may be
considered as uncorrelated. This means that when we are calculating tunneling
rates for one junction the other junction may be viewed as a capacitor.
Especially in the blockade region where our approach predicts no flow of
current, higher order perturbation theory leads to important corrections. These
are due to virtual transitions involving simultaneous tunneling through both
junctions. This so-called co-tunneling which is not hindered by the Coulomb
interaction is relevant if the tunneling resistances are no longer large
compared to the resistance quantum $R_K$.\cite{AveOdin} For a detailed discussion see
Chap.~6.

The basic rule which will be needed for the network analysis is the
transformation between the Thevenin and Norton configurations shown in Fig.~18.
\begin{figure}
\begin{center}
\includegraphics{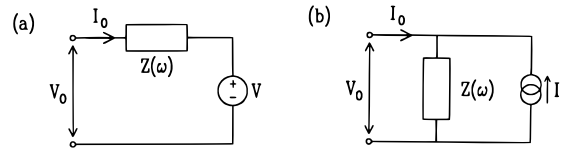}
\end{center}
\vspace{-5pt} 
\caption{(a) Thevenin configuration: A voltage source $V$ in series with the
impedance $Z(\omega)$. (b) The equivalent Norton configuration: A current
source $I(\omega)=V(\omega)/Z(\omega)$ in parallel with the impedance
$Z(\omega)$.}
\end{figure}
The two configurations form two-terminal devices through which a current
$I_0(\omega)$
flows if a voltage $V_0(\omega)$ is applied. From the outside the two
configurations
appear as equivalent if the same voltage $V_0$ leads to the same current $I_0$.
In the Thevenin configuration the voltage drop is given by $V_0(\omega) =
I_0(\omega)
Z(\omega) + V(\omega)$ where the current and the voltages may in general be
frequency-dependent. On the other hand, the current flowing into the 
Norton configuration
is given by $I_0(\omega) = -I(\omega)+ V_0(\omega)/Z(\omega)$ if the current of
the current source in Fig.~18b
is flowing upwards. These two equations lead to the relation $V(\omega) =
Z(\omega) I(\omega)$ between the voltage and
current sources in the two configurations.

We introduce the method of network analysis by applying it to the single
tunnel junction\index{junction!single}. In a
first step we separate the tunneling junction into a tunneling element
in parallel with a capacitor describing
the junction capacitance. The tunneling
element transfers electrons through the circuit which appears as the
two-terminal device depicted in Fig.~19a.
\begin{figure}
\begin{center}
\includegraphics{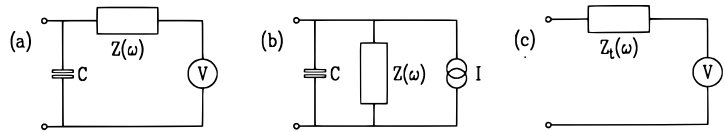}
\end{center}
\vspace{-5pt} 
\caption{Transformation of a single junction\protect\index{junction!single} 
circuit into an equivalent
effective circuit. (a) Original circuit as seen from the junction. (b)
Equivalent Norton configuration. (c) Effective single junction circuit.}
\end{figure}
We simplify the circuit by
transforming it into the Norton configuration shown in Fig.~19b. The current
source is given by $I(\omega) = V(\omega)/Z(\omega)$. While transforming
circuits we always keep the frequency dependence which is especially important
when capacitors are involved as is the case for the double 
junction. Only at
the end we account for the fact that we have a dc voltage source by taking the
limit $\omega\to 0$. In Fig.~19b the capacitance $C$ and the external impedance
$Z(\omega)$ are seen to form the total impedance $Z_t(\omega)$ defined in
(\ref{eq:I12}).
Returning to the Thevenin configuration of Fig.~19c we get a voltage source
$V(\omega) Z_t(\omega)/Z(\omega)$ which reduces to the original voltage $V$ in
the limit $\omega\to 0$. Electrons are now transferred through the effective
circuit. This leads to the work $eV$ done by the voltage source. The effective
impedance seen by the tunneling element is the total impedance $Z_t(\omega)$
containing the capacitance $C$ and the impedance $Z(\omega)$ in parallel. For
very low impedances the capacitor is thus shortened out and charging effects
become unimportant. In contrast, for a very large impedance the capacitor
remains and charging effects become apparent unless they are smeared out by
thermal fluctuations.  This picture fits our earlier considerations very well.

Having gained confidence in this method we apply it to the double 
junction
system shown in Fig.~20.
\begin{figure}
\begin{center}
\includegraphics{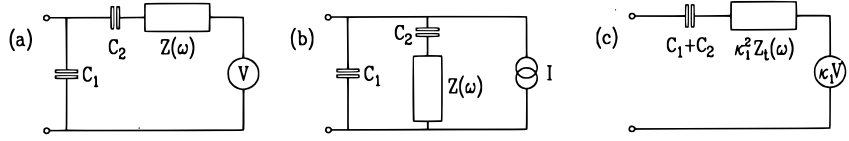}
\end{center}
\vspace{-5pt} 
\caption{Transformation of a double junction
circuit into an equivalent
effective circuit. (a) Original circuit as seen from the first junction.
The second junction is treated as a capacitor. (b) Equivalent Norton
configuration. (c) Effective circuit for tunneling through the first junction.}
\end{figure}
We consider tunneling through the first junction and
therefore treat the second junction as a capacitor thereby disregarding the
possibility of electron tunneling through the latter junction. We then arrive
at the Thevenin configuration shown in Fig.~20a. For sake of simplicity we will
not keep track of the charges during the transformations we are going to
perform. While in principle this would be possible, it will turn out that we
know
the charges on the capacitors of the effective circuit from our considerations
in the previous subsection. It is straightforward to apply the transformation
to the Norton configuration (Fig.~20b) and back to the Thevenin configuration
(Fig.~20c) as we did for the single junction. The new circuit contains a
capacitance, an effective impedance, and an effective voltage source which can
all be simply interpreted.

The voltage source has an effective voltage
$\kappa_1V$ where
\begin{equation}
\kappa_i = C/C_i\ \ \ (i=1,2).
\label{eq:D14}
\end{equation}
Since the total capacitance
$C$ is always smaller than the smallest capacitance $C_i$, the ratio $\kappa_i$
is always less than one. The effective voltage source can easily be interpreted
in terms of the work $\kappa_1 eV$ done by the source when an electron is
transferred by the tunneling element. After an electron has tunneled, charge is
transferred in the original circuit through the voltage source in order to
reestablish electrostatic equilibrium according to (\ref{eq:D10})--(\ref{eq:D12}).
For $C_G=0$ we indeed find
that the transferred charge is $\kappa_1 e$. The charge which has to be
transferred through the voltage source after an electron has tunneled through
one junction is smaller than an elementary charge. Only after the electron has
also tunneled through the other junction, the two charges transferred through
the voltage source add up to an elementary charge. This is in agreement with
$\kappa_1+\kappa_2=1$ which follows directly from (\ref{eq:D1}) and
(\ref{eq:D14}).

The effective impedance $\kappa_1^2 Z_t(\omega)$ with
\begin{equation}
Z_t(\omega) = {1\over i\omega C + Z^{-1}(\omega)}
\label{eq:D15}
\end{equation}
has the same structure as for the single junction if one replaces the single
junction capacitance by the total capacitance (\ref{eq:D1}) seen by the
external circuit.
In addition, there is again a reduction factor which for the impedance seen
by the first junction is $\kappa_1^2$. As a consequence, the influence of the
external circuit is
reduced. For a system consisting of $N$ junctions of about the same capacitance
one finds as a generalization that the effective impedance is reduced by a
factor of $1/N^2$. This means that one may apply the global rule
for circuits containing many junctions. However, one 
should bear in mind
that for sufficiently large voltages one will always find a crossover to the
local rule due to the sum rules satisfied by $P(E)$ unless the external
impedance vanishes. This crossover will occur at voltages which are about a
factor
$1/\kappa_i$ larger for a double junction system than 
for a single junction.
How can one understand the reduced environmental influence? From a physical
point of view the tunnel junction is to a certain extent decoupled from the
external circuit by the other junction. More formally, one has two
equivalent sets of charges $\{Q_1, Q_2\}$ and $\{Q, q\}$ for which one
introduces
the canonically conjugate phases $\{\varphi_1, \varphi_2\}$ and $\{\varphi,
\psi\}$. Here, $\varphi_1$ and $\varphi_2$ are defined as straightforward
generalizations of the single junction phase according to (\ref{eq:phase}). 
Now, $\{\varphi, \psi\}$ are related to $\{\varphi_1, \varphi_2\}$ by
\begin{equation}
\psi = \kappa_2\varphi_1 - \kappa_1\varphi_2
\label{eq:D16}
\end{equation}
and
\begin{equation}
\varphi = \varphi_1 + \varphi_2.
\label{eq:D17}
\end{equation}
The nonvanishing commutators between phases and charges are
\begin{equation}
[\varphi_1,Q_1] = ie,\ \ \ \ [\varphi_2,Q_2] = ie
\label{eq:D18}
\end{equation}
and
\begin{equation}
[\varphi,Q] = ie,\ \ \ \ [\psi,q] = ie.
\label{eq:D19}
\end{equation}
In the spirit of the tunneling Hamiltonian (\ref{eq:HT1}) for a single junction
we write for the tunneling Hamiltonian of the first junction
\begin{equation}
H_{T,1} = \sum_{kq\sigma}T_{kq}c_{q\sigma}^{\dagger}
c_{k\sigma}^{\phantom{\dagger}} \exp(-i\varphi_1) + \mbox{H.c.}
\label{eq:D20}
\end{equation}
where we may express the operator changing the charge on the first
junction as
\begin{equation}
\exp(-i\varphi_1) = \exp(-i\kappa_1\varphi-i\psi).
\label{eq:D21}
\end{equation}
Since $\psi$ is conjugate to $q$ the operator $\exp(-i\psi)$ describes the
change of the island charge by one elementary charge. The operator
$\exp(-i\kappa_1\varphi)$ couples the tunneling
process to the environment. It is the factor $\kappa_1$ appearing there which
leads to the reduction of the environmental coupling by $\kappa_1^2$.

Finally, in our effective circuit of Fig.~20c we have a capacitance $C_1+C_2$
which is related to the charging energy of the island $(ne)^2/2(C_1+C_2)$.
The charging energy corresponding to the total charge $Q$ may become 
irrelevant if
the external impedance is small. In contrast, the capacitor in series with the
total impedance is always affected by an electron which is transferred through
the effective circuit and thus the charging energy of the island will affect
the rate for any environment.

Having applied network analysis, we have now a rather clear picture of the
relevant quantities governing the dynamics of double junctions. Hence, we 
are in a position to immediately 
write down the expressions for
the double junction tunneling rates which will be discussed in the next
subsection. We only mention that network considerations become especially
useful when considering more complicated circuits. A straightforward extension
of the double junction is the one-dimensional array of junctions which will be
discussed briefly in Sec.~6.9.\ and in more detail in Chap.~7. Another
application is the single electron transistor if gate and stray capacitances
are taken into account.\cite{Wyrowski}
%
%
\subsection{Tunneling rates in a double junction system}
The changes in the expressions for the tunneling rates in double junction
systems compared with those for single junctions can be motivated by taking into
account the discussion in the previous subsection. We emphasize again that
the results we are going to discuss could as well be obtained from an explicit
calculation of the rates in second order perturbation theory.\cite{multizphys}

In the previous sections we
have found that the environmental influence on tunneling rates in normal
as well as superconducting single junctions may be described by means of the
probability $P(E)$ of energy exchange between the tunneling electron and the
external circuit. For double junction systems we have to account for the reduced
coupling to the environment, and the probability to transfer energy to the
environmental modes is given by
\begin{equation}
P(\kappa_i,E) = {1\over 2\pi\hbar}\int_{-\infty}^{+\infty} {\rm d}t
\exp\left[\kappa_i^2 J(t) + {i\over\hbar}Et\right].
\label{eq:D22}
\end{equation}
The correlation function $J(t)$ is defined as for the single junction in
(\ref{eq:I51}) provided
the capacitance $C$ appearing in the total impedance is the total capacitance
(\ref{eq:D1}) of the double junction system.

The energy difference for elastic tunneling of an electron through the $i$-th
junction onto the island is given by
\begin{eqnarray}
E_i(V,q) &=& \kappa_i eV + {q^2\over 2(C_1+C_2)} - {(q-e)^2\over
2(C_1+C_2)}\nonumber\\
&=& \kappa_i eV + {e(q-e/2)\over C_1+C_2}\ \ \ \ (i=1,2)
\label{eq:D23}
\end{eqnarray}
where the effective island charge $q=ne$ for a double junction and $q=ne+Q_0$ 
for a SET transistor in the limit $C_G\to 0$.
For practical purposes it is often useful to express this energy difference in
terms of quantities of the $i$-th junction only. Using (\ref{eq:D2}) and
(\ref{eq:D3}) we may rewrite (\ref{eq:D23}) to obtain
\begin{equation}
E_i(Q_i) = {e\over C_i}(Q_i-Q_i^c).
\label{eq:D24}
\end{equation}
Here, we have introduced a critical charge\index{critical charge}
\begin{equation}
Q_i^c = {e\over 2} (1-\kappa_i).
\label{eq:D25}
\end{equation}
Although (\ref{eq:D24}) contains only quantities of the $i$-th junction it 
still describes the change of electrostatic energy for the entire circuit.

It is now straightforward to write down the forward tunneling rate through the
first junction \cite{multizphys,triest}
\begin{equation}
\GF_1(V,q) = {1\over e^2R_1}\int_{-\infty}^{+\infty} {\rm d}E {E\over
1-\exp(-\beta E)}P(\kappa_1, E_1(V,q)-E).
\label{eq:D26}
\end{equation}
Here, $R_1$ is the tunneling resistance of the first junction.
The forward and backward tunneling rates for the first junction are connected
by
\begin{equation}
\GB_1(V,q) = \GF_1(-V,-q)
\label{eq:D27}
\end{equation}
since in the backward tunneling process the electron is tunneling from the
island opposite to the direction favored by the applied voltage.
As for the single junction rates there exists a detailed balance symmetry which
now connects rates for different island charges
\begin{equation}
\GB_1(V,q-e) = \exp[-\beta E_1(V,q)]\GF_1(V,q).
\label{eq:D28}
\end{equation}
The forward and backward tunneling rates through the second junction are
obtained from the respective rates for the first junction by exchanging the
indices 1 and 2 and by changing $q$ into $-q$. For the forward tunneling
rate one thus finds
\begin{equation}
\GF_2(V,q) = {1\over e^2R_2}\int_{-\infty}^{+\infty} {\rm d}E {E\over
1-\exp(-\beta E)}P(\kappa_2, E_2(V,-q)-E).
\label{eq:D29}
\end{equation}
The relations corresponding to (\ref{eq:D27}) and (\ref{eq:D28}) now read
\begin{equation}
\GB_2(V,q) = \GF_2(-V,-q)
\label{eq:D30}
\end{equation}
and
\begin{equation}
\GB_2(V,q+e) = \exp[-\beta E_2(V,-q)]\GF_1(V,q).
\label{eq:D31}
\end{equation}
We end this general part on tunneling rates by noting that for a symmetric
double junction system with equal capacitances $C_1=C_2$ and equal tunnel
resistances $R_1=R_2$ all tunneling rates are related to each other by
$\GF_1(V,q) = \GB_1(-V,-q) = \GF_2(V,-q) = \GB_2(-V,q)$.
%
%
\subsection{Double junction in a low impedance environment}
\index{environment!electromagnetic|(}
For explicit analytical results of the environmental influence on electron 
tunneling
rates we restrict ourselves to the limits of very low and very high impedance
environments. The first case is of relevance for most practical cases because
of the reduced effective impedance. The high impedance case, on the other hand,
determines the behavior at very large voltages.

In the limit of vanishing external impedance tunneling is elastic and we have
like in the single junction case $P(\kappa_i,E) = \delta(E)$. Then, the
tunneling rates may easily be evaluated and we get from (\ref{eq:D26}) for the
forward tunneling rate through the first junction
\begin{equation}
\GF_1(V,q) = {1\over e^2 R_1} {E_1(V,q)\over 1-\exp[-\beta E_1(V,q)]}.
\label{eq:D32}
\end{equation}
For zero temperature this reduces to
\begin{equation}
\GF_1(V,q) = {1\over e^2 R_1} E_1(V,q) \Theta(E_1(V,q))
\label{eq:D33}
\end{equation}
where $\Theta(E)$ is the unit step function. Therefore, at zero temperature
$\GF_1(V,q)$ is only different from zero if $E_1(V,q)>0$. This
justifies that we call $Q_i^c$, which was defined in (\ref{eq:D24}) and
(\ref{eq:D25}), a critical charge.\index{critical charge} The effective 
charge on the island
through which electron tunneling is considered has to exceed the
critical charge to allow for a finite tunneling rate. Together with
(\ref{eq:D23})
we now find the following conditions under which the rates are nonvanishing:
\begin{eqnarray}
&\GF_1(V,q):\ \ &V + {1\over C_2}\left(q-{e\over 2}\right) > 0
\label{eq:D34}\\
&\GB_1(V,q):\ \ &V + {1\over C_2}\left(q+{e\over 2}\right) < 0 \label{eq:D35}\\
&\GF_2(V,q):\ \ &V - {1\over C_1}\left(q+{e\over 2}\right) > 0 \label{eq:D36}\\
&\GB_2(V,q):\ \ &V - {1\over C_1}\left(q-{e\over 2}\right) < 0.\label{eq:D37}
\end{eqnarray}
The heavy lines in Fig.~21a
\begin{figure}
\begin{center}
\includegraphics{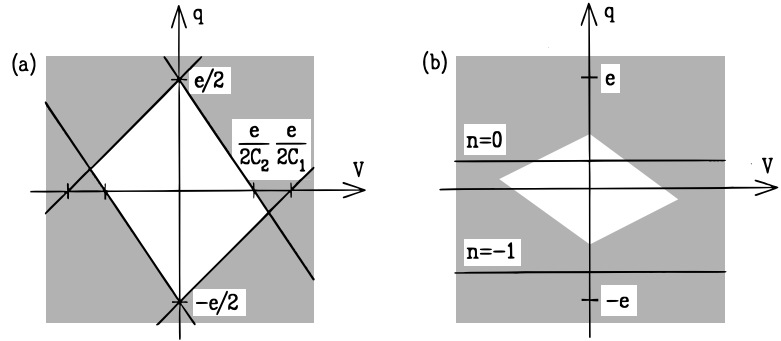}
\end{center}
\caption{(a) Low impedance stability diagram for the effective island charge 
$q$ in dependence on the
transport voltage $V$. In the non-shaded area the state $n=0$ is stable while
in the shaded area one or more tunneling rates are non-vanishing. (b) Stability
diagram for a transistor with $Q_0=e/4$. The possible effective island charges
are shifted accordingly.}
\end{figure}
indicate the parameter region where one of the 
Eqs.~(\ref{eq:D34})--(\ref{eq:D37}) is fulfilled as an
equality. The area inside these lines is the region where the island charge
$n=0$ is stable because all tunneling rates vanish. In the shaded areas one or
more rates are different from zero. Suppose now that we have an
ideal double junction system without offset charges so that the effective 
island charge
is given by $q=ne$. If $\vert V\vert < \min(e/2C_1,e/2C_2)$ and $n\ne 0$ then
the rates force the electrons to tunnel in such a way that after some time
the island charge is zero. The state $n=0$ is stable in this voltage regime
since all rates vanish. As a consequence at zero temperature there is no
current if the absolute value of the voltage is below $\min(e/2C_1,e/2C_2)$ and
we find a Coulomb gap\index{Coulomb gap|(} even in the low impedance case. 
This important difference
as compared to the case of a single junction is due to the charging energy 
related to the island charge.

If we now apply an offset charge\index{offset charge} either by placing 
a charge 
near the island or by using a transistor setup according to Fig.~16 with
$Q_0=C_GV_G$, the Coulomb gap will be affected. The possible effective
island charges are then no longer given by $q=ne$ but by $q=ne+Q_0$. With this
replacement we immediately get the tunneling rates for the single electron
transistor from the double junction rates (\ref{eq:D26}), (\ref{eq:D27}),
(\ref{eq:D29}), and (\ref{eq:D30}). The transistor rates then depend on $V$,
$n$, and $V_G$. Note that going from forward to backward tunneling rates now
not only involves a change in sign of $V$ and $n$ but also of $V_G$.
To obtain the influence of the offset charge on the Coulomb gap
we have to consider only charges in the range $-e/2 < Q_0 <
e/2$ since an integer number of elementary charges can always be absorbed in
$n$. This means that for a transistor the stable island charge is not
necessarily given by $n=0$ but depends on the gate voltage. However, 
if $Q_0$ is
in the range just mentioned the stable state will be $n=0$. The situation for
$Q_0=e/4$ is shown in Fig.~21b. It becomes clear from this figure that the
range
in which $n=0$ is stable is decreased as compared to $Q_0=0$. For $Q_0=e/2$
there is no voltage range for which an island charge is stable. This means
that in the low impedance limit the Coulomb gap will vanish for $Q_0=e/2$. The
dependence of the Coulomb gap on the offset charge\index{offset charge|(} 
is shown in Fig.~22.
\begin{figure}
\begin{center}
\includegraphics{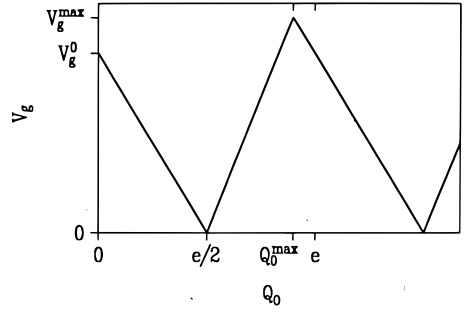}
\end{center}
\caption{Low impedance Coulomb gap for a single electron transistor as function
of the offset charge $Q_0=C_GV_G$. If $C_1<C_2$ one has $V_g^0=e/2C_2$,
$V_g^{\rm max}=e/(C_1+C_2)$, and $Q_0^{\rm max}=e/2+\kappa_2e$.}
\end{figure}
For
larger offset charges the picture is continued periodically with period $e$
according to the argument given above. Finally, we mention that the results for
a low impedance environment may easily be generalized for a transistor with
finite gate capacitance. If the transport voltage is divided symmetrically as
shown in
Fig.~16 one finds from (\ref{eq:D13}) that the replacement $C_1 \to C_1 +
C_G/2$ and $C_2 \to C_2 + C_G/2$ will account for the gate 
capacitance.\cite{Wyrowski}
%
%
\subsection{Double junction in a high impedance environment}
The similarity of the tunneling rates for single and double junctions shows 
also in the rate expressions for a double junction in a high impedance
environment. For finite temperatures we obtain
\begin{equation}
P(\kappa_i,E) = {1\over \sqrt{4\pi\kappa_i^2E_ck_BT}} \exp[-{(E-\kappa_i^2
E_c)^2\over 4\kappa_i^2 E_c k_BT}]
\label{eq:D38}
\end{equation}
which differs from the corresponding single junction result (\ref{eq:I73}) only
by the factor $\kappa_i^2$\vadjust{\break} in front of $E_c$ which is due to 
the reduced coupling of the double junction to the environment. 
Accordingly, at zero temperature (\ref{eq:D38}) reduces to
$P(\kappa_i, E) = \delta(E-\kappa_i^2 E_c)$. Together with (\ref{eq:D26}) we
then find for the forward tunneling rate through the first junction
\begin{equation}
\GF_1(V,q) = {1\over e^2R_1}[E_1(V,q)-\kappa_1^2E_c]\Theta(E_1(V,q)-
\kappa_1^2E_c)\ \ \ \ \hbox{for}\ T=0.
\label{eq:D39}
\end{equation}
By rewriting the energy difference as
\begin{equation}
E_1(V,q) - \kappa_1^2E_c = {Q_1^2\over 2C_1} - {(Q_1-e)^2\over 2C_1}
\label{eq:D40}
\end{equation}
it becomes clear that the local rule determines
the zero temperature tunneling rates for a high impedance environment as it is
the case for single junctions. Note that one may rewrite (\ref{eq:D40}) in the 
local form (\ref{eq:D24}) with a critical
charge $e/2$. This high impedance critical charge is unaffected by the reduced
coupling to the environment since the local rule knows only about the capacitor
of the junction through which the electron is tunneling. As for the low
impedance environment we give the conditions under which the four rates are
nonvanishing at zero temperature:
\begin{eqnarray}
&\GF_1(V,q):\ \ &V + {q\over C_2} - {e\over 2C} > 0\label{eq:D41}\\
&\GB_1(V,q):\ \ &V + {q\over C_2} + {e\over 2C} < 0\label{eq:D42}\\
&\GF_2(V,q):\ \ &V - {q\over C_1} - {e\over 2C} > 0\label{eq:D43}\\
&\GB_2(V,q):\ \ &V - {q\over C_1} + {e\over 2C} < 0.\label{eq:D44}
\end{eqnarray}
These conditions are of course equivalent to the requirement that the charge
on the junction should be larger than the critical charge $e/2$. Again
(\ref{eq:D41})--(\ref{eq:D44}) define a region of stability for the state $n=0$
which is shown in
Fig.~23.
\begin{figure}
\begin{center}
\includegraphics{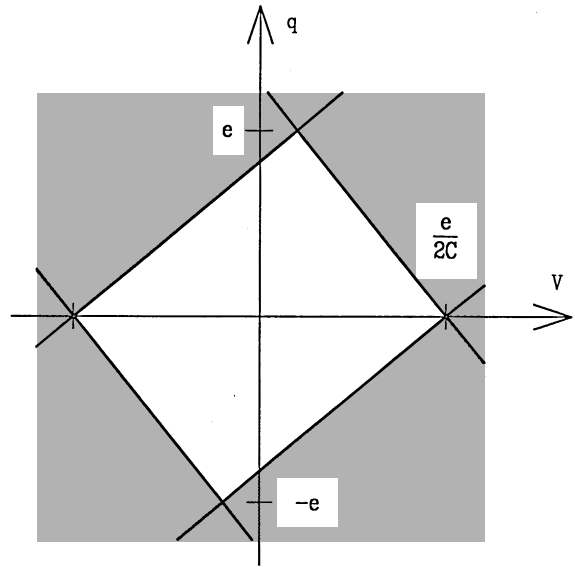}
\end{center}
\caption{High impedance stability diagram for the effective island charge $q$ in
dependence on the
transport voltage $V$. In the non-shaded area the island charge $q=0$ is 
stable while in the shaded area one or more tunneling rates are non-vanishing.}
\end{figure}
In comparison with the low impedance case this state is stable here
for a wider range of parameters. 
In the absence of an offset charge the Coulomb gap
is given by $e/2C$ which always exceeds the low impedance Coulomb gap because
$C<C_1, C_2$. If offset charges are present we may apply the same arguments as
for the low impedance case. We observe that in $q$-direction the stable region
extends over a range exceeding one elementary charge. As a consequence, one 
finds a Coulomb gap even for an offset charge $Q_0=e/2$ where the gap vanishes
for low impedance
environments. The high impedance gap as a function of the offset charge is
shown in Fig.~24.
\begin{figure}
\begin{center}
\includegraphics{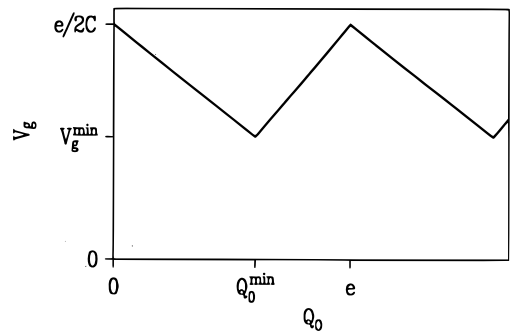}
\end{center}
\caption{High impedance Coulomb gap for a single electron transistor as
function of the offset charge $Q_0=C_GV_G$. For $C_1<C_2$ one has $V_g^{\rm
min}=e/2C-e/(C_1+C_2)$ and $Q_0^{\rm min}= \kappa_1 e$.}
\end{figure}
Another consequence of this wide stability region in
$q$-direction is that for certain voltages two states with different island
charges may be stable. Which one of the two states is realized depends on how
the stability region is reached. Such a multistability may also result from a
capacitor in series with the tunnel junction thereby producing
a high impedance environment. This situation is discussed in more detail in
Chap.~3 in connection with the single electron trap and related devices.
\index{offset charge|)}\index{Coulomb gap|)}
\index{environment!electromagnetic|)}
%
%
\subsection{Current-voltage characteristics of a double junction}
For a single junction the current could be calculated by subtracting the
backward from the forward tunneling rate and multiplying this result by the
elementary charge. This was possible since every tunneling process contributed
to the current in the respective direction. For a double junction the situation
is more complicated because the tunneling rates depend on previous tunneling
processes which lead to a certain island charge. As we know from subsection
6.3.\ the tunneling rates depend on the external voltage $V$ and the effective
island
charge $q$. The external voltage is taken to be constant and it is always
assumed that electrostatic equilibrium at the double junction is established
before the next electron tunnels. Then the state of the double junction is
characterized by the number $n$ of electrons on the island. Neglecting
correlations between different tunneling processes we may write down a master
equation which connects states with different island charge. The
probability to find the double junction in the state $n$ may change by leaving
this state or by coming into this state from the states $n-1$ or $n+1$
\begin{equation}
\dot p_n = \Gamma_{n,n+1}p_{n+1} + \Gamma_{n,n-1}p_{n-1} - (\Gamma_{n+1,n} +
\Gamma_{n-1,n})p_n.
\label{eq:D45}
\end{equation}
Here, $\Gamma_{k,l}$ is the rate for a transition from state $l$ to state $k$.
Since each tunneling process changes the island charge by $e$ the states $l$
and $k$ are neighbors on the island charge ladder with $\vert l-k\vert=1$.
There exist two independent possibilities to change the island charge, namely
by tunneling through the first or through the second junction. Accordingly, the
two rates have to be summed up yielding
\begin{eqnarray}
\Gamma_{n+1,n} &=& \GB_1(n) + \GF_2(n)\label{eq:D46}\\
\Gamma_{n-1,n} &=& \GF_1(n) + \GB_2(n)\label{eq:D47}
\end{eqnarray}
where we suppressed the dependence on the external voltage. In verifying these
two relations one should keep in mind that the island charge is defined
as $ne$, and hence an additional
electron on the island decreases the island charge and thereby $n$. Since we
are not interested in the transient behavior we calculate the stationary
probabilities $p_n$ by requiring $\dot p_n=0$. It is easy to see that
probabilities satisfying the detailed balance condition
\begin{equation}
\Gamma_{n,n+1}p_{n+1} = \Gamma_{n+1,n}p_n
\label{eq:D48}
\end{equation}
are a solution of the master equation (\ref{eq:D45}). Since only nearest
neighbor states
are connected by nonvanishing rates it can be shown that this solution where
the upward flow equals the downward flow is the only nontrivial solution.
Starting from a neutral island one finds from (\ref{eq:D48}) the stationary
solution
\begin{equation}
p_n = p_0 \prod_{m=0}^{n-1} {\Gamma_{m+1,m}\over \Gamma_{m,m+1}}
\label{eq:D49}
\end{equation}
and
\begin{equation}
p_{-n} = p_0 \prod_{m=-n+1}^0 {\Gamma_{m-1,m}\over \Gamma_{m,m-1}}
\label{eq:D50}
\end{equation}
with $n>0$ in both formulas. The only free parameter left is $p_0$ which
is determined by the normalization condition
\begin{equation}
\sum_{n=-\infty}^{+\infty} p_n = 1.
\label{eq:D51}
\end{equation}

Knowing the stationary probability to find the island charge $ne$ we may now
calculate the current-voltage characteristics for a double junction from

\break
\begin{equation}
I = e\sum_{n=-\infty}^{+\infty} p_n (\GF_1(n) - \GB_1(n)) = e
\sum_{n=-\infty}^{+\infty} p_n(\GF_2(n)-\GB_2(n)).
\label{eq:D52}
\end{equation}
Because of current conservation it does not matter for which junction
we calculate the current. The equality of the second and third expression in
(\ref{eq:D52}) is ensured by the detailed balance condition (\ref{eq:D48}).

While the above considerations are valid also for finite temperatures we will
restrict ourselves to zero temperature to further illustrate the calculation of
current-voltage characteristics. In the simplest case the voltage is below the
gap voltage. According to our earlier discussions the state $n=0$ then is
stable and only rates leading to a decrease of the absolute value of the island
charge are nonvanishing. This means that the stationary solution of the master
equation is given by $p_0=1$, $p_n=0$ for $n\ne 0$. Since all rates vanish for
$n=0$ we find from (\ref{eq:D52}) that indeed the current vanishes in the
blockade
region. Let us now increase the voltage beyond the gap voltage for a
double junction with different capacitances $C_1<C_2$ in a low impedance
environment (cf.\ Fig.~21a). We begin by considering voltages satisfying
$e/2C_2<V<e/2C_1$. Setting $n=0$, Fig.~21a tells us that tunneling of
electrons through the first junction onto the island is allowed while tunneling
through the second junction is forbidden. Being at $n=-1$ the rates only allow
the transition back to $n=0$ by tunneling through the second junction.
Consequently, two states, namely $n=0$ and $n=-1$, are involved. From
(\ref{eq:D50}) and (\ref{eq:D51}) one readily gets
\begin{equation}
p_0 = {\GF_2(V,-e)\over \GF_1(V,0)+\GF_2(V,-e)}
\label{eq:D53}
\end{equation}
and
\begin{equation}
p_{-1} = {\GF_1(V,0)\over \GF_1(V,0)+\GF_2(V,-e)}.
\label{eq:D54}
\end{equation}
The probabilities $p_0$ and $p_{-1}$ together with (\ref{eq:D52}) yield
for the current
\begin{equation}
I=e\Gamma(V)
\label{eq:D55}
\end{equation}
where the effective rate $\Gamma(V)$ is given by
\begin{equation}
{1\over\Gamma(V)} = {1\over\GF_1(V,0)} + {1\over\GF_2(V,-e)}.
\label{eq:D56}
\end{equation}
Since the two tunneling processes occur one after the other, the rates are
added inversely and the total rate is dominated by the slower rate. We note
that $\Gamma(V)$ here is really only an effective rate because the two step
tunneling process does not lead to a purely exponential time dependence.
Still $1/\Gamma(V)$ is the average time between tunneling events across the
double junction system.

Let us now increase the voltage to the regime where $e/2C_1<V<3e/2C_2$.
Assuming $1<C_2/C_1<3$ we are sure that it is not possible that the island is
charged with two electrons. According to Fig.~21 we now have two possibilities
to
leave the state $n=0$. Either an electron may tunnel through the first junction
to the right or it may tunnel through the second junction to the right.
Depending on what actually happens, the island charge is then either $-e$ or
$e$. The island may not be charged further. Therefore, in the next step an
electron has to tunnel through the other junction restoring the neutral island.
We thus have two competing processes, namely $n=0\to e\to 0$ and $n=0\to -e\to
0$. These two mechanisms now allow for two subsequent tunneling processes
occurring at the same junction. The sequence of tunneling through the two
junctions is no longer fixed but contains a statistical element.

It is obvious that by increasing the voltage the situation will become more and
more complicated. In the following section we will derive some properties of
the current-voltage characteristic which also hold at higher voltages. In
general, however, one has to resort to numerical techniques. It is rather
straightforward to use (\ref{eq:D49})--(\ref{eq:D51}) to determine the
stationary probabilities
$p_n$ and calculate the current-voltage characteristics by means of
(\ref{eq:D52}).
%
%
\subsection{Coulomb staircase}
In the first part of this subsection we will calculate the current through a
double junction at zero temperature for special
values of the voltage for the limits of low and high impedance environments. We
will assume that the capacitances $C_1$ and $C_2$ of the tunnel junctions are
equal $C_1 = C_2 = C_J$ while the tunneling resistances $R_1$ and $R_2$ may
take arbitrary values. For the positive voltages $V_m = (e/C_J)(m+1/2), (m = 0,
1, 2,\ldots)$ in the low impedance case and $V_m = (e/C_J)(m+1), (m = 0, 1,
2,\ldots)$ in the high impedance case the tunneling rates are given by
\begin{eqnarray}
\GF_1(m,n) &=& {1\over eC_J(R_1+R_2)}\left(1+{R_2\over R_1}\right)
{m+n\over 2}\Theta(m+n)\label{eq:D57}\\
\GB_1(m,n) &=& -{1\over eC_J(R_1+R_2)}\left(1+{R_2\over R_1}\right) {m+n\over
2}\Theta(-m-n)\label{eq:D58}\\
\GF_2(m,n) &=& {1\over eC_J(R_1+R_2)}\left(1+{R_1\over R_2}\right) {m-n\over
2}\Theta(m-n)\label{eq:D59}\\
\GB_2(m,n) &=& -{1\over eC_J(R_1+R_2)}\left(1+{R_1\over R_2}\right) {m-n\over
2}\Theta(-m+n).\label{eq:D60}
\end{eqnarray}
Here, $m$ and $n$ correspond to voltages $V_m$ and island charges $ne$, 
respectively. We
calculate the occupation probabilities of the $n$-th state by starting from
$n=0$. For $n>0$ one immediately finds from (\ref{eq:D58}) that $\GB_1(m,n)$
vanishes.
Furthermore, we find from (\ref{eq:D59}) that $\GF_2(m,n)$ vanishes for $n\geq
m$ and
thus, according to (\ref{eq:D49}), $p_{m+k}(m)=0$ for $k>0$. Together with
(\ref{eq:D60}) this
means that $\GB_2(m,n)$ vanishes for all $n$ for which $p_n\ne 0$. The detailed
balance condition (\ref{eq:D48}) thus yields
\begin{equation}
p_{n+1}(m) = p_n(m) {\GF_2(m,n)\over \GF_1(m,n+1)}.
\label{eq:D61}
\end{equation}
From similar considerations for $n<0$ one finds
\begin{equation}
p_{n-1}(m) = p_n(m) {\GF_1(m,n)\over \GF_2(m,n-1)}.
\label{eq:D62}
\end{equation}
Making use of the rates (\ref{eq:D57}) and (\ref{eq:D59}), the two equations
(\ref{eq:D61}) and (\ref{eq:D62}) may be recast into
\begin{equation}
p_n(m) = \left({R_1\over R_2}\right)^n {(m!)^2\over (m-\vert n\vert )!
(m+\vert n\vert )!}p_0(m).
\label{eq:D63}
\end{equation}
Exploiting properties of binomial coefficients, we find for the normalization
condition
\begin{equation}
\sum_{n=-m}^{m} p_n = p_0(m) {(m!)^2\over (2m)!} \left({R_1\over
R_2}\right)^m \left(1+{R_2\over R_1}\right)^{2m} = 1.
\label{eq:D64}
\end{equation}
This determines $p_0(m)$ and we finally get the probabilities
\begin{equation}
p_n(m) = {(2m)!\over (m-\vert n\vert)!(m+\vert n\vert)!}{(R_1/R_2)^{n+m}\over
(1+R_1/R_2)^{2m}}.
\label{eq:D65}
\end{equation}
When this is combined with (\ref{eq:D52}), we obtain for the current
\begin{equation}
I = {e\over C_J(R_1+R_2)}{1\over 2}\left(1+{R_2\over R_1}\right)
\sum_{n=-m}^m (m+n)p_n(m).
\label{eq:D66}
\end{equation}
Here, we have evaluated the current through the first junction and taken 
into account that the backward
rate does not contribute. The first term in the sum is obtained from the
normalization condition (\ref{eq:D51}) while the second term is given by
\begin{equation}
\sum_{n=-m}^m n p_n(m) = m {R_1-R_2\over R_1+R_2}.
\label{eq:D67}
\end{equation}
The latter result may be derived by viewing $p_n(m)$ as a function of $R_1/R_2$
and applying the same trick used to derive (\ref{eq:I41}). From
(\ref{eq:D66})
and (\ref{eq:D67}) we get our final result for special points of the
current-voltage characteristic \cite{multizphys}
\begin{equation}
I(V_m) = {e\over C_J(R_1+R_2)}m\ \ \ \ \ \ (m= 0, 1, 2,\ldots)
\label{eq:D68}
\end{equation}
at voltages
\begin{equation}
V_m = {e\over C_J} (m+{1\over 2})\ \ \ \ \ \ {\rm (low\ impedance\
environment)}
\label{eq:D69}
\end{equation}
or
\begin{equation}
V_m = {e\over C_J} (m+1)\ \ \ \ \ \ {\rm (high\ impedance\ environment)}.
\label{eq:D70}
\end{equation}
Thus, for certain voltages the current-voltage characteristic touches an Ohmic
current-voltage characteristic with resistance $R_1+R_2$ which is shifted by
the gap voltage $e/2C_J$ in the low impedance case and by $e/C_J$ in the high
impedance case.

Let us now discuss the current-voltage characteristics between
the voltage points for which we just calculated the current. To keep things
as simple as possible we assume the tunneling resistance
$R_2$ of the second junction to be very large compared to $R_1$. Then the
island will be charged through the first junction up to a maximum charge.
Occasionally an electron will tunnel through the second junction
resulting in a current through the double junction. From the condition
(\ref{eq:D34}) for a non-vanishing rate through the first junction, one finds
that the maximum island charge is
given by $n_{\rm max}e = - e[C_JV/e-1/2]$, where $[\ldots ]$ denotes the 
largest
integer smaller or equal to the argument. In the limit $R_2\gg R_1 $ the current
in the presence of a low impedance environment then reads
\begin{equation}
I(V) = e\GF_2(V) = {1\over 2R_2}\left(V-\frac{e}{C_J}\left(n_{\rm max}+
\frac{1}{2}\right)\right).
\label{eq:D71}
\end{equation}
At the voltages given by (\ref{eq:D69}) we recover
the current (\ref{eq:D68}). Increasing the voltage we observe a jump in current
by
$e/(2R_2C_J)$ because the maximum island charge is increased by $e$. This jump
is followed by a linear current-voltage characteristic with differential
resistance $2R_2$. This ensures the validity of (\ref{eq:D68}) and
(\ref{eq:D69}). One may apply
the same arguments for the high impedance case where the whole picture is just
shifted in voltage by $e/2C_J$.

The steplike structure which we have found is called the Coulomb 
staircase.\cite{stair1}--\cite{stair4}
It is very distinct if the ratio of the tunnel resistances is very
different from one. 
For $R_1\approx R_2$ the steps are barely visible. 
For general parameters
one has to evaluate (\ref{eq:D49})--(\ref{eq:D52}) numerically. 
In Fig.~25
\begin{figure}
\begin{center}
\includegraphics{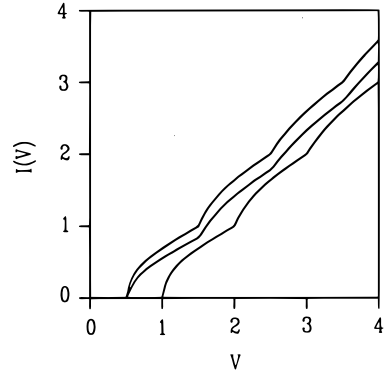}
\end{center}
\caption{Zero temperature current-voltage characteristics of a double junction
in a low impedance environment (left curve), an environment with Ohmic
resistance $R_K/5$ (middle curve), and a high impedance environment (right
curve). The junction parameters are $C_2=C_1$ and $R_2=10R_1$. Voltage is given
in units of $e/2C$ and current is given in units of
$e/(2C(R_1+R_2))$.}
\end{figure}
current-voltage characteristics are shown for low and high impedance
environments as well as for the case of an Ohmic resistance
$Z(\omega)=R_K/5$.
As for the single junction one finds for such a low conductance a
crossover from the low impedance characteristic to the high impedance
characteristic. For $C_1\ne C_2$, the staircase need not be as
regular as it appears for a double junction with equal capacitances. 

In this section on double junction systems we mentioned in several places how
the results have to be generalized to account for an offset charge $Q_0$. It is
now straightforward to calculate current-voltage characteristics for different
offset charges. As an example we present in Fig.~26 
\begin{figure}
\begin{center}
\includegraphics{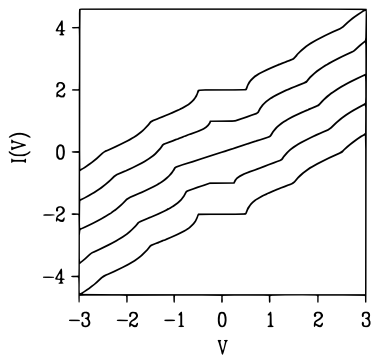}
\end{center}
\caption{Zero temperature current-voltage characteristics of a single electron
transistor in a low impedance environment. The junction parameters are
$C_1=C_2=2C$  and $R_2=10R_1$. The offset charge is increased from $Q_0=0$ for
the lowest curve to $Q_0=e$ in the highest curve in steps of $e/4$. The voltage
is given in units of $e/2C$ and the current is given in units of
$e/(2C(R_1+R_2))$.}
\end{figure}
the zero temperature current-voltage characteristics of a
single electron transistor in a low impedance environment. This figure shows
the Coulomb staircase as well as the dependence
of the Coulomb gap on the offset charge. The junctions chosen here have equal
capacitances but a ratio of tunneling resistances $R_2/R_1=10$ to produce a
marked Coulomb staircase. The offset charge changes from $Q_0=0$ in the lowest
curve to $Q_0=e$ in the uppermost curve. As we have discussed already, 
the Coulomb
gap depends on the offset charge. The middle curve with $Q_0=e/2$ does 
not show a Coulomb gap as is expected from our earlier considerations. 
Furthermore, the
characteristics for an offset charge different from 0 and $e/2$ exhibit an
asymmetry. As we found earlier all orientations of tunneling processes are
reversed if we make the replacement $V\to -V$ and $q\to -q$. As $-e/4$ and 
$3e/4$ are
equivalent offset charges we find that the characteristics for $V>0$ and
$Q_0=e/4$ should be identical to the characteristics for $V<0$ and $Q_0=3e/4$
and vice versa. This can clearly be seen in Fig.~26.
%
%
\subsection{SET-transistor and SET-electrometer}\index{electrometer|(}
\index{SET transistor|(}\index{offset charge|(}
In a SET-transistor setup like the one shown in Fig.~16 the current through the
junctions depends on both the transport voltage $V$ and the gate voltage $V_G$.
So far we have mainly concentrated on current-voltage characteristics $I(V)$.
In this section we will keep $V$ fixed and discuss how the current changes with
the offset charge $Q_0$. The offset charge may be due to a gate voltage coupled
capacitively to the island or due to some other mechanism. The dependence of the
current on the offset charge can be exploited in two ways. By means of the gate
voltage one may control the current thereby realizing a 
transistor.\cite{ALREV91} On the
other hand, one may use the current to measure the offset charge. In this case
one uses the circuit as a very sensitive electrometer.\cite{Fulton} For the 
practical aspects of these devices we refer the reader to Chaps.~3 and 9.  
Here, we want to apply the results obtained above to the calculation of
$I$-$Q_0$-characteristics. As in the previous sections it will not be possible
to give a closed analytical expression for arbitrary transport voltages. We
therefore restrict ourselves to the regime below the Coulomb gap voltage. Since
the performance of the transistor and electrometer reaches an optimum when 
biased at the gap voltage, this is the regime of practical interest.

To be more specific, let us choose a setup with equal junction capacitances
$C_1=C_2=2C$ but arbitrary ratio of tunneling resistances $R_1/R_2$. In
practice, the assumption of a low impedance environment will be well satisfied.
Furthermore, we restrict ourselves to the case of zero temperature. From
the discussion of the current-voltage characteristics of a double junction it
is clear how to generalize the calculation to finite temperatures. However,
in this case one has to resort to numerical methods.

As already mentioned, we consider transport voltages below the low impedance gap
$e/4C$. The offset charge is assumed to satisfy $0\leq Q_0 <e$. Then, once
a stationary situation is reached, only two island charge
states are occupied (cf.\ Fig.~21b).
Taking the transport voltage to be positive we find from (\ref{eq:D35}) and
(\ref{eq:D37}) that the backward tunneling rates $\GB_1$ and $\GB_2$ vanish if
the island charge takes one of the two allowed values. When initially $n=0$ and
thus $q=Q_0$ the tunneling rate $\GF_1$ through the
first junction is nonvanishing while $\GF_2$ is zero 
according to (\ref{eq:D34}) and (\ref{eq:D36}). 
After an electron has
tunneled through the first junction $q$ has changed to $q-e$. Now, $\GF_1$
vanishes and $\GF_2$ is different from zero allowing the electron to tunnel
from the island. We now have the same situation as described in Sec.~6.6.
The corresponding Eqs.~(\ref{eq:D53}) and (\ref{eq:D54}) read
\begin{equation}
p_0 = \frac{\GF_2(V,Q_0-e)}{\GF_1(V,Q_0)+\GF_2(V,Q_0-e)}
\label{eq:D72}
\end{equation}
and
\begin{equation}
p_{-1} = \frac{\GF_1(V,Q_0)}{\GF_1(V,Q_0)+\GF_2(V,Q_0-e)}.
\label{eq:D73}
\end{equation}
Inserting these probabilities for the two possible island charge states we find
with (\ref{eq:D52}) for the current
\begin{equation}
I(V,Q_0) = e\frac{\GF_1(V,Q_0)\GF_2(V,Q_0-e)}{\GF_1(V,Q_0)+\GF_2(V,Q_0-e)}.
\label{eq:D74}
\end{equation}
According to (\ref{eq:D26}) and (\ref{eq:D29}), for equal capacitances 
$C_1=C_2$ the rates through the first and second junction are related by
\begin{equation}
\GF_2(V,q) = \frac{R_1}{R_2}\GF_1(V,-q).
\label{eq:D75}
\end{equation}
Together with (\ref{eq:D33}) and (\ref{eq:D23}) we finally obtain for the
current at fixed transport voltage
\begin{eqnarray}
I(Q_0)=\frac{1}{2}\frac{\left({\displaystyle\frac{Q_0-e/2}{2C}}\right)^2-V^2}
{(R_1-R_2)
{\displaystyle\frac{Q_0-e/2}{2C}}-(R_1+R_2)V}&&\nonumber\\
&&\hspace{-5truecm}\times\Theta(Q_0-\frac{e}{2}+2CV)\,\Theta(-Q_0+
\frac{e}{2}+2CV).
\label{eq:D76}
\end{eqnarray}
The two step-functions $\Theta(x)$ are a consequence of the Coulomb gap. For
very low transport voltages one needs an offset charge rather close to $e/2$ to
obtain a current. As a function of the gate voltage the zero bias conductance 
thus displays a sequence of peaks at $Q_0 =C_GV_G=e(k+1/2)$, $k$ integer. For
semiconductor nanostructures these Coulomb blockade oscillations\index{Coulomb
blockade!oscillation} are discussed in great detail in Chap.~5.
With increasing transport
voltage the range of offset charges leading to a nonvanishing current becomes
larger. Finally, if the transport voltage equals the gap voltage, the current
only vanishes for $Q_0=0$ as expected. 
Above this voltage the control of the current by the gate rapidly decreases and
the performance of the transistor or 
electrometer is reduced as can be seen from Fig.~27.
\begin{figure}
\begin{center}
\includegraphics{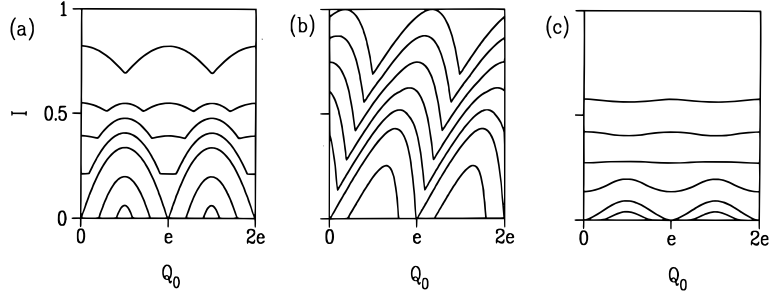}
\end{center}
\caption{
$I$-$Q_0$ characteristics at zero temperature for SET transistors with gate
capacitance $C_G=2C$ and Ohmic environment $Z_1(\omega)=Z_2(\omega)=R/2$. (a)
Symmetric transistor with $R/R_K=0.05$. The transport voltages in units of the
gap voltage $V_g^0=V_g(Q_0=0)=e/C_{\Sigma}$ are from bottom to top $V=0.2, 0.6,
1.0, 1.2, 1.4, 1.6, 2.0$. (b) Asymmetric transistor with $R_1/R_2=10$ and
external resistance $R/R_K=0.05$. The transport voltages are
$V=0.6, 1.0, 1.2, 1.4, 1.6, 1.8, 2.0$. (c) Symmetric transistor
with $R/R_K=1$. The transport voltages are $V=0.6,1.0,1.6,2.0,2.4,2.8.$ The
current is given in units of $V_g^0/(R_1+R_2)$.}
\end{figure}

For equal tunneling resistances $R_1=R_2$ the $I$-$Q_0$-characteristics at
transport voltages below the gap voltage are given by parabolas symmetric to
$Q_0=e/2$. This behavior is shown in Fig.~27a together with some curves at
higher transport voltages. The environmental impedance is taken to be Ohmic
and rather small ($R/R_K=0.05$). If the tunneling resistance $R_1$ is much
larger than $R_2$ the $I$-$Q_0$-characteristic (\ref{eq:D76}) becomes
\begin{equation}
I(Q_0)=\frac{1}{2R_1}\left(\frac{Q_0-e/2}{2C}+V\right)\,
\Theta(Q_0-\frac{e}{2}+2CV)\, \Theta(-Q_0+\frac{e}{2}+2CV)
\label{eq:D77}
\end{equation}
which at the gap voltage reduces to
\begin{equation}
I(Q_0)=\frac{Q_0}{4R_1C}.
\label{eq:D78}
\end{equation}
In contrast to the parabolic characteristic obtained for the symmetric
transistor we now have a sawtooth-like characteristic with the slope
determined by the larger tunneling resistance. For the opposite case, $R_2\gg
R_1$, one finds from (\ref{eq:D76}) at the gap voltage
\begin{equation}
I(Q_0) = \frac{e-Q_0}{4R_2C},
\label{eq:D79}
\end{equation}
i.e.\ a reversed sawtooth characteristic. Numerical results for a ratio of
tunneling resistances $R_1/R_2=10$ and low Ohmic damping are presented in
Fig.~27b. Note that by choosing an asymmetric transistor one may obtain a
very high sensitivity\index{electrometer!sensitivity} for a certain range of 
offset charges. Fig.~27c shows
numerical results for a rather large
Ohmic impedance $R=R_K$ of the environment. Obviously the sensitivity on the
offset charge is
drastically reduced as compared with an electrometer embedded in 
a low impedance environment.\index{junction!double|)}\index{electrometer|)}
\index{SET transistor|)}\index{offset charge|)}
%
%
\subsection{Other multijunction circuits}\index{junction!multi|(}
The methods we have discussed so far in this section are not only applicable to
double junction systems but also to circuits containing more than two
junctions. Such multijunction systems can exhibit interesting physical behavior
and are therefore discussed extensively in other chapters of this book. We
mention systems containing few tunnel junctions (Chaps.~3 and 9),
one-dimensional arrays (Chap.~7), and two-dimensional arrays (Chap.~8).

The network analytical approach introduced in Sec.~6.2.\ may be applied to
a general multijunction circuit. To calculate tunneling rates across a given
junction the circuit can be reduced to an
effective single junction circuit containing a tunneling element, an
environmental impedance and an effective voltage source. Such a reduction
becomes possible by applying the Norton-Thevenin transformation discussed 
earlier. To
disentangle complex circuits one usually will also need the transformation
between a star-shaped and a triangle-shaped network, the so-called
T-$\pi$-transformation.\cite{network} Generally, the effective impedance will
contain a contribution diverging like $\omega^{-1}$ for small frequencies.
By splitting off this pole one separates the effective impedance into a
capacitance related to the charging energy and an impedance describing the
environmental influence. In this way, one directly finds expressions for the
tunneling rates through the individual junctions as long as simultaneous
tunneling through more than one junction is neglected. For a discussion of
phenomena arising if co-tunneling is taken into account, we refer the reader to
Chap.~6.

For the remainder of this section, we concentrate on the influence of the 
environment and choose as an
example a one-dimensional array\index{array!one-dimensional|(} of tunnel 
junctions as shown in Fig.~28.
\begin{figure}
\begin{center}
\includegraphics{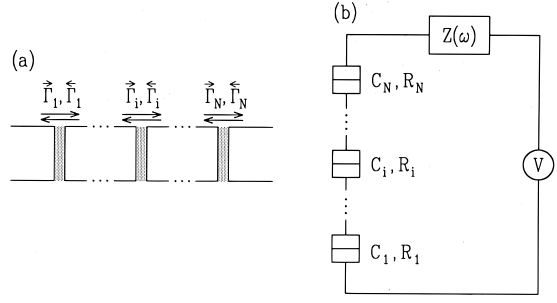}
\end{center}
\caption{(a) Schematic drawing of a one-dimensional $N$-junction array. The
arrows indicate forward and backward tunneling through the barriers. (b)
Circuit containing a one-dimensional $N$-junction array with capacitances
$C_i, (i=1,\ldots, N)$
and tunneling resistances $R_i$ coupled to a voltage source $V$ via the
external impedance $Z(\omega)$.}
\end{figure}
Although the network analysis for an array is straightforward, we shall first
present here a more microscopic approach which brings out the underlying
physics.
For simplicity, we neglect a capacitive coupling to ground which may be present
in a real setup and which is of importance for the description of charge
solitons in one-dimensional arrays (cf.\ Chap.~7). For our purposes it is
sufficient to consider the
capacitances associated with the $N$ tunnel junctions carrying the charges $Q_k\
(k=1,\ldots,N)$. As for the single and double junction systems we introduce
phases $\varphi_k\ (k=1,\ldots,N)$ satisfying the commutation relations
\begin{equation}
[\varphi_j,Q_k]=ie\delta_{jk}.
\label{eq:D80}
\end{equation}
To describe the environmental influence it is convenient to introduce another
set of charges and phases. From the surrounding circuit the array of tunnel
capacitors may be\vadjust{\break} viewed as a single capacitor with total capacitance
\begin{equation}
C=\left(\sum_{k=1}^N \frac{1}{C_k}\right)^{-1}
\label{eq:D81}
\end{equation}
carrying the total charge
\begin{equation}
Q = C\sum_{k=1}^N\frac{Q_k}{C_k}.
\label{eq:D82}
\end{equation}
In addition there are $N-1$ island charges
\begin{equation}
q_k = Q_k -Q_{k+1}\ \ (k=1,\ldots,N-1).
\label{eq:D83}
\end{equation}
This set of charges $\{Q,q_k\}$ is associated with a set of phases $\{\varphi,
\psi_k\}$ satisfying the commutation relations
\begin{equation}
[\varphi,Q]=ie,\ [\psi_k,q_k]=ie
\label{eq:D84}
\end{equation}
with all other commutators vanishing. The two sets of phases are related by
\begin{eqnarray}
\varphi_1 &=&\psi_1 +\frac{C}{C_1}\varphi\label{eq:D85}\\
\varphi_k &=& \psi_k -\psi_{k-1} + \frac{C}{C_k}\varphi\ \ \ \
(k=2,\ldots,N-1)\label{eq:D86}\\
\varphi_N &=& -\psi_{N-1} + \frac{C}{C_N}\varphi
\label{eq:D87}
\end{eqnarray}
as for the double junction system in Sec.~6.2.\ The
operator $\exp(-i\varphi_k)$ describes tunneling through the $k$-th junction.
According to (\ref{eq:D86}) this may be decomposed into an electron leaving the
$k$--1-th island $(-\psi_{k-1})$ and an electron entering the $k$-th island
$(\psi_k)$. In (\ref{eq:D85}) and (\ref{eq:D87}) only one $\psi$-operator
occurs since tunneling through the first or last junction affects only one 
island charge. The operators $\kappa_k\varphi$ are associated with the change of
the total charge $Q$ seen by the environment. Here, $\kappa_k=C/C_k$ is the 
obvious
generalization of (\ref{eq:D14}) to the multijunction case. 
The relations
(\ref{eq:D85})--(\ref{eq:D87}) essentially contain all information we need to
know about tunneling in a one-dimensional array. The change in electrostatic 
energy connected
with a tunneling process consists of contributions arising from the change of
island charges and a contribution from the work done by the voltage source to
restore the total charge.  The latter is given by $\kappa_k eV$ since according
to (\ref{eq:D85})--(\ref{eq:D87}) the charge transferred to the voltage source
after an electron has tunneled through the $k$-th junction is 
$\kappa_k e$. As in (\ref{eq:D22}) for the double
junction, the factor $\kappa_k$ leads to a reduced influence of the environment
due to a decoupling of the $k$-th tunnel junction from the environment
by the other junctions.
The total impedance is thus effectively reduced by a factor $\kappa_k^2$.
Assuming that the capacitances $C_k$ of the tunnel junctions are all of the
same order, we find from (\ref{eq:D81}) that the total capacitance is smaller by
a factor $1/N$. As a consequence, the influence of the environment on a
$N$-junction system is reduced by a factor $1/N^2$ as compared to a single
junction. The assumption of a low impedance environment is therefore usually
well satisfied.

To determine the tunneling rates we first derive an explicit expression for the
internal charging energy due to the island charges. Solving
(\ref{eq:D82}) and (\ref{eq:D83}) for the charges on the junctions we find
\begin{eqnarray}
Q_1 &=& Q + C\sum_{i=1}^N\sum_{k=1}^{i-1}\frac{q_k}{C_i}\nonumber\\
Q_n &=& Q + C\sum_{i=1}^N\sum_{k=1}^{i-1}\frac{q_k}{C_i}-\sum_{k=1}^{n-1}q_k
\ \ \ (n=2,\ldots, N).
\label{eq:D88}
\end{eqnarray}
After some algebra one obtains for the internal charging energy of the array
\index{electrostatic energy of array}
\begin{equation}
\varepsilon(q_1,\ldots, q_{N-1})=\sum_{i=1}^N \frac{Q_i^2}{2C_i} -
\frac{Q^2}{2C} = \sum_{k,l=1}^{N-1} \frac{1}{2}(C^{-1})_{kl}q_kq_l
\label{eq:D89}
\end{equation}
with
\begin{equation}
(C^{-1})_{kl} = C\sum_{m=1}^{\min(k,l)}\frac{1}{C_m}\sum_{n=\max(k,l)+1}^N
\frac{1}{C_n}.
\label{eq:D90}
\end{equation}
Here, $(C^{-1})_{kl}$ is the inverse of the capacitance matrix
\index{capacitance matrix} 
\begin{equation}
C_{kl} =\left\{\begin{array}{ll}
                C_k+C_{k+1}&\ \ \ \mbox{for $l=k$}\\
                -C_{k+1}&\ \ \ \mbox{for $l=k+1$}\\
                -C_k&\ \ \ \mbox{for $l=k-1$}\\
                0 &\ \ \ \mbox{otherwise}.
               \end{array}\right.
\label{eq:D91}
\end{equation}
We note that the structure of (\ref{eq:D89}) is also valid for more complicated
multijunction circuits (cf.\ Chap.~3). Only the explicit form of the
capacitance matrix will differ.

As a generalization of the energy difference (\ref{eq:D23}) for electron
tunneling in a double junction system we have for the one-dimensional array
the changes in electrostatic energy
\begin{eqnarray}
E_1(V,q_1,\ldots,q_{N-1}) &=& \kappa_1eV + \varepsilon(q_1,\ldots,q_{N-1}) -
\varepsilon(q_1-e,q_2,\ldots,q_{N-1}),\nonumber\\
E_i(V,q_1,\ldots,q_{N-1}) &=& \kappa_ieV + \varepsilon(q_1,\ldots,q_{N-1})
\nonumber\\
&&\hspace{1cm}- \varepsilon(q_1,\ldots,q_{i-2},q_{i-1}+e,q_i-e,q_{i+1},
\ldots,q_{N-1})\label{eq:D92}\\
&&\hspace{5cm}(i=2,\ldots, N-1),\nonumber\\
E_N(V,q_1,\ldots,q_{N-1}) &=& \kappa_NeV + \varepsilon(q_1,\ldots,q_{N-1}) -
\varepsilon(q_1,\ldots,q_{N-2},q_{N-1}+e)\nonumber
\end{eqnarray}
containing contributions from the change in the internal charging energy as
well as from the work done by the voltage source. While the general structure
of $E_i$ becomes very apparent if expressed in the set of variables $\{V, q_1,
\ldots, q_{N-1}\}$ the explicit form is rather complicated. On the other hand,
we may as well choose the set of charges $\{Q_1, \ldots, Q_N\}$. Making use of
(\ref{eq:D82}) and (\ref{eq:D83}) we then find from (\ref{eq:D92}) together
with (\ref{eq:D89}) the very simple result
\begin{equation}
E_i(Q_i) = \frac{e}{C_i}(Q_i-Q_i^c)
\label{eq:D93}
\end{equation}
with the critical charge\index{critical charge}
\begin{equation}
Q_i^c = \frac{e}{2}(1-\kappa_i).
\label{eq:D94}
\end{equation}
This is a straightforward generalization of the double junction result
(\ref{eq:D24}) and (\ref{eq:D25}).

Before turning to the tunneling rates we will shortly outline how the result
(\ref{eq:D93}) can be obtained by network analysis. Considering electron
tunneling through the $i$-th junction, we may combine the other junctions into
one capacitor of capacitance
\begin{equation}
\tilde C = \left(\sum_{j\ne i} \frac {1}{C_j}\right)^{-1} = \frac{CC_i}{C_i-C}
\label{eq:ctilde}
\end{equation}
carrying the charge
\begin{equation}
\tilde Q = \frac{C_iC}{C_i-C} \sum_{j\ne i}\frac{Q_j}{C_j}.
\label{eq:qtilde}
\end{equation}
We thus have reduced the one-dimensional array to an effective double junction
system. We now view the capacitances $C_i$ and $\tilde C$ as the two
capacitances in a double junction system. Accordingly, $Q_i$ and $\tilde Q$ are
the charges sitting on the two junction capacitors. Using our results from
Sec.~6.2.\ we find an effective single junction circuit like the one depicted
in Fig.~20c. The capacitor with effective capacitance
\begin{equation}
C_{\rm eff} = C_i + \tilde C = \frac{C_i^2}{C_i-C}
\label{eq:ceff} 
\end{equation}
carries the charge
\begin{equation}
q_{\rm eff} = Q_i-\tilde Q = \frac{C_i}{C_i-C}(Q_i-CV)
\label{eq:qeff}
\end{equation}
which corresponds to the island charge of a double junction. As already
discussed above the effective impedance is given by $\kappa_i^2 Z_t(\omega)$
and the effective voltage is $\kappa_iV$. The difference in electrostatic
energy $E_i$ is now
readily obtained as
\begin{equation}
E_i(Q_i) = \frac{q_{\rm eff}^2}{2C_{\rm eff}} - \frac{(q_{\rm eff}-e)^2}
{2C_{\rm eff}} + \kappa_i eV = \frac{e}{C_i}(Q_i-Q_i^c)
\label{eq:ei}
\end{equation}
in agreement with (\ref{eq:D93}) and (\ref{eq:D94}).

We are now in a position to write down the tunneling rates for a
one-dimensional array.
Using the same kind of
reasoning as for the single junction or double junction we obtain for the
forward tunneling rate through the $i$-th junction \cite{multizphys}
\begin{eqnarray}
\GF_i(V,q_1,\ldots,q_{N-1}) &=& {1\over e^2R_i}\int_{-\infty}^{+\infty}
{\rm d}E {E\over 1-\exp(-\beta E)}\nonumber\\
&&\hspace{2cm}\times P(\kappa_i, E_i(V,q_1,\ldots,q_{N-1})-E).
\label{eq:D95}
\end{eqnarray}
The backward tunneling rate is related to the forward tunneling rate by
\begin{equation}
\GB_i(V,q_1, \ldots, q_{N-1}) = \GF_i(-V,-q_1, \ldots, -q_{N-1})
\label{eq:D96}
\end{equation}
and the detailed balance symmetry
\begin{displaymath}
\GB_1(V,q_1-e,q_2,\ldots,q_{N-1}) =
\exp[-\beta E_1(V,q_1,\ldots,q_{N-1})]\GF_1(V,q_1,\ldots,q_{N-1}),\nonumber
\end{displaymath}
\begin{eqnarray}
\GB_i(V,q_1,\ldots,q_{i-2},q_{i-1}+e,q_i-e,q_{i+1},\ldots,q_{N-1})&&\\
&&\hspace{-7cm}= \exp[-\beta E_i(V,q_1,\ldots,q_{N-1})]
\GF_i(V,q_1,\ldots,q_{N-1})\ \ \ (i=2, \ldots, N-1),\nonumber
\label{eq:D97}
\end{eqnarray}
\begin{eqnarray}
\GB_N(V,q_1,\ldots,q_{N-2},q_{N-1}+e)&&\nonumber\\ 
&&\hspace{-3.6cm}=
\exp[-\beta E_N(V,q_1,\ldots,q_{N-1})]\GF_N(V,q_1,\ldots,q_{N-1}).\nonumber
\end{eqnarray}

It was pointed out above that the influence of the environment on the
tunneling of electrons in a $N$-junction array is reduced by a factor of the
order of $1/N^2$. This means that for most applications the low impedance
limit will be correct. The rate expression (\ref{eq:D95}) then reduces to the
global rule result
\begin{equation}
\GF_i(V,q_1,\ldots,q_{N-1}) = \frac{1}{e^2R_i}\frac{E_i(V,q_1,\ldots,q_{N-1})}
{1-\exp[-\beta E_i(V,q_1,\ldots,q_{N-1})]}
\label{eq:D98}
\end{equation}
which at zero temperature yields
\begin{equation}
\GF_i(V,q_1,\ldots,q_{N-1}) = \frac{1}{e^2R_i} E_i(V,q_1,\ldots,q_{N-1})
\Theta\Big(E_i(V,q_1,\ldots,q_{N-1})\Big).
\label{eq:D99}
\end{equation}
As for the simpler circuits, the zero temperature rate is only different from
zero if $E_i(V,q_1,\ldots,q_{N-1})>0$, i.e., if the
charge $Q_i$ on the $i$-th junction exceeds the critical charge $Q_i^c$ given
by (\ref{eq:D94}). This
local formulation of the blockade criterion involving only the charge $Q_i$,
the capacitance $C_i$ of the $i$-th junction, and the capacitance of the
surrounding junctions is of great importance for the understanding of
few-junction systems (cf.\ Chap.~3).

Let us finally discuss the size of the Coulomb gap.\index{Coulomb gap} 
As for the double junction
system the stable state at voltages below the gap voltage is characterized by
vanishing island charges $q_i\equiv 0$. According to (\ref{eq:D88}) the charges
on the junctions are then all given by $Q_i=CV$. Tunneling at the $i$-th
junction can therefore occur if $\vert V\vert > (e/2C)(1-\kappa_i)$. No current
will flow if the voltage across the array is smaller than
$(e/2C)\min_i(1-\kappa_i)$. Since $\kappa_i=C/C_i$, the gap voltage is
determined by the junction with the largest capacitance. For an array with
equal capacitances, i.e.\ $C_i=C_J=NC$, the low impedance Coulomb gap is given 
by $(1-1/N)(e/2C)=(N-1)(e/2C_J)$. 
However, this gap and the Coulomb offset observed at large
voltages where the high impedance gap $e/2C=N(e/2C_J)$ appears, differ by 
$e/2C_J$.\index{junction!multi|)}\index{array!one-dimensional|)}

\vspace{2\baselineskip}
\noindent
{\ackn Acknowledgements}. 
We have benefitted from the interaction with numerous colleagues of whom we can
name only few here. One of us (G.-L.~I.) has enjoyed a fruitful collaboration
with H.~Grabert and would like to acknowledge discussions with P.~Wyrowski and
the members of the Quantronics Group at Saclay. The other author (Yu.~V.~N.) 
is grateful to
T.~Claeson, D.~Haviland, and L.~S.~Kuzmin for discussions on experimental
aspects of single charge tunneling and to A.~A.~Odintsov and M.~Jonson for
critically reading the manuscript on the microscopic foundation. He enjoyed the
hospitality of the Fachbereich Physik at Essen where this chapter was written.
We thank G.~Falci for granting us permission to reproduce Figs.~13 and 14.
 
%
%
%
%
\appendix
\section{Microscopic foundation}\index{environment!electromagnetic|(}
\subsection{Introduction}
        One may note that in the previous   sections   the
tunnel junction was  treated  as  the  primary  object.  The  whole
electrodynamics was reduced to the network theory and the only
trace  of  solid  state  physics   was   the   one-electron  Fermi
distribution function. Therefore, one may  call  it  `phenomenological
approach'. It is difficult  to  underestimate  its  importance  for
applications. Nevertheless, there are some reasons to go deeper and
to discuss the microscopic foundation of the method applied. First, the
microscopic
derivation is always a good way to test and probably  to  confirm
the phenomenology. Second, the range of the  applicability  becomes
clearly visible. Third, the  interesting  links  between  different
approaches and different phenomena can be comprehended and  some
new effects can be described.

In this appendix we move from the  microscopic  description  of
the  metals  on both sides of  the   tunnel   junction  towards  the
phenomenology step by step. We encounter  interesting  physics
on every step and have a general look on it.

We are starting with the  formulation  of  the  problem  in  a
general way: how does the electromagnetic interaction affect the
tunneling rate? We develop a semiclassical approach to electron
motion and derive the basic formula  which  allows us to  evaluate  the
effect in terms of Maxwell  electrodynamics  and a Boltzmann  master
equation  description  for  the  electrons.  Then  we discuss  the
fundamental relation between the voltage applied  to  the  junction
and the effective scales at which the tunneling electron  feels the
electrodynamic environment.

After that we apply our approach to Altshuler's  diffusive
anomalies in the density of states.\cite{Yuli1} Within this framework these
anomalies and the Coulomb blockade appear to be two sides of one coin. While
proceeding we  consider
the most important case when the field induced by the electron moves
faster than the electron itself. It allows us to forget about  this
electron. Moving along this way we remind the  reader  of some  simple
electrodynamics  in  a metal-thin insulator-metal system  and
consider  fingerprints  of  this  electrodynamics  on  the   tunnel
junction $I$-$V$ curve. In the very end we show how one can move
from the continuous electrodynamics to the network theory finishing
the  consequent  microscopic  derivation  of  the  phenomenological
approach.
\subsection{General problem}
 As it is widely known, it is easier to answer general
questions than specific ones so that we try to formulate the problem
in a most general way. Let us answer the question: How does
the electromagnetic interaction affect the electron tunneling rate?

First, of course, this interaction forms the crystal lattice of metals
and of the insulator layer through which the electrons tunnel. It determines
the energy spectrum of elementary excitations in these condensed media
and thus provides every decoration of the scene for this solid state
physics performance. Everything is made here by the electromagnetic fields
acting on the length, energy, and time scales of the order of atomic values.
The question how it makes these things is in fact the basic problem of
condensed matter physics and we are not going to solve it just here. We come
on the scene formed and we are interested in the part of the electromagnetic
field which is: \topsep 0pt
\begin{enumerate}  \parsep 0pt \itemsep 0pt
\item slow enough not to change the electron energy on an atomic scale
\item weak enough not to turn electrons from its trajectories
\item described by linear electrodynamic equations
\item basically uniform on the atomic scale.
\end{enumerate}
\topsep 60pt
We call it the low-frequency part of the electromagnetic interaction and will
deal only with this part. Also we assume the voltage applied to the
junction to be much less than the typical atomic energy or the Fermi energy
for electrons. So we will operate on scales which are
much larger than the atomic ones. One may call this formulation a mesoscopic
problem but we prefer not to do it.
        The evaluation of the effect under consideration has been done in
Ref.~\cite{Yuli2} by using the standard formalism  of  field  trajectory  integrals.
This approach is
inconvenient due to the use of an imaginary time representation and the
relative complexity.
Here we use another way to derive it which is based mostly on physical
reasoning. This way can be compared with the one used in Ref.~\cite{Yuli3}.

        First let us note that on this scale a semiclassical approach to the
electron motion is valid
or, in simple terms, we can consider electrons as classical particles that
are scattered and jump through the tunnel barrier. Switching off the
low-frequency part of the electromagnetic interaction we introduce a
probability $w(y,k_{\rm in},k_f)$ to jump through the barrier at the point
$y$ on the barrier
surface with initial electron wave vector $k_{\rm in}$ and wave vector $k_f$
after tunneling. Only the electrons near the Fermi
surface can tunnel so we need to know this probability only on this surface.
The tunneling is completely elastic because the interaction is switched off.
The total tunneling rate then is given by
\begin{equation}
\Gamma(V) = \int\! {\rm d}^2 y {\rm d}^2k_{\rm in} {\rm d}^2k_f\, \nu_1 \nu_2
w(y,k_{\rm in},k_f) \int\!
{\rm d}\epsilon {\rm d}\epsilon'\,
f(\epsilon) [1-f(\epsilon'-eV)]  \delta(\epsilon-\epsilon').
\label{eq:Mic1}
\end{equation}
Here we integrate over $y$ belonging to the junction area, the two-dimensional
wave vectors $k_{\rm in}$ and $k_f$ parametrize the Fermi surface, the
$\delta$-function ensures
the tunneling to be elastic, $\nu_{1,2}$ denote the densities of states
per
energy interval at the Fermi surface
for the two metal banks, respectively, and $f(\epsilon)$ is the Fermi
distribution function.

        Let us consider now the problem from the quantum mechanical point of
view. We introduce the electron
propagation amplitude $K$ connecting the electron wave functions $\psi$ at
different times by \cite{Yuli3}
\begin{equation}
\psi(x,t) = \int {\rm d}^3x' K(x,t;x',t')\psi(x',t').
\label{eq:Mic2}
\end{equation}
Within the semiclassical approach the different classical trajectories
contribute to\break $K(x,t;x',t')$ without interfering with each other. If we
are interested in evaluating the tunneling
rate at a given point $y$ with given
$k_{\rm in}$ and $k_f$ there is a unique trajectory determined
by these parameters which contributes. The phase of the
propagation
amplitude  is proportional to the
classical action along this trajectory:
\begin{equation}
K(x,t;x',t') \sim \exp(iS/\hbar).
\label{eq:Mic3}
\end{equation}

   This is the way to take the interaction into account. Let us note that
electromagnetic interaction can be treated as the exchange of photons and as
a first step  consider the electron motion in the photon field.  In
accordance with the correspondence principle one should add to the total action
the interaction term and thus obtain the propagation amplitude in the presence
of a field:
\begin{equation}
K_{\rm field} (x',t';x'',t'') = K(x',t';x'',t'') \exp(iS_{\rm int}/\hbar)
\end{equation}
with
\begin{equation}
S_{\rm int} = e \int_{t''}^{t'} {\rm d}t
\left[\frac{v^{\alpha}\Big(k(t)\Big)
A^{\alpha}\Big(x(t),t\Big)}{c} + \phi\Big(x(t),t\Big)\right].
\label{eq:Mic4}
\end{equation}
Here, $e$ is the electron charge, $A^{\alpha}(x,t)$
and
$\phi(x,t)$ are the vector and scalar potential of the electromagnetic
field, respectively, and $x(t)$ and $k(t)$ are trajectory
parameters.  On a large time scale the
electron
is scattered many times by impurities and the metal surface and the trajectory
is extremely complicated.  This is a reason to characterize this trajectory by
the probability $p(x,k,t)$ for the electron to be in the point $x$ with wave 
vector $k$ at time $t$. It is worth
to emphasize that the probability is
conditional:  the electron must jump at the point $y$ with certain $k_{\rm in}$ and $k_f$
at time $0$.
Therefore, it is more convenient to rewrite (\ref{eq:Mic4}) in the
form
\begin{equation}
S_{\rm int} = e \int_{t''}^{t'} {\rm d}t \left[\frac{j^{\alpha}(x,t)
A^{\alpha}\Big(x(t),t\Big)}{c} +
\rho(x,t) \phi(x,t)\right]
\label{eq:Mic5}
\end{equation}
where
\begin{equation}
j^{\alpha}(x,t) = \int\! {\rm d}^2k\, v^{\alpha}(k) p(x,k,t-T_1-t'')
\end{equation}
and
\begin{equation}
\rho(x,t) = \int\! {\rm d}^2k\, p(x,k,t-T_1-t'').
\end{equation}
Here, $T_1$ is the time needed to move
from $x''$ to $y$ along the trajectory.
A careful analysis shows that there is also a probability flow
\begin{equation}
j^{\alpha}(x,t)
= v^{\alpha}(k_{\rm in})N^{\alpha}p(y,k_{\rm in},t-T_1-t'') =
v^{\alpha}(k_{f})N^{\alpha}p(y,k_{f},t-T_1-t'')
\end{equation}
at an $x$ belonging to the oxide barrier volume, where $N$ is a vector normal to the
junction surface.
 The necessity of this term
results from the conservation of probability flow.  The barrier thickness is of
the order of a few atomic sizes and at first look the contribution of this term to the
action is much smaller than that of the probability flow in the metal.
However, the electromagnetic
field in the oxide barrier may be much larger than in the surrounding metals and therefore
one has to keep this term.

Performing the field quantization we replace classical potentials by
appropriate Heisenberg time-dependent operators.  The propagation amplitude
also becomes an operator.  The tunneling rate at given energies $E_{\rm in}$,
$E_{f}$ in initial and final states is proportional to the square of the Fourier
transform of the propagation amplitude
\begin{equation}
\Gamma (E_{\rm in},E_f) \sim \langle\hat
K(E_{\rm in},E_f) \hat K^+(E_{\rm in},E_f)\rangle
\label{eq:Mic6}
\end{equation}
where $\langle\ldots\rangle$ denotes the average
over the equilibrium  density matrix of the electromagnetic field and
\begin{equation}
\hat K(E_{\rm in},E_f) = \int\! {\rm d}t' {\rm d}t''\,
\exp\left[\frac{i}{\hbar}(E_ft''-E_{\rm in}t')\right] \hat K(t',x';t'',x'').
\label{eq:Mic7}
\end{equation}
In this equation $x''$ and $x'$ are chosen to be very far from the
jump point $y$ in the one bank and in the other one, respectively. They can
be characterized by the two traversal times $T_{1}$ needed to move from $x''$ to
$y$  and $T_2$ needed to move from $y$ to $x'$. For $T_{1,2}
\rightarrow \infty $ the square of the
propagation amplitude reaches a certain
limiting value related to the rate.

        If we are not interested in relativistic effects
concerning the electromagnetic field propagation we are able to omit the vector
potential in (\ref{eq:Mic4}) and (\ref{eq:Mic5}). Using (\ref{eq:Mic5}) we
then obtain

\break
\begin{eqnarray}
\hat K(E_{\rm in},E_f) &\sim&
\int\! {\rm d}t'' \exp\left[\frac{i}{\hbar}
(E_f-E_{\rm in})t''\right]\nonumber\\ 
&&\hspace{2cm}\times\exp\!\left[\frac{i}{\hbar}\!\int^{t''}_{t''-T_1-T_2} 
\!\!\! {\rm d}t\int\! {\rm d}^3x\, \rho(t-t''-T_1,x) \hat\phi(t,x)\right].
\label{eq:Mic7a}
\end{eqnarray}
According to the previous considerations we can consider $\hat\phi(t,x)$ as the sum
of a large number of boson operators as it has been considered within the
phenomenological approach in Sec.~2.3.
It allows us to use the simple rules for operator multiplication
when calculating the square. For example,
\begin{eqnarray}
\Big\langle\exp\! \Big(i\!\int\! {\rm d}t\, C_1(t) \hat\phi(t,x)\Big)  \exp\! 
\Big(-i\!\int\! {\rm d}t\, C_2(t)
\hat\phi(t,x)\Big)\Big\rangle&&\nonumber\\
&&\hspace{-5cm}= \exp\Big(-\int\!{\rm d}t_1 {\rm d}t_2\, F(t_1,t_2)
\langle\hat\phi(t_1) \hat\phi(t_2)\rangle\Big)
\end{eqnarray}
with
\begin{equation}
F(t_1,t_2) = {C_1(t_1)C_1(t_2)+ C_2(t_1)C_2(t_2)\over 2}
-C_1(t_1)C_2(t_2).
\end{equation}
        After some algebra one gets
\begin{equation}
\Gamma(E_{\rm in},E_f) \sim
P(E_{f}-E_{\rm in})
\label{eq:Mic8}
\end{equation}
where we introduced
\begin{equation}
P(E) = \int {{\rm d}\tau\over 2\pi}\ \exp\left[\frac{i}{\hbar}E\tau\right] 
\exp(-Y(\tau))
\label{eq:Mic9}
\end{equation}
with
\begin{equation}
Y(\tau) = {e^2\over\hbar^2} \int
{{\rm d}\omega\over 2\pi} (1-e^{-i\omega\tau})
\int {\rm d}^3x {\rm d}^3x' \rho_{\omega}(x)\rho_{-\omega}(x')
D^{+-}(\omega;x,x').
\label{eq:Mic10}
\end{equation}
The latter can be expressed in terms of the Fourier
transforms of $\rho$ and the field correlation function
$\langle\hat\phi(0) \hat\phi(t)\rangle$:
\begin{equation}
\rho_{\omega}(x) = \int\! {\rm d}t\, \rho(t,x)
e^{i\omega t}
\end{equation}

\begin{equation}
D^{+-}(\omega;x,x') = \int\!{\rm d}t\,\langle\hat\phi(0,x) \hat\phi(t,x')\rangle
e^{i\omega t}.
\end{equation}

        The coefficient of
proportionality in (\ref{eq:Mic8}) can be determined if we
switch off the electromagnetic interaction.
Thus we obtain for the total
tunneling rate as a nice generalization of
(\ref{eq:Mic1})
\begin{eqnarray}
{\Gamma}(V) = \int\! {\rm d}^2 y {\rm d}^2k_{\rm in} {\rm d}^2k_f\, \nu_1 \nu_2
w(y,k_{\rm in},k_f)&& \nonumber\\
&&\hspace{-3cm}\times\int\! {\rm d}\epsilon {\rm d}\epsilon'
f(\epsilon) [1-f(\epsilon'-eV)]
P(\epsilon'-\epsilon;y,k_{\rm in},k_f).
\label{eq:Mic11}
\end{eqnarray}

        Now we should discuss what we have done. The physical picture
in fact is rather simple. In (\ref{eq:Mic1}) we replace the $\delta$-function,
which reflects the fact that in the absence of the interaction the
tunneling is elastic,
by the probability $P(E)$ to have a certain energy change when
tunneling
in the presence  of  the  interaction.  This  change  is  provided  by
emission or
absorption of photons. Due  to the semiclassical nature of electron
motion and due
to the small energy transfer for every emission/absorption act all of these
acts happen independently. Thus for every photon energy the
probability to emit a
certain number of photons obeys the Poissonian statistics and (\ref{eq:Mic9})
and (\ref{eq:Mic10}) are simply a
generalized mathematical formulation of this fact. Within the
phenomenological approach this was discussed in Sec.~4.1. Within
the microscopic theory we were able to express these
photon emission/absorption probabilities in terms of the electron motion along
the classical trajectory and the field correlation function. Thus we
managed
to simplify the problem considerably. Now, we can calculate the
tunneling rate if we
know all about classical electron motion and the electrodynamics in the region
around the junction. Let us now express our results in a form
appropriate for practical use.

        First we express the field correlation function in terms of the
response function by means of the fluctuation-dissipation theorem
\begin{equation}
D^{+-}(\omega;x,x') = - \left(\coth({\hbar\beta\omega\over 2})-1\right)\,
\hbox{Im} D^A_{\omega}(x,x').
\label{eq:Mic12}
\end{equation}
Here $D^A_{\omega}(x,x')$ is the advanced response function
\begin{equation}
\Phi_{\omega}(x) = -{e\over\hbar}\ \int {\rm d}^3x' D^A_{\omega}(x,x')
\rho_{\omega}(x')
\label{eq:Mic13}
\end{equation}
which determines the field response $\Phi_{\omega}(x)$ to an external
charge placed
at point $x'$. It is convenient to consider $e\rho_{\omega}(x)$ as
this external
charge. $\Phi$ is determined by the electrodynamic equations in a
medium
with a given source. It is easier to solve these equations than to
completely calculate the response
function $D_{\omega}^A$. The result (\ref{eq:Mic10}) may be rewritten in
terms of $\Phi$ as
\begin{equation}
Y(\tau) = {e\over\hbar} \int {{\rm d}\omega\over 2\pi} (1-e^{i\omega\tau})
\int {\rm d}^3x
\hbox{Im} [\Phi_{\omega}(x)\rho_{-\omega}(x)]
\left[\coth({\hbar\beta\omega\over 2})+1\right].
\label{eq:Mic14}
\end{equation}

To calculate the probability $p(x,t,k)$
which characterizes the electron motion
one may use the standard Boltzmann master equation approach. For
example, for the probability to be on one of
the banks after tunneling the Boltzmann equation reads
\begin{eqnarray}
{\partial p(k)\over\partial t} = v^{\alpha}(k){\partial
p(k) \over \partial {x_{\alpha}}} +\int {\rm d}^2k'\,W(k,k',x)
\Big(p(k')-p(k)\Big)&&\nonumber\\
&&\hspace{-3.5cm}+\delta(t)\delta^3(x-y)\delta^2(k-k_f).
\label{eq:Mic15}
\end{eqnarray}
Here, $v^{\alpha}(k)$ is the electron
velocity in the state with a wave vector
$k$, $W(k,k',x)$ is the scattering
rate from the state with $k$ to the
state with $k'$ due to the impurities
and the metal surface, and the source term
describes the electron arrival from the other bank at $t=0$.
        Often we need to describe the
electron motion only on a time scale
which is much larger than the time
between the scattering events. In this
case one may use the diffusion equation
for the electrons:
\begin{equation}
{\partial \rho\over\partial t} =
D\Delta \rho +
\delta(t)\delta^3(x-y)
\label{eq:Mic16}
\end{equation}
with appropriate boundary conditions.

For the derivation of (\ref{eq:Mic11}) we assumed a non-relativistic field.
Strictly speaking this means that we are not able, for example, to
describe
the inductance in electric circuits. This is why it is worth to emphasize that
all the field relativistic effects can be treated in the same manner. The most
convenient gauge choice is $\phi(x,t)=0$ and the field is
described by the vector potential only.
Acting along the same lines we express
the answer
\begin{equation}
Y(\tau) = {e^2\over\hbar^2} \int {{\rm d}\omega\over 2\pi} (1-e^{-i\omega\tau})
\int\! {\rm d}^3x {\rm d}^3x'\,
j_{\omega}^{\alpha}(x)j_{-\omega}^{\beta}(x')
D^{+-}_{\alpha\beta}(\omega;x,x')
\label{eq:Mic16a}
\end{equation}
in terms of the vector
potential correlation function
\begin{equation}
D^{+-}_{\alpha\beta} (\omega;x,x') = \int\! {\rm d}t e^{i\omega t}
\frac{\langle\hat A^{\alpha}(0,x) \hat A^{\beta}(t,x')\rangle}{c^2}.
\end{equation}
Here $\vec j$ is the probability flow
introduced earlier. It is also
possible to simplify this form by
introducing the vector potential
response $A^{\alpha}(x,t)$ on the external
current $e \vec j(x,t)$ yielding
\begin{equation}
Y(\tau) = {e\over\hbar c} \int {{\rm d}\omega\over 2\pi}
(1-e^{i\omega\tau})\left[\coth({\hbar\beta\omega \over 2})+1\right]
\int\! {\rm d}^3x\, \hbox{Im} [A_{\omega}^{\alpha}(x) j^{\alpha}_{-\omega}(x)].
\label{eq:Mic17}
\end{equation}
        Now we are in a position to apply these general results
to some illustrative examples.
\subsection{Time of tunneling}
        Before making these applications it is important to discuss the
relevant
time scale on which we are going to operate. As far as we consider tunneling
rates  this relevant time should be the time of tunneling.
        The problem is not transparent and sometimes it leads to
misunderstanding. To feel that let us consider the noninteracting electrons
 on a scale much larger then the atomic one. The electrons rush along
the metal and sometimes jump through the tunnel
barrier. The thickness of this barrier is of the atomic order and
a jump takes no
time. On the other hand, the tunneling is elastic and the energy
loss equals
zero. According to quantum mechanics, the energy uncertainty $\Delta
E$ and the time for tunneling $t$ obey the relation
$\Delta E\cdot t \simeq \hbar$. It means that the time for elastic
tunneling is infinitely long.

        So we have a real choice: from zero to infinity. This is
natural since quantum
mechanics always gives rise to duality. The answer  depends on the way
how the
time of tunneling is introduced or, in practical terms, it depends on
the quantity measured.

If we are sure that our tunneling is elastic there are many ways to introduce
the traversal time (see \cite{Yuli6} for a review) and sometimes to measure it
\cite{Yuli7}. In
this case the time is determined by the properties of the electron's motion
under the barrier
and there is some interesting physics due to the virtual nature of this
motion. In contrast, for inelastic tunneling the voltage  applied to
the
tunnel barrier and/or the temperature impose strict restrictions on the
frequency $\omega$ of the radiation emitted/absorbed: $\hbar\omega\leq
\hbox{max} \lbrace eV, k_BT \rbrace$. This frequency determines the time scale
as can be
seen from the previous equations. If the most part of the tunneling events are
inelastic it also determines the time of tunneling. If it is not so
this argument is applicable only for inelastic events.

Thus the tunneling may be characterized by two time scales. The
case when
these scales are of the same order is described in Refs.~\cite{Yuli3,Yuli8}.
Note that in this
case the tunnel barrier is suppressed by the applied voltage.

        The frequency scale specifies also the length scale since we
are able to
estimate the distance which the electromagnetic field or electrons propagate
for a given time. It allows us to find the effective geometry of the
junction or, in other terms, to determine whether it can be considered
as a point
or as an interface of two semi-infinite metal banks. The length scale
increases
with decreasing frequency so that the lower the voltage and temperature the
further away is the horizon which the junction  sees. However, we
should  emphasize that the
length scale can not be determined unambiguously. The electrons and
the field
propagate with different velocity. Moreover, different types of
electromagnetic
excitations differ in its velocities. This makes the length scale and
effective geometry dependent on the inelastic process under
consideration.
\subsection{One-photon processes: anomalies and fingerprints}
        Now we try to find the scale for the strength of the effect considered.
To characterize this strength it is convenient to introduce the effective
frequency dependent impedance which the electron feels when tunneling:

\begin{equation}
Z_{\rm eff}(\omega) = \frac{i \omega}{e} \int {\rm d}^3x \,\Phi_{\omega}(x)
\rho_{-\omega}(x)
\label{eq:Mic18}
\end{equation}
or for the other gauge choice (\ref{eq:Mic17}):
\begin{equation}
Z_{\rm eff}(\omega) = \frac{i \omega}{ec}\int{\rm d}^3x\,A_{\omega}^{\alpha}(x)
j^{\alpha}_{-\omega} (x).
\label{eq:Mic19}
\end{equation}

The value of this impedance in the frequency region considered governs the
deviations from Ohm's law. If this value exceeds the quantum unit of
resistance $R_{K}=h/e^2$ the probability for many-photon processes is
significant, the deviations are large, and the tunneling
rate is strongly suppressed in comparison to Ohm's law. If not, the most part
of the tunneling events are elastic and there are only  small deviations from
an Ohmic behavior which are due to one-photon processes.
        Very roughly the effective impedance can be estimated as the resistance
of a metallic piece on an appropriate length scale. Usually, on the microscopic
scale this resistance is small in comparison with $R_K$ and the deviations from
Ohm's law are small. For our illustrative applications we use this fact and we
will consider mostly one-photon processes.

  Let us express the probability $P(E)$ in terms of the impedance. To do this we
expand the exponent in (\ref{eq:Mic9}) with respect
to $Z_{\rm eff}$. The first order dependence of $P(E)$ on $Z_{\rm eff}$ will be
\begin{equation}
\delta P(E) = P_1(E) - \delta(E)\int
{\rm d}E'P_1(E')
\end{equation}
where $P_1(E)$ is the probability to emit one photon in a unit energy interval
\begin{equation}
P_1(E) = {2\hbox {Re}
Z_{\rm eff}(E/\hbar)\over E R_K} {1\over
1-\exp(-\beta E)}.
\label{eq:Mic20}
\end{equation}
The simple formula
\begin{equation}
R_T {\partial^2 I \over \partial V^2} = {2Z_{\rm eff}(eV/\hbar)\over R_K
V}
\label{eq:Mic21}
\end{equation}
valid at zero temperature reflects the influence of one-photon processes on the
$I$-$V$ curve.
        By the way there is an answer how one can observe this small deviations
on the background of the main effect. These deviations are clearly visible on
the differential conductance-voltage curve or on the second derivative of the
current in the region of small voltages because Ohm's law is linear. It is
convenient to divide the observable deviations into two classes:
\begin{enumerate} \parsep 0pt \itemsep 0pt
\item anomalies: the anomalous power law is displayed in some voltage region,
\item fingerprints: the deviation is localized near a certain voltage.
\end{enumerate}
\subsection{Diffusive anomalies}
        So-called zero-voltage anomalies were observed in tunnel junction
experiments from the early sixties. As it is comprehended now they were caused
by different mechanisms. The early explanation ascribed the whole effect to
the scattering by paramagnetic impurities.\cite{Yuli10} It was confirmed that
this can sometimes produce such anomalies \cite{Yuli11}, but there was also an
effect in the absence of these impurities.

In 1975, Altshuler and Aronov proposed a more fundamental mechanism to be
responsible for zero-voltage anomalies.\cite{Yuli12} They calculated the
interelectron
Coulomb interaction effect on the carrier density of states near the Fermi
level. Although the contribution to the density of states was found to be small
it influences the observed anomalies because it depends nonanalytically on the
distance from the Fermi level. A relation of the same kind as
Eq.~(\ref{eq:Mic1}) was
applied to calculate the tunnel current. The densities of states $\nu_{1,2}$
in this relation were allowed to have a small energy-dependent part. This
results in a square-root contribution to the conductance of a junction and in
the transition from the square-root to the logarithmic dependence as the
thickness of the electrodes is reduced. These dependencies were perfectly
confirmed by the experiments of Refs.~\cite{Yuli13} and \cite{Yuli14}.

        In our opinion, these results are correct but there are two points of
criticism concerning the link between the tunneling rate and the density of
states. A discussion of these points will probably allow to better understand
the physics involved.

        First, the relation used assumes the tunneling rate to be proportional
to the density of states in the banks of the junction. This assumption is
undoubtedly correct within a one-particle theory where only elastic tunneling
is
possible, but it is not satisfied when the  interelectron interaction
is included.
If the interaction affects the density of states, it is not consistent to
neglect it when calculating the tunnel current.

        Second, we are quite pessimistic about the principal possibility to
measure the electron density of states in the presence of interaction excluding
only few cases. The density of states in the presence of interaction is defined
and calculated as being proportional to the probability to annihilate an
electron
with a given energy at a given point. Under realistic circumstances the number
of electrons is conserved. We therefore can not annihilate an electron but only
pull
it out at a point and then measure the energy of this electron. In
fact, this is the way how the experimental methods work, from X-ray to
tunneling
methods. But if there is an interaction by which the electron may loose or gain
energy when
being pulled out, there is a fundamental restriction on the resolution of
these measurements.

        This is why we prefer to discuss the effect of the electromagnetic
interaction
on the tunneling rate but not on the density of states. Now, we obtain
anomalies involved in the framework of the method presented above and we will
find them to be due to inelastic tunneling.

        Let us first consider two semi-infinite metals separated by an
insulating layer. We assume the frequency scale related to the voltage and
temperature to be less then the  inverse electron momentum relaxation time
${1/ \tau_{\rm imp}}$. This assumption allows us to use a diffusion equation
for
describing the electron motion. In order to evaluate the effective impedance
we first calculate the Fourier transform of the conditional probability
$ \rho_{\omega} $. Solving Eq.~(\ref{eq:Mic16}) we obtain
\begin{equation}
\rho_{\omega}(r) = \pm {1 \over 2 \pi D r} \exp (-\sqrt{\frac{i(\omega \pm
i0)}{D}}r).
\label{eq:Mic22}
\end{equation}
Here, different signs refer to the different banks and $r$ is the distance
from the point of tunneling. Now we should evaluate $\Phi_{\omega}(r)$. To do
this, we solve the electrostatic equation
with external charge $e\rho_{\omega}(r)$:
\begin{equation}
\Delta  \Phi_{\omega}(r) = 4 \pi \Big(q(r)+e \rho_{\omega}(r)\Big).
\label{eq:Mic23}
\end{equation}
Here, $q(r)$ is the charge density formed by the metal electrons. As it is
known
metals are electroneutral and the sum on the right hand side of
(\ref{eq:Mic23}) equals zero
in the metal. The only point where it is not zero is the mere point of the
tunneling. The potential difference produces a current in the metal with
density $\vec j= -\sigma \vec \nabla
\Phi$, $\sigma$ being the metal
conductivity. Due to electroneutrality
the total current through the point of
tunneling must be equal to the flow of
external charge through this point with
the inverted sign. It allows  us to obtain
$\Phi_{\omega}(r)$:
\begin{equation}
\Phi_{\omega}(r)= \pm {e\over 2 \pi
\sigma r}.
\end{equation}
After integration over space we find for the effective
impedance
\begin{equation}
\hbox{Re} Z_{\rm eff}(\omega) = {1\over \pi \sigma} \sqrt{{\omega\over 2
D}}.
\end{equation}
With the aid of (\ref{eq:Mic21}) we obtain for the anomaly of the tunneling
current
\begin{equation}
R_T \delta I = {4e^2\over 3 \pi^2
\sigma \hbar} \sqrt{\frac{eV}{2D\hbar}}\ V.
\label{eq:Mic24}
\end{equation}
 The  result is the same as in \cite{Yuli12} if taking into account the surface
effect \cite{Yuli15} but now we can ascribe it to inelastic tunneling.
       The length scale is of the order $(eV/D\hbar)^{-1/2}$ and the
impedance
can be estimated as the resistance of a metal piece of that size. The
resistance
increases with decreasing size provided the size is less than the electron mean
free path $l$. The maximum value of
resistance would be of the order of $\ R_K
(k_Fl) \ll R_K$ for a reasonable
metal ($k_F$ is the electron wave vector
at the Fermi surface) and it ensures that the deviation is never comparable
with the main current.

Now we change the effective geometry and let the electrodes be films
of thickness $d$. The above consideration is valid if $(eV/D\hbar)^{-1/2} \gg
d$ and we now investigate the opposite limiting case $(eV/D\hbar)^{-1/2} \ll
d$. In this case
we can treat the probability and voltage as to be approximately constant across
the film. It is convenient to use Fourier transformation here with the wave
vectors $\vec q$ along the film plane. For $\rho_{\omega}(q)$ we obtain
\begin{equation}
\rho_{\omega}(q) = \pm {1 \over \pm i\omega + D q^2}.
\end{equation}
The calculation of $\Phi_{\omega}$ is a
little bit more complex. The films separated by the insulating layer can be
considered as a large capacitor with $C_0$ being the capacitance per unit area.
The voltage difference between the electrodes produces the finite density of
charge per unit area $\tilde q = C_0 \Delta \phi$. Due to the symmetry of the
system, $\Phi_{\omega}(x)$ has different signs but equal magnitude on the
different electrodes. This allows us to write down the conservation law for the
charge density in the following form:
\begin{equation}
{\partial \tilde q \over \partial t}- \hbox{div}(\sigma \nabla \phi) = e
\delta(t)\delta^2(x).
\end{equation}
 As the next step we obtain
\begin{equation}
\Phi_{\omega}(q)= \pm {e\over i\omega + D^* q^2}.
\end{equation}
$D^*=\sigma d/2 C_0$ may be interpreted as the diffusivity of the electric
field. As a rule $D^*\gg D$ and the field propagates faster then the electron.
To obtain the impedance we integrate over $\vec q$ and note that the dominant
contribution to the integral comes from a wide region of $q$: $(\omega/D)^{1/2}
\gg q \gg (\omega/D^*)^{1/2}$. We can ascribe the effect neither to field nor
to electron
propagation: there is something in between. It is convenient to express the
answer introducing the film sheet resistance $R_{\Box} = (\sigma d)^{-1}$

\begin{equation}
R_T {\partial^2 I \over \partial V^2} = {2R_{\Box} \over \pi R_K V}
\ln (D^*/D).
\label{eq:Mic25}
\end{equation}
It differs from the one derived from the density of states by a logarithmic
factor. This
is natural because the approach of Refs.~\cite{Yuli1,Yuli12,Yuli15} does not
take into account the
field propagation induced by the tunneling electron.

  As far as we know we considered here all observable anomalies related to
electron diffusion. Considering the variety of other anomalies and the effect
of the external circuit we
needed not to take into account electron motion at all. The reason for
that is the following:
\subsection{Field moves faster than the electrons}
Indeed this is the usual case as we already have seen when we considered
field propagation along resistive films: the field diffuses faster than the
electrons. If the resistivity of the electrodes is lower we encounter different
types of electromagnetic excitations which can be as fast as the light and they
effectively overtake electrons.
Due to this fact the length scale for the electromagnetic field is much larger
than
that one for electron propagation. Thus the electromagnetic field is constant
on the length scale for electrons. It means we can
use the simplest expression for $\rho_{\omega}(x)$:

\begin{equation}
\rho_{\omega}(x) = \pm {\delta(x) \over i\omega \pm 0}.
\end{equation}
Different signs correspond to different banks. Integrating over $x$ in
Eq.~(\ref{eq:Mic18})
we obtain the simple but promising result:
\begin{equation}
Z(\omega) = \Phi^{(1)}_{\omega}(0) - \Phi^{(2)}_{\omega}(0).
\label{eq:Mic26}
\end{equation}
Here, the superscripts (1) and (2) refer to different banks. Now, we are
allowed to omit the
subscript `eff' because we have the honest electrodynamic impedance defined
as voltage difference between banks at point 0 provided that this is the
voltage
response to the current produced by the electron jumping over the tunnel
barrier. We have made one more step towards the phenomenology.

As an application of Eq.~(\ref{eq:Mic26}) we consider the
influence of the
undamped electromagnetic excitations which can propagate along the junction
interface. We will assume that the junction is large enough so that the typical
time
scale defined by voltage/temperature  is much smaller than the time needed
for the electromagnetic excitation to cross the junction. The existence of
these
undamped excitations is provided by the insulating layer which separates the
metallic banks and which can be considered as an infinite capacitor
characterized
by the capacitance $\tilde C$ per unit area. To calculate the impedance we
write down the balance  equation  for  the charge  density  of  this
capacitor

\begin{equation}
{\partial \tilde q \over \partial t} = J_z + \delta(x) \delta(t)
\end{equation}
where $J_z$ is the volume density of the electrical current taken on the
metal surface ($z$ is normal
to the junction interface). Performing the Fourier transform in time and in
space
coordinates along the interface and expressing all in terms of the voltage
difference we can rewrite the previous equation as
\begin{equation}
\Big(i\omega \tilde C  - iB(\omega,q)\Big)\Phi_{\omega}(q) = 1.
\label{eq:Mic27}
\end{equation}
Here, we introduced the non-local link between the normal current and
the voltage  on  the junction: $ J_z (q) =
iB(\omega,q)\Phi_{\omega}(q)$. From the previous equation we  have for the
impedance

\begin{equation}
Z(\omega) = -i\int {{\rm d}^2 q
\over {(2 \pi)}^2}{1 \over \tilde C \omega - B(\omega,q)}.
\end{equation}
 We need only the real part of the impedance to evaluate its effect on the
$I$-$V$ curve.
While the excitations are undamped the poles of the impedance at every $q$ 
lie on the
real axis and only these poles contribute to the real part. Thus for the real
part of the impedance we obtain

\begin{equation}
\hbox{Re} Z(\omega) = {\pi \over \tilde C} \int  {{\rm d}^2 q \over
{(2\pi)}^2}\delta \Big(\omega - \Omega(q)\Big)
\label{eq:Mic28}
\end{equation}
where $\Omega(q)$ is the
spectrum of electromagnetic excitations. To evaluate the effect we need to know
only the capacitance and this spectrum.

Let us first consider plasma excitations. For a bulk metal the plasmons
can not have an energy less than the plasma frequency $\omega_p$. Nevertheless
there are  low energy plasma excitations localized on the junction interface.
We consider them in some detail. The electrical field in the metal produces the
current of electron plasma
\begin{equation}
\vec J = { \omega_p^2 \over 4 \pi i \omega } \vec E.
\end{equation}
The electrostatic potential in the metal obeys Coulomb's law $\Delta \phi = 0$.
It means that
\begin{equation}
\Phi(z,q) = \Phi(0,q)\exp(-qz)
\end{equation}
and the electrical field on the metal surface is $E_z(0,q)=q\Phi(0,q)$.
Combining these relations together with
(\ref{eq:Mic27}) we obtain for the spectrum
\begin{equation}
\Omega(q)=\omega_p \sqrt {q/8 \pi \tilde C}
\end{equation}
and for the deviation of the tunnel current
\begin{equation}
V{\partial^2 I\over \partial V^2} =
\hbox{sign}(V) {16 \pi \tilde C \over R_K
\omega_p} {({V/ \omega_p})}^3.
\label{eq:Mic29}
\end{equation}
We write `sign' here in order to emphasize the non-analytical behavior of this
deviation. As a rule the dimensionless factor on the right hand side of
(\ref{eq:Mic29}) is less
than unity. For a reasonable thickness of the tunnel barrier of a few atomic
lengths this coefficient is of the order $E_F/\omega_p$. This ratio is less
than unity for most metals.

One may note that the velocity of the plasma excitations increases with
decreasing frequency so that at lower frequencies we should take into account
relativistic effects omitted in the previous  consideration.  Actually, an
electrical
current in the plasma produces a magnetic field and an alternating magnetic
field induces an electric one. Due to this fact the field penetration depth is
restricted by the value of
$c/\omega_p$. By taking this fact into account the
spectrum of electromagnetic excitations is given by

\begin{equation}
\Omega(q) = \omega_p \ \sqrt{{q^2 \over 8 \pi \tilde C
\sqrt{(\omega_p/c)^2+q^2 }}}
\label{eq:Mic30}
\end{equation}
which describes the crossover between high frequency plasma
excitations and low frequency Swihart waves. In terms of voltage this crossover
occurs at

\break
\begin{equation}
eV \sim \omega_p/\hbar \sqrt{\omega_p / c \tilde C} 
\end{equation}
which is about 100 mV for aluminium.
At lower voltages, we obtain for the deviation

\begin{equation}
R_T\delta ({{\partial I\over\partial V}}) = 16 \pi V {e^3 \over \hbar^2
v_{\rm sw}^2 \tilde C}
\label{eq:Mic31}
\end{equation}
where the velocity of the Swihart waves is $v_{\rm sw}=\sqrt{\omega_p/8\pi c 
\tilde C}$.

        Usually, the electrodes are thin metallic films. For low voltages when
for a typical $q$ we have $qd \simeq 1$,
where $d$ is the film thickness, the previous
results should be modified.  The velocity of electromagnetic waves
in this system can be compared with the
speed of light, so that Eq.~(\ref{eq:Mic28}) should also be modified. We
present the result for the case $qd \ll 1$ \cite{Yuli16}
\arraycolsep 0pt
\begin{eqnarray}
&&v_1= \omega_p^2 d/8\pi\tilde C,\ \ \ \ \ 
v=v_1/\sqrt{1+(v_1/c^*)^2},\nonumber\\
&&\hbox{Re} Z(\omega) = {\pi \over \tilde C} \int {{\rm d}^2 q \over
{(2\pi)}^2}\delta (\omega - vq) {1\over 1+(v_1/c^*)^2}\\
&&R_T\delta ({{\partial I\over\partial V}}) = 16 \pi V {e^3 \over \hbar^2
v_{1}^2 \tilde C}.\nonumber
\label{eq:Mic32}
\end{eqnarray}
\arraycolsep 1.4pt
Here, $v_1$ is the speed of electromagnetic excitations  without  taking
into account
relativistic effects, $v$ is the real speed, $c^*$ is the speed of light in the
insulating layer and the Lorentz factor describes the relativistic effects.

        There is a variety of different regimes of field propagation in this
large area junctions and we will not review this matter here. Some of the cases
were described in Refs.~\cite{Yuli16} and \cite{Yuli9}. Now we return to 
small-area junctions which are
mostly discussed in this book and consider the influence of finite size on
the whole picture described above.
\subsection{Junction-localized oscillations}
We now concentrate our attention on the 100-10000 angstroms size
junction that are under experimental investigation now. The time of
electromagnetic excitation propagation along the whole junction corresponds to
a voltage in the region from ten to several hundred microvolts. The excitations
are practically not damped in this frequency region.

        There are basically two ways to connect the junction with the contact
wires: first, to interrupt the thin wire by the tunnel barrier; second, to form
this barrier by overlapping of the two films. The junction area corresponds to
the wire cross-section in the first case and may be much larger than the latter
in the second case.

        Consider now the electromagnetic excitation propagation along the
junction. The first and the second case correspond to two-dimensional and
three-dimensional geometries of the previous subsection, respectively. In the
simplest case of a rectangular junction the boundary conditions permit only
discrete values of $\vec q$. The modes with $\vec q \ne 0$ are
junction-localized oscillations that can be excited by electron tunneling. The
frequencies of these oscillations are simply $\omega_m = \Omega(\vec q_m)$,
where
$\vec q_m$ are permitted values of the wave vector. These frequencies evidently
remain discrete regardless of the actual junction form.

We can make use of the formulae in the previous subsection for the effect of
excitations on
the junction $I$-$V$ curve if we replace the integration over $\vec q$ by the
summation over discrete values $\vec q_m$:

\begin{equation}
\int {{\rm d}^2 q\over (2\pi)^2} \ \rightarrow \ {1\over S}
\sum_{\vec q_m}
\end{equation}
where $S$ is the junction area.
This yields the very simple expression for the effect at $eV \gg k_BT$

\begin{equation}
{\partial^2 I\over \partial V^2} R_T = E_c \sum_{\vec q_m} \frac{1}{V}
\delta\Big(eV-\hbar \omega(\vec q_n)\Big).
\label{eq:Mic33}
\end{equation}
Here, $E_c=e^2/2C$ is the charging energy of the whole tunnel capacitor
having the capacitance $C$. Eq.~(\ref{eq:Mic33}) is valid only if the speed of
excitations is
much smaller than the speed of light. Otherwise it should be multiplied by the
Lorentz factor of Eq.~(\ref{eq:Mic32}). Thus, every oscillation makes a
fingerprint on the
$I$-$V$ curve at the corresponding voltage.

        It is worth to  compare these results with the predictions of the
phenomenological theory. According to that one at voltages larger than the
inverse time of discharge through the leads attached to the junction we have a
linear $I$-$V$ curve with offset $E_c/e$. From (\ref{eq:Mic33}) one may see
that it is valid only if the voltage does not exceed the lowest energy of 
the oscillation spectrum.
At the voltages which are corresponding to the oscillation energies the junction
conductance is jumping. The magnitude of the jump can be determined from the
fact that the $I$-$V$ curve is gaining the additional offset  $E_c/e$ at this
point. If
this frequency is $n$-fold degenerate, the additional offset is $nE_c/e$. At
high voltages a large number of oscillations can be excited by the tunneling
electron and the asymptotic law is determined by the appropriate expression for
the infinite-area junction.

        If we take into account the oscillation damping and/or the finite
temperature, the jumps gain a finite width. The appropriate expressions one may
find in Ref.~\cite{Yuli16}.
       Therefore the offset obviously increases with increasing voltage. It
shows how the applicability of the phenomenology is restricted.

 Now we  finish by opening
\subsection{A gateway into networks}
In the previous subsection we have said nothing about the mode with $\vec
q = 0$. The existence of this mode is a straightforward consequence of charge
conservation: if the junction is included into some electrical circuit there
should be a possibility for charge to go out of the junction, and modes with
$\vec q_m \ne 0$ do not provide this possibility. In this zero mode the voltage
difference is constant along the whole junction. Therefore, the mode dynamics
does not depend upon the nearest junction environment but is determined by
the external circuit. The phenomenological equations derived in the previous
sections are completely valid for this dynamics, so that there is a gateway from
the microscopic world to the networks created by
human beings.\index{environment!electromagnetic|)}
%
%
%
%
 
\arraycolsep 5pt
\vspace{2\baselineskip}

\end{document}